\documentclass[a4paper,11pt]{report}
\usepackage[margin=2cm]{geometry}
\usepackage{amsmath}
\usepackage{braket}
\usepackage{quantikz}
\usepackage[normalem]{ulem}
\usepackage{environ}
\newcommand{\myqctmp}[2][0.25]{\Qcircuit @C=#2em @R=#1em @!R}
\NewEnviron{myqcircuit}[1][0.25]{\vcenter{\myqctmp[#1]{0.5} {\BODY}}}
\usepackage{amsmath}
\usepackage{amssymb}
\usepackage{mathtools}
\usepackage{nicefrac}
\usepackage{fancyvrb}
\RecustomVerbatimEnvironment{Verbatim}{Verbatim}{fontsize=\small}
\PassOptionsToPackage{hyphens}{url}
\usepackage[colorlinks,urlcolor=blue,linkcolor=black,citecolor=blue,breaklinks=true]{hyperref}
\usepackage[capitalise,nameinlink]{cleveref}
\usepackage[utf8]{inputenc}
\usepackage{pmboxdraw}
\usepackage{textcomp}
\usepackage{multicol}
\usepackage{multirow}
\usepackage{pifont}   
\usepackage{listings}
\usepackage{graphicx} 
\usepackage{tikz}
\usetikzlibrary{positioning, calc, decorations.pathmorphing, shapes.geometric,arrows.meta}
\usepackage{caption}
\usepackage{subcaption}
\usepackage{booktabs}
\usepackage{siunitx}
\usepackage[table]{xcolor}
\usepackage{diagbox}
\usepackage{subcaption}
\usepackage[style=numeric-comp, maxcitenames=1, maxbibnames=3, url=true, doi=true, isbn=false, sorting=nty,giveninits=true]{biblatex}
\usepackage{tcolorbox}  

\newtheorem{finding}{Finding}

\newtcolorbox{findingbox}{
  colback=gray!15,    
  colframe=gray!50,   
  sharp corners,      
  boxrule=0.5pt,      
  left=5pt, right=5pt, top=5pt, bottom=5pt, 
}

\makeatletter
\renewenvironment{finding}
{\par\refstepcounter{finding}%
 \begin{findingbox}%
 \textbf{Finding~\thefinding.} }%
{\end{findingbox}\par}
\makeatother

\newcommand{\perf}{\mathcal{P}}
\newcommand{\estimate}[1]{\textcolor{blue}{#1$^{?}$}}

\definecolor{colorA}{HTML}{FF6B6B}
\definecolor{colorB}{HTML}{4ECDC4}
\definecolor{colorC}{HTML}{45B7D1}
\definecolor{colorD}{HTML}{96CEB4}
\definecolor{colorE}{HTML}{FFEAA7}
\definecolor{colorF}{HTML}{DDA0DD}
\definecolor{timecolor1}{HTML}{c8f5c8}  
\definecolor{timecolor2}{HTML}{a0faa0}  
\definecolor{timecolor3}{HTML}{6cc46c}  
\definecolor{timecolor4}{HTML}{2f872f}  
\definecolor{timecolor5}{HTML}{065c06}  
\definecolor{timecolor6}{HTML}{f5f23b}  
\definecolor{timecolor7}{HTML}{f5aa3b}  
\definecolor{timecolor8}{HTML}{f5413b}  

\usepackage{tcolorbox}

\newcommand{\roundcell}[3][1.5mm]{%
  \tcbox[on line, boxsep=1pt, left=2pt, right=2pt, top=1pt, bottom=1pt,
         colback=#2, colframe=#2, arc=#1, boxrule=0pt]{#3}%
}

\usepackage{pgfplotstable}
\usepackage{pgfplots}
\pgfplotsset{compat=1.18}
\newcommand{\throughputplot}{%
\begin{tikzpicture}[trim axis right]
\begin{axis}[
  scale only axis,
  width=6cm,
  height=5.6cm,
  xmode=log,
  xmin=0.1,
  xmax=1000000000,
  ymin=-0.5,
  ymax=8.5,
  xlabel style={font=\small},
  xtick={0.1,1,10,100,1000,10000,100000,1000000,10000000,100000000,1000000000},
  xticklabels={$10^{-1}$,$10^0$,$10^1$,$10^2$,$10^3$,$10^4$,$10^5$,$10^6$,$10^7$,$10^8$,$10^9$},
  xtick style={font=\tiny},
  ytick=\empty,
  axis y line*=none,
  axis x line*=bottom,
 ]

\fill[blue!60] (0.1,8.2-0.3) rectangle (315.57600000000025,8.2+0.3);           
\fill[blue!60] (0.1,7.2-0.3) rectangle (315575.9999999999,7.2+0.3);           
\fill[blue!60] (0.1,6.2-0.3) rectangle (315575999.9999998,6.2+0.3);           
\fill[orange!70] (0.1,5.1-0.3) rectangle (3.155760000000001,5.1+0.3);           
\fill[orange!70] (0.1,4.1-0.3) rectangle (3155.760000000003,4.1+0.3);           
\fill[orange!70] (0.1,3.1-0.3) rectangle (3155760.0000000023,3.1+0.3);          
\fill[red!65] (0.1,2-0.3) rectangle (0.4632022216791023,2+0.3);          
\fill[red!65] (0.1,1-0.3) rectangle (463.20222167910237,1+0.3);          
\fill[red!65] (0.1,0-0.3) rectangle (463202.2216791027,0+0.3);           

\end{axis}
\end{tikzpicture}%
}

\usepackage[toc, acronym, nopostdot]{glossaries}
\usepackage{etoolbox}

\makeglossaries

\newcommand{\newglossaryentrywithacronym}[3]{
    \newglossaryentry{#1_gls}{
        name={#2},
        description={#3}
    }

    \newglossaryentry{#1}{
        type=\acronymtype,
        sort={#1},
        name={\glslink{#1_gls}{#1}},
        description={\glslink{#1_gls}{#2}},
        first={#2 (\glslink{#1_gls}{#1})}
    }
}

\newcommand{\newacronymentry}[2]{
    \newglossaryentry{#1}{
        type=\acronymtype,
        name={#1},
        description={#2},
        first={#2 (#1)}
    }
}

\newcommand*{\GlsToExclude}{CPU,GPU,NERSC}
\makeatletter 
\newcommand{\myglsreset}{%
  \glsresetall 
  \@for\tmp:=\GlsToExclude\do{%
    \glsunset{\tmp}%
  }%
}
\makeatother 

\newacronymentry{ODE}{Ordinary Differential Equation}
\newacronymentry{PDE}{Partial Differential Equation}
\newacronymentry{PIC}{Particle-In-Cell}
\newacronymentry{HPC}{High Performance Computing}
\newacronymentry{HEP}{High Energy Physics}
\newacronymentry{LQCD}{Lattice Quantum Chromodynamics}
\newacronymentry{QCD}{Quantum Chromodynamics}
\newacronymentry{DFT}{Density Functional Theory}
\newacronymentry{CPU}{Central Processing Unit}
\newacronymentry{GPU}{Graphics Processing Unit}
\newacronymentry{DOE}{U.S. Department of Energy}
\newacronymentry{SC}{Office of Science}
\newacronymentry{NERSC}{National Energy Research Scientific Computing Center}
\newacronymentry{NESAP}{NERSC Science Acceleration Program}
\newacronymentry{AI}{Artificial Intelligence}
\newacronymentry{IRI}{Integrated Research Infrastructure}
\newacronymentry{ASCR}{Advanced Scientific Computing Research}
\newacronymentry{BES}{Basic Energy Sciences}
\newacronymentry{BER}{Biological and Environmental Research}
\newacronymentry{FES}{Fusion Energy Science}
\newacronymentry{NP}{Nuclear Physics}
\newacronymentry{ML}{Machine Learning}
\newacronymentry{QRW}{Quantum Relevant Workload}
\newacronymentry{VQE}{Variational Quantum Eigensolver}
\newacronym
    [longplural={Variational Quantum Algorithms}, shortplural={VQAs}]
    {VQA}
    {VQA}
    {Variational Quantum Algorithm}
\newacronymentry{QC}{Quantum Computing}
\newacronym
   [longplural={Quantum Processing Units}, shortplural={QPUs}]
   {QPU}
   {QPU}
   {Quantum Processing Unit}

\newacronymentry{QCCD}{Quantum Charge-Coupled Device}
\newacronymentry{SPAM}{State Preparation and Measurement}
\newacronymentry{SSP}{Sustained System Performance}
\newacronymentry{SSI}{Sustained System Improvement}
\newacronymentry{THC}{Tensor Hypercontraction}
\newacronymentry{SQSP}{Sustained Quantum System Performance}

\newglossaryentrywithacronym{CNOT}{Controlled-NOT}{
    An controlled-NOT gate (also known as a controlled-$X$ gate) is a two-qubit entangling gate. It applies a NOT (or Pauli-$X$) operation on the target qubit when the control qubit is in the $\ket{1}$-state. Its action is summarized as $\ket{a}\ket{b} \mapsto \ket{a}\ket{b \oplus a}$.
}

\newglossaryentrywithacronym{QEM}{Quantum Error Mitigation}{
    Techniques to reduce the impact of errors in NISQ devices without the full overhead of quantum error correction. Methods include zero-noise extrapolation, probabilistic error cancellation, and symmetry verification. While these approaches cannot eliminate errors entirely, they can significantly improve computation accuracy at the cost of increased sampling or classical post-processing.
}

\newglossaryentrywithacronym{QEC}{Quantum Error Correction}{
    The process of protecting quantum information from decoherence and operational errors by encoding logical qubits into entangled states of multiple physical qubits. Error correction codes detect and correct errors without directly measuring the protected information, enabling fault-tolerant quantum computation when error rates fall below the code threshold.
}

\newglossaryentrywithacronym{QED}{Quantum Error Detection}{
     The process of identifying when errors have occurred in a quantum system without necessarily correcting them. Error detection requires fewer resources than full error correction and can be used to post-select successful runs or trigger re-computation. This approach is particularly useful for quantum communication and certain NISQ applications where modest error rates are acceptable.
}

\newglossaryentrywithacronym{PEC}{Probabilistic Error Cancellation}{
    A quantum error mitigation technique that reduces the effect of noise by running multiple circuit variants with inserted recovery operations. The results are combined with appropriate weights to cancel out error contributions statistically. While this method can significantly improve accuracy, it incurs an exponential sampling overhead that limits its applicability to moderate circuit depths.
}

\newglossaryentrywithacronym{FTQC}{Fault-Tolerant Quantum Computing}{
    A quantum computing paradigm where logical operations can be performed reliably despite the presence of errors in physical components. This requires quantum error correction codes, fault-tolerant gate implementations, and error rates below the code threshold to ensure that errors do not propagate faster than they can be corrected.
}

\newglossaryentrywithacronym{NISQ}{Noisy Intermediate Scale Quantum}{
    Noisy Intermediate-Scale Quantum devices, referring to current-generation quantum computers with 50-1000 qubits that lack full error correction. NISQ devices can potentially demonstrate quantum advantage for specific problems but are limited by noise, short coherence times, and restricted circuit depths, requiring specialized algorithms and error mitigation techniques.
}

\newglossaryentrywithacronym{qLDPC}{Quantum Low Density Parity Check}{
    A class of quantum error correction codes characterized by sparse parity check matrices, where each qubit participates in only a few syndrome measurements. These codes can achieve better encoding rates than surface codes while maintaining good error correction properties, though they typically require non-local connectivity that poses implementation challenges on current hardware.
}

\newglossaryentrywithacronym{GSEE}{Ground State Energy Estimation}{
    A fundamental quantum computing application that determines the lowest energy eigenvalue of a quantum system's Hamiltonian. This problem is central to quantum chemistry and materials science, as ground state energies determine molecular properties and reaction rates. Quantum algorithms like variational quantum eigensolver (VQE) and quantum phase estimation can solve this problem with potential advantages over classical methods.
}

\newglossaryentrywithacronym{QPE}{Quantum Phase Estimation}{
    A fundamental quantum algorithm that extracts eigenvalue information from unitary operators by encoding phase differences into measurable probability amplitudes. This algorithm underlies many quantum applications including Shor's factoring algorithm and quantum chemistry simulations, offering exponential speedup for certain problems but requiring deep circuits.
}

\newglossaryentrywithacronym{LCU}{Linear Combination of Unitaries}{
    An important algorithm in developing block encoding for arbitrary operators by first decomposing them in terms of unitary operators, such as the Pauli basis. In a second stage, the applications of prepare, select, and unprepare primitives allows for the construction of the block encoding of the target operator. 
     This algorithm can serve as the preparatory step before QSVT algorithm is applied.
}

\newglossaryentrywithacronym{QSVT}{Quantum Singular Value Transformation}{
    A unified framework for quantum algorithms that transforms singular values of block-encoded matrices through polynomial approximations. This technique generalizes many quantum algorithms including phase estimation, amplitude amplification, and Hamiltonian simulation, providing optimal query complexity for a broad class of matrix transformation problems.
}

\newglossaryentrywithacronym{QSP}{Quantum Signal Processing}{
    A framework for designing quantum algorithms by constructing polynomial transformations through sequences of rotation gates and signal operators. QSP provides optimal quantum algorithms for a wide range of problems by encoding desired polynomial functions into alternating sequences of signal operators and signal processing operators. This technique forms the foundation for quantum singular value transformation and achieves optimal query complexity for tasks including Hamiltonian simulation, matrix inversion, and amplitude amplification.
}

\newglossaryentrywithacronym{QIS}{Quantum Information Science}{
    The interdisciplinary field studying how quantum mechanical systems can be used to store, process, and transmit information. QIS encompasses quantum computing, quantum communication, quantum cryptography, and quantum sensing, exploring both fundamental limits of information processing and practical applications. The field combines physics, computer science, mathematics, and engineering to develop quantum technologies that can surpass classical limitations in computation, communication security, and measurement precision.
}

\newglossaryentry{qubit}{
    name={Qubit}, 
    text={qubit}, 
    description={
        The fundamental unit of quantum information, analogous to a classical bit but capable of existing in superposition states $\ket{0}$, $\ket{1}$, or any linear combination $\alpha\ket{0} + \beta\ket{1}$. Qubits enable quantum parallelism and entanglement, the key resources for quantum computational advantage, but require careful isolation from environmental noise to maintain coherence.
    }
}

\newglossaryentry{qutrit}{
    name={Qutrit}, 
    text={qutrit}, 
    description={
        A three-level quantum system that extends the qubit concept to a superposition of three basis states $\ket{0}$, $\ket{1}$, $\ket{2}$. Qutrits can provide advantages in certain quantum algorithms and error correction schemes by offering higher information density and more efficient gate decompositions, though they are generally more challenging to control and maintain than qubits.
    }
}

\newglossaryentry{physical qubit}{
    name={Physical Qubit}, 
    text={physical qubit}, 
    description={
        The fundamental hardware unit of quantum information in a quantum computer, such as a superconducting circuit, trapped ion, or quantum dot. Physical qubits are subject to decoherence and operational errors, typically maintaining quantum states for microseconds to seconds depending on the platform, necessitating error correction for practical quantum computing.
    }
}

\newglossaryentry{logical qubit}{
    name={Logical Qubit}, 
    text={logical qubit}, 
    description={
        An encoded qubit protected by quantum error correction, formed from multiple physical qubits working together to store quantum information redundantly. Logical qubits can maintain coherence far longer than individual physical qubits by detecting and correcting errors, enabling reliable quantum computation at the cost of increased resource requirements.
    }
}

\newglossaryentry{quantum circuit}{
    name={Quantum Circuit}, 
    text={quantum circuit}, 
    description={
        A sequence of quantum gates applied to qubits, representing a quantum algorithm or computation. Circuits are typically expressed as directed acyclic graphs where nodes represent gates and edges indicate qubit connections. Circuit depth, width, and gate count are key metrics that determine resource requirements and feasibility on a quantum computer.
    }
}

\newglossaryentry{quantum gate}{
    name={Quantum Gate}, 
    text={quantum gate}, 
    description={
        A reversible operation on one or more qubits that forms the basic building block of quantum circuits. Common gates include single-qubit rotations (X, Y, Z, Hadamard) and two-qubit entangling gates (CNOT, CZ). Gate fidelity, duration, and connectivity constraints determine the practical implementation of quantum algorithms on physical hardware.
    }
}

\newglossaryentry{code distance}{
    name={Code Distance}, 
    text={code distance}, 
    description={
        A measure of a quantum error correction code's robustness, defined as the minimum number of physical qubit errors required to cause a logical error. A code with distance $d$ can detect up to $d-1$ errors and correct up to $\lfloor(d-1)/2\rfloor$ errors. Higher code distances provide stronger error protection but require more physical qubits per logical qubit.
    }
}

\newglossaryentry{code threshold}{
    name={Code Threshold}, 
    text={code threshold}, 
    description={
        The critical physical error rate below which quantum error correction becomes beneficial, allowing logical error rates to decrease exponentially as more resources are added. Above this threshold, adding more error correction actually increases the logical error rate. For surface codes, the threshold is approximately 1\% error rate per operation.    
    }
}

\newglossaryentry{error syndrome}{
    name={Error Syndrome}, 
    text={error syndrome}, 
    description={
        The pattern of measurement outcomes from ancilla qubits that indicates which errors have occurred in a quantum system without directly measuring the data qubits. Syndrome extraction allows errors to be identified and corrected without collapsing the quantum state, forming the basis of quantum error correction protocols.
    }
}

\newglossaryentry{surface code}{
    name={Surface Code},
    text={surface code},
    description={
         The leading quantum error correction code for near-term implementation, arranging physical qubits on a 2D lattice where data and measurement qubits alternate in a checkerboard pattern. Surface codes require only nearest-neighbor interactions, tolerate error rates up to $\sim1\%$, and scale favorably, making them the primary choice for many quantum computing architectures despite their high qubit overhead.
    }
}

\newglossaryentry{shot}{
    name={Shot},
    text={shot},
    description={
        A single execution of a quantum circuit on a quantum computer, producing one measurement outcome from the probabilistic quantum state. Since quantum measurements are inherently probabilistic, multiple shots are typically required to estimate expectation values or probability distributions. The number of shots needed scales with desired statistical accuracy, with standard error decreasing as $\mathcal{O}(N_s^{-1/2})$ for $N_s$ shots.
    }
}

\newglossaryentry{qubitization}{
    name={Qubitization},
    text={qubitization},
    description={
         A quantum algorithm technique that encodes the eigenvalues of a Hamiltonian into the phases of a unitary operator, enabling efficient quantum simulation and eigenvalue estimation. This method block-encodes the Hamiltonian into a larger unitary matrix, allowing quantum phase estimation to extract eigenvalues with optimal query complexity. Qubitization often provides better scaling than Trotterization for Hamiltonian simulation problems.
    }
}

\newglossaryentry{Trotter}{
    name={Trotter},
    text={Trotter},
    description={
        A method for simulating quantum time evolution by decomposing the exponential of a sum of non-commuting operators into a product of simpler exponentials. While conceptually simple and widely used, Trotterization typically requires more gates than advanced methods like qubitization for achieving the same accuracy, though it has the advantage of requiring fewer ancilla qubits, making it more suitable for near-term quantum devices with limited qubit counts.
    }
}

\newglossaryentry{first quantization}{
    name={First Quantization},
    text={first quantization},
    description={
         A formulation of quantum mechanics where particles are treated as quantum mechanical objects with wavefunctions, while fields remain classical. In this approach, the Schr\"odinger equation describes the evolution of $N$-particle wavefunctions in configuration space, with the wavefunction amplitude giving the probability of finding particles at specific positions. While conceptually straightforward, first quantization becomes computationally challenging for many-body systems due to the exponential growth of the configuration space with particle number.
    }
}

\newglossaryentry{second quantization}{
    name={Second Quantization},
    text={second quantization},
    description={
         A formulation of quantum mechanics using creation and annihilation operators to describe many-body quantum systems in terms of occupation numbers rather than individual particle coordinates. This approach naturally handles variable particle numbers and identical particle statistics (bosons/fermions), making it essential for quantum chemistry and condensed matter physics. Second quantization provides the foundation for efficient quantum simulation algorithms, as fermionic or bosonic operators can be mapped to qubit operations through transformations like Jordan-Wigner or Bravyi-Kitaev.
    }
}
\usepackage[misc]{ifsym}

\begin{document}

\begin{titlepage}
    \noindent
    \begin{minipage}[t]{0.5\textwidth}
        \includegraphics[height=2cm]{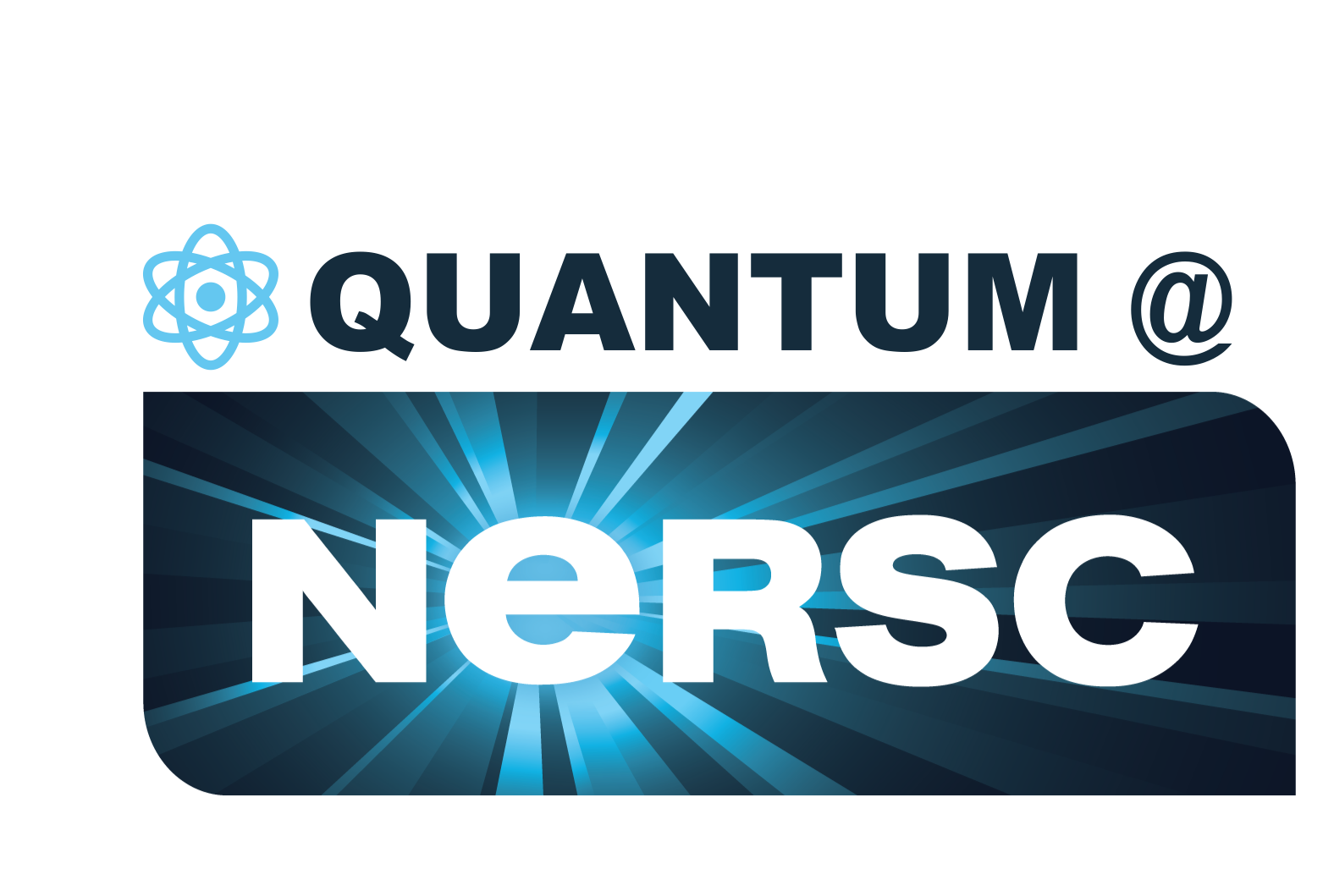}
    \end{minipage}%
    \begin{minipage}[t]{0.5\textwidth}
        \raggedleft
        {\large Report No. LBNL-2001699}
    \end{minipage}

    \vspace{3cm}

    \centering
    
    {\huge
     \textbf{
        Quantum Computing Technology Roadmaps and Capability Assessment for Scientific Computing
     }
    }\\[0.5cm]

    {\LARGE
    An analysis of use cases from the NERSC workload
    }\\[2cm]
    
    {
    \Large Daan Camps\hfill Ermal Rrapaj\hfill Katherine Klymko\hfill Hyeongjin Kim\hfill Kevin Gott\\
    \vspace{0.1cm}
    Siva Darbha\hspace{1cm} Jan Balewski\hspace{1cm} Brian Austin\hspace{1cm} Nicholas J. Wright
    }\\[1cm]

    {
    \Large \setlength{\baselineskip}{0pt}
    National Energy Research Scientific Computing Center (NERSC)\\
    \vspace{0.1cm}
    Lawrence Berkeley National Laboratory (LBNL)\\
    \vspace{0.2cm}
    Berkeley, CA, USA
    }\\[1cm]
    
    {\large\today}
    
    \vfill

    \begin{flushleft}
    \Letter \hspace{0.1cm} \texttt{\{dcamps;ermalrrapaj;baustin;njwright\}@lbl.gov}\hfill Version 1.0
    \end{flushleft}
    
\end{titlepage}

\begin{abstract}
    The National Energy Research Scientific Computing Center (NERSC), as the high-performance computing (HPC) facility for the Department of Energy's Office of Science, recognizes the essential role of quantum computing in its future mission. In this report, we analyze the NERSC workload and identify materials science, quantum chemistry, and high-energy physics as the science domains and application areas that stand to benefit most from quantum computers. These domains jointly make up over 50\% of the current NERSC production workload, which is illustrative of the impact quantum computing could have on NERSC's mission going forward.
    We perform an extensive literature review and determine the quantum resources required to solve classically intractable problems within these science domains.    
    This review also shows that the quantum resources required have consistently decreased over time due to algorithmic improvements and a deeper understanding of the problems. At the same time, public technology roadmaps from a collection of ten quantum computing companies predict a dramatic increase in capabilities over the next five to ten years. Our analysis reveals a significant overlap emerging in this time frame between the technological capabilities and the algorithmic requirements in these three scientific domains. We anticipate that the execution time of large-scale quantum workflows will become a major performance parameter and propose a simple metric, the~\gls{SQSP}, to compare system-level performance and throughput for a heterogeneous workload. 
\end{abstract}

\tableofcontents
\newpage

\chapter*{Executive Summary}
\label{chap:executive_summary}
\addcontentsline{toc}{chapter}{Executive Summary}

The \gls{NERSC} is charged with accelerating scientific discovery at the \gls{DOE} \gls{SC} through \gls{HPC} and data analysis. As the \gls{DOE} \gls{SC} mission \gls{HPC} user facility, \gls{NERSC} pushes the envelope of \gls{HPC} by strategically deploying new supercomputers on approximately a five year life cycle. %

\textbf{Quantum computers may fundamentally alter the \gls{HPC} landscape} and profoundly advance the \gls{DOE} \gls{SC} mission by the end of the decade. \gls{QC} is expected to \textit{(i)} lead to exponential speedups for important scientific problems that classical \gls{HPC} cannot solve, \textit{(ii)} do so in an energy-efficient manner that is decoupled from Moore's law, and \textit{(iii)} offer opportunities and challenges to develop new paradigms for \gls{HPC}-\gls{QC} integration. The analysis in this report shows that \textbf{at least 50\% of the \gls{DOE} \gls{SC} production workload, which aggregates the computational needs of more than 12,000 \gls{NERSC} users across the \gls{DOE} landscape, is allocated to solving problems for which exponential improvements in terms of problem complexity, time-to-solution, and/or solution accuracy are anticipated with the advent of \gls{QC}.} %

We focus on \textbf{three major scientific domains} of strategic relevance to \gls{NERSC}'s mission ---\textit{materials science, quantum chemistry, and high energy physics}--- and collect over 140 end-to-end resource estimates for benchmark problems from the scientific literature. We draw \textbf{three main conclusions} from this analysis. First, \textbf{Hamiltonian simulation} of quantum mechanical systems, also known as real-time evolution, is \emph{the} key quantum algorithmic primitive upon which many known applications and speedups rely. 
Second, the \textbf{end-to-end resource estimates differ significantly by scientific domain}. Condensed matter physics offers a prime candidate for early quantum advantage as model problems relevant to materials science are more easily mapped to quantum computers and thus the least quantum resources are required at a given problem scale. Problems in quantum chemistry, including electronic structure, require an intermediate number of resources as the encoding overhead is larger. High energy physics problems, including lattice gauge theories, show the most dramatic scaling due to the encoding overhead encountered in including both fermionic and gauge degrees of freedom. Third, we show that, while asymptotically optimal algorithms exist, constant factor \textbf{algorithmic improvements over the past five years have reduced the quantum resources} needed to compute the ground state energy of a strongly-correlated molecule \textbf{by orders of magnitude}. We expect this trend to continue going forward and advance applications in all aforementioned domains. We show that current estimates of quantum resources required for scientific quantum advantage start at about 50 to a 100 \glspl{qubit} and a million \gls{quantum gate} operations and go up from there. %

In addition, we analyzed and combined \textbf{public technology roadmaps} from ten different quantum vendors and observe that all vendors \textbf{predict an exponential increase in quantum computer performance over the next decade by up to nine orders of magnitude}. When realized, this will unlock unprecedented capabilities that will accelerate scientific discovery at \gls{DOE} \gls{SC} and enable breakthroughs that were previously out of reach. %

\begin{figure}[ht!]
\begin{center}
\includegraphics[width=0.8\linewidth]{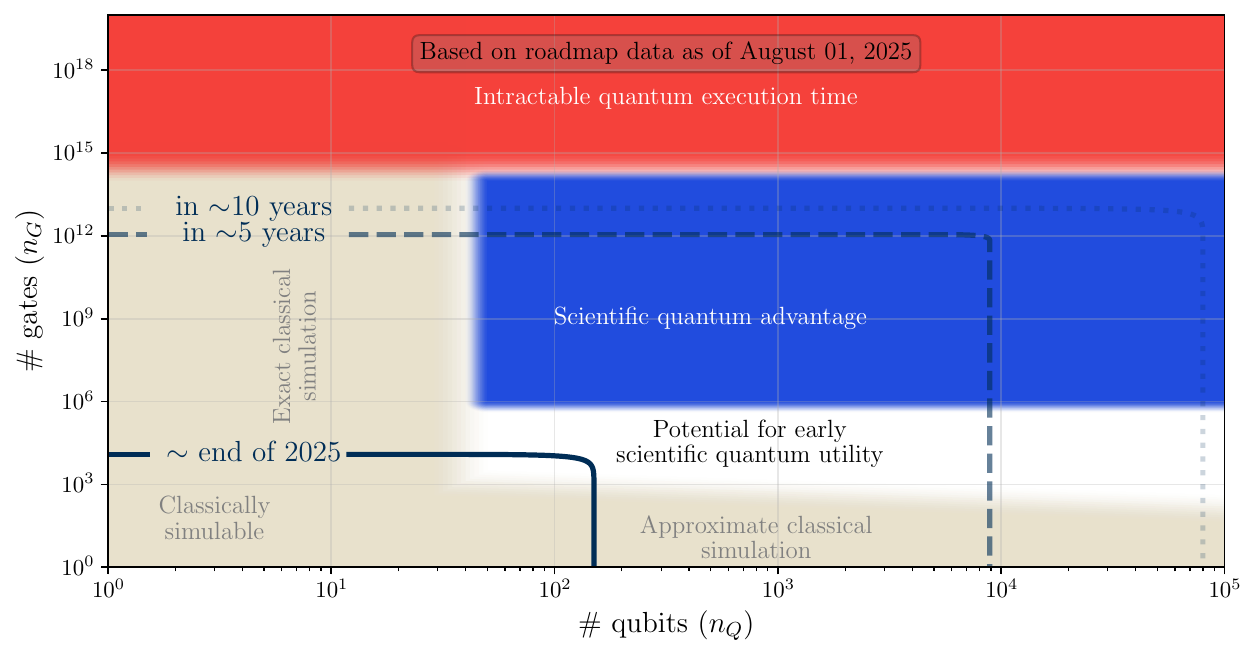}
\end{center}
\caption{Schematic overview of quantum resources in terms of \gls{quantum gate} count (y-axis) and \gls{qubit} count (x-axis). The figure shows 4 different regions of \gls{quantum circuit} volume. (1) \glspl{quantum circuit} in the light yellow region are considered classically simulable as they are either too narrow (i.e.~at low \gls{qubit} count) and can be simulated exactly, or too shallow (i.e.~low \gls{quantum gate} depth) and can be simulated with approximate methods such as tensor networks. (2) The region in blue starting at 50 qubits and 1 million quantum gates is the area where we expect scientific quantum advantage of relevance to \gls{NERSC} based on the resource estimate data we collected. (3) The intermediate region shown in white might offer early quantum advantage but we have not identified resource estimates in this area. (4) We label \gls{quantum gate} depths larger than $10^{14}$ in red to indicate that we expect this region to be unfeasible in the next ten years based on impractical quantum execution time estimates. The lines demarcate the highest performance quantum computers expected by the end of 2025, in five years from today, and in ten years from today.}
\label{fig:roadmap_regions}
\end{figure}

\Cref{fig:roadmap_regions} summarizes the vendor roadmap milestones by showing the highest performance \gls{QC} systems expected at the end of 2025, in five years and ten years from today. The anticipated technological progress shows excellent overlap with the application requirements for known scientific quantum advantage of interest to \gls{DOE} \gls{SC} mission (area shown in blue), with potential for early scientific quantum utility within the next two years (area shown in white). %

Finally, we discuss execution time estimates for quantum applications at scientific advantage scale and describe eight timescales ---from less than a second up to more than a year--- and their relevance to end users and system administrators. Quantum computer clock speed estimates presented in the literature range from sub-kHz up to GHz speeds and will determine the execution time. Based on the available data, \textbf{we construct a simple model, called the \gls{SQSP}, to estimate the throughput} for nine possible quantum computer specifications on a prototypical workload of eight applications. Our model shows that the number of applications that can be run in a one year window can be as low as one and go up to tens of millions, with \textbf{most scenarios predicting a throughput of hundreds to (tens of) thousands of applications per year}. Our methodology can be helpful to \textbf{inform decision-making processes for \gls{QC} system procurements}. %

\newpage

\chapter{Introduction}
\label{chap:intro}
\myglsreset

\gls{NERSC}~\cite{nersc}, located at Lawrence Berkeley National Laboratory, is the \gls{DOE} \gls{SC} mission \gls{HPC} facility. An ever-growing community, currently made up of over 12,000 scientists across the \gls{DOE} research complex, makes use of leading-edge compute resources deployed at \gls{NERSC}. \gls{NERSC} has an outstanding track record of successfully bringing advanced compute technologies to a large user community by supporting pathfinding projects such as the \gls{NESAP}~\cite{nesap}, and by providing comprehensive training and support. %

To remain at the leading-edge of computing, \gls{NERSC} has strategically deployed new supercomputers on approximately a five year life cycle. The current \emph{Perlmutter} system~\cite{perlmutter}, deployed in 2021, is the ninth generation \gls{NERSC} supercomputer and the first to combine a heterogeneous architecture offering both \gls{CPU} and \gls{GPU} compute nodes. Perlmutter has advanced \gls{NERSC}'s ability to support  numerical modeling, simulation, and data analysis. Furthermore, Perlmutter has unlocked new capabilities for \gls{AI} for scientific applications, a trend that \gls{NERSC} expects to expand further with its tenth generation system. The NERSC-10 system, named \emph{Doudna}~\cite{doudna} in honor of Dr.~Jennifer Doudna, was announced in May 2025 and is expected to become available to users in the 2026 time frame. Doudna will provide at least 10$\times$ the performance of Perlmutter, and fundamentally expand the AI and workflow-based compute capabilities at \gls{NERSC}. Doudna will offer real-time data processing to experimental facilities that rely on \gls{NERSC} for their compute needs, seamlessly linking to \gls{DOE}'s experimental and observational scientific user facilities as part of the \gls{IRI}~\cite{iri}. %

To ensure that \gls{NERSC} continues to fulfill its central role in the \gls{DOE} \gls{SC} mission as we move toward the 2030s, it is essential that \gls{NERSC} understands the applicability of quantum computing to its mission. 
Concurrently, a 2023 \gls{DOE} workshop report~\cite{doe_brn_qc} identifies critical research priorities for advancing quantum technology toward practical utility in scientific applications. Drawing from a decade of \gls{DOE} investments in \gls{QC} software and hardware, the report outlines a grand challenge to ``\emph{demonstrate a rigorously quantifiable, end-to-end quantum advantage relative to state-of-the-art classical counterparts, particularly for problems with practical or scientific significance for which asymptotic exponential quantum advantages have been established},'' along with five priority research directions (respectively in software, algorithms, benchmarking, error handling, and networks) that require collaborative efforts among computer scientists, mathematicians, and physicists to achieve advances across the full technology stack. The workshop emphasized ``\emph{the need for end-to-end demonstrations of disruptive quantum advantages for scientific applications, positioning \gls{QC} as a transformative technology that will complement and potentially revolutionize \gls{HPC} capabilities for \gls{DOE}'s scientific mission areas}.'' %

Building on this research-focused report, \gls{DOE} proposed a~\gls{QIS} Applications Roadmap~\cite{DOE_roadmap} in 2024 that outlines a strategic vision for \gls{QC} development that positions \gls{DOE} at the forefront of this transformative technology. The roadmap charts a clear progression through four distinct eras over an estimated 20 year window: from current \gls{NISQ} devices and small demonstrations of \gls{QEC} (0-5 years), to small quantum computers with \gls{QEC} (5-10 years), large quantum computers with \gls{QEC} (10-20 years), and finally very large \gls{FTQC} (20+ years). Key technology milestones include demonstrating 1,000 \glspl{physical qubit} with error rates ten times below \gls{code threshold}, scaling to 10,000 \glspl{physical qubit} while maintaining low error rates, and developing quantum interconnects for modular architectures. The roadmap highlights promising application areas including chemistry and materials science, where quantum simulations could eliminate uncontrolled approximations that afflict classical methods and enable revolutionary scientific discoveries. The report says that within the next five years, calculations on specific molecules or materials that are on the boundary of classical tractability might cross that threshold, potentially augmented by \gls{HPC} resources. \gls{DOE}'s national laboratories and user facilities are uniquely positioned to lead this effort through interdisciplinary collaborative research, software stack development, and the establishment of widely accessible quantum user facilities that will accelerate the development and deployment of quantum computers for these breakthrough scientific applications. %

Consequently, \gls{NERSC} sees \gls{QC} as a crucial and rapidly growing pillar of its ten year strategic plan for 2024-2034~\cite{nersc_strat}. These reports, roadmaps and strategies are motivated by the observation that quantum computers offer a fundamentally different paradigm of computing. By leveraging the principles of quantum mechanics, they have the potential to solve certain problems asymptotically faster than any computer based on the principles of classical physics is expected to. An important caveat is that proving that a quantum algorithm is faster than any classical algorithm is an extremely challenging task. Instead, researchers typically compare quantum algorithms against the best classical methods \emph{currently known} -- arguably, the strongest evidence of a exponential speedup using this approach exists for Shor's prime factoring algorithm~\cite{Schor1,Schor}. One significant drawback of this approach is that if a superior classical algorithm is discovered later, claimed quantum advantages may disappear. Key areas (besides Shor's algorithm) for which exponential asymptotic quantum speedups are expected to be exponential include simulation of quantum system dynamics~\cite{Zalka_1996,wiesner1996,childs04,aharonov2003}, computing static properties of quantum systems~\cite{kitaev1995}, and solving linear systems of equations~\cite{harrow09}. Other areas, such as search and optimization could see low-degree polynomial speedups on quantum computers in certain regimes~\cite{grover}. We refer to~\cite{QuantumAlgorithmZoo,dalzell2023} for an overview of known quantum algorithms and applications. %

In the context of these trends, this report examines how \gls{QC} is expected to impact \gls{NERSC}'s mission and workload over the next ten years. We synthesize findings from academic literature with the forward-looking projections of \gls{QC} companies. We observe that the projections and roadmaps that have been proposed by industry follow an accelerated trajectory compared to \gls{DOE}'s roadmap~\cite{DOE_roadmap} and predict very large-scale systems to become available within ten years. We remark that our endeavor necessitates various assumptions and approximations, which, compounded by the inherent uncertainty of technological forecasting, leads to a significant degree of uncertainty in our estimates. As this field is characterized by rapid evolution, we will continue to monitor its development, refine our analysis and adjust our strategy accordingly.
Therefore, we intend this to be a living document, and plan to release updated versions when appropriate.

The remainder of this report is structured as follows. We begin by reviewing the workload that \gls{NERSC} users are running on the current Perlmutter system in~\Cref{chap:nerscworkload} and identify the three main scientific domains of interest for which quantum computers will deliver speedups. \Cref{chap:qc_intro} provides a primer on key \gls{QC} concepts that are mentioned throughout the report. \Cref{chap:apps} collects and reviews the resources required to solve more than 100 target problems across the scientific domains of interest to \gls{NERSC} that were identified in~\Cref{chap:nerscworkload}. We base our analysis on estimates available in the scientific literature that go beyond the asymptotic speedups and account for the end-to-end resource requirements. We distill these results into a single figure of merit which captures the space-time cost of a quantum application. Next, in \Cref{chap:vendorroadmaps}, we analyze various quantum technology roadmaps that different vendors have put forward for the next decade. We uniformize the different vendor milestones with respect to the previously introduced figure of merit. This allows us to bound and compare the capabilities of forthcoming quantum computers. We also compare these bounds to the resources required to solve applications in the various scientific domains and observe a significant overlap emerging in the next five to ten years. We comment on the execution time as a key constraint for large-scale quantum applications in~\Cref{chap:exec_time_throughput} and present a simple model, called the~\gls{SQSP} metric, that allows us to compare system-level performance and throughput for a heterogeneous workload. \Cref{chap:conclusion} concludes the report by summarizing the main findings. A \nameref{sec:acronyms} and \nameref{sec:glossary} is included after the conclusion.%

\chapter{NERSC Workload}
\label{chap:nerscworkload}
\myglsreset

The workload at \gls{NERSC} is inherently varied and complex with a large community of users coming from six different program offices in the \gls{DOE} \gls{SC}, including \gls{ASCR}, \gls{BES}, \gls{BER}, \gls{FES}, \gls{HEP}, and \gls{NP}. %

\gls{NERSC} users develop and run a breadth of scientific computing codes at a large range of scales and \gls{GPU}-readiness levels. This means that \gls{NERSC} compute resources must be versatile as well as highly performant. The Perlmutter system provided a \gls{CPU} as well as a \gls{GPU} partition in an era when multiple relevant scientific codes were still in the process of being ported to \gls{GPU}s. At the same time, \gls{NERSC} provided extensive support to enable many more codes to benefit from \gls{GPU} acceleration. %
 
\begin{figure}[htp]
    \centering
    \includegraphics[width=0.75\linewidth]{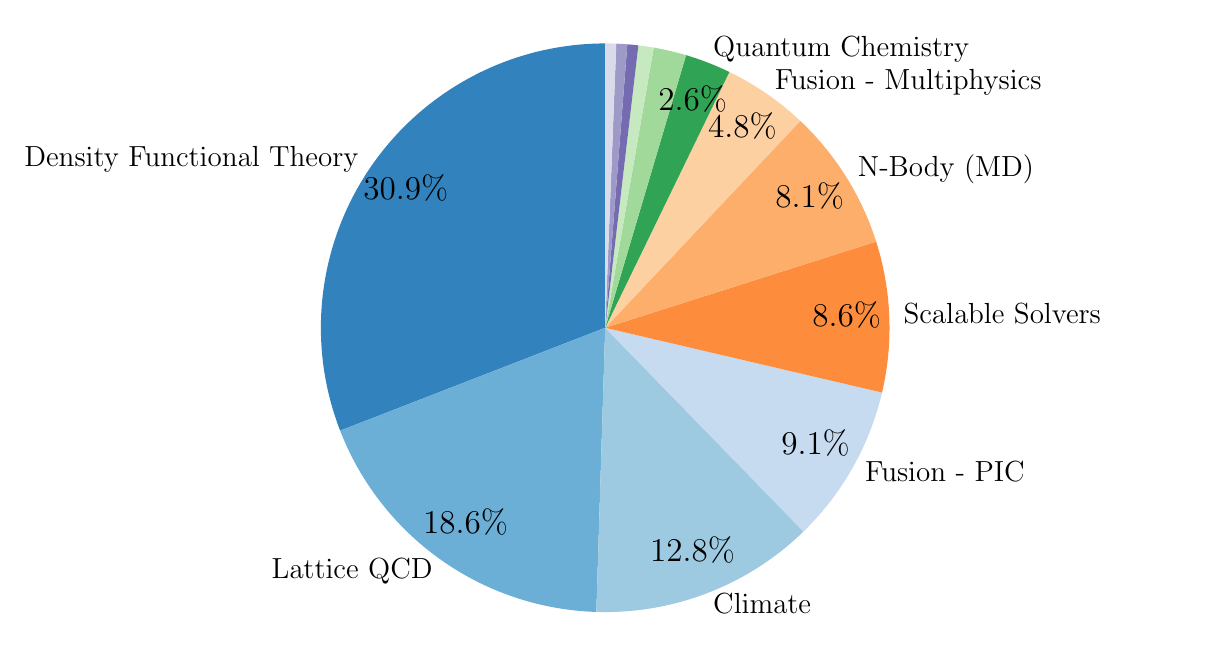}
    \caption{Distribution of the compute cycles used at \gls{NERSC} across different categories of algorithms and application areas.}
    \label{fig:nersc-workload}
\end{figure}

\Cref{fig:nersc-workload} summarizes the resource allocation for various workloads on \gls{NERSC} systems.
Thanks to \gls{NERSC}'s ongoing efforts, for example through the aforementioned \gls{NESAP} program, most of Perlmutter compute performance is stemming from the \gls{GPU} partition.

\gls{DFT} stands as a cornerstone of computational \textbf{materials science} and \textbf{quantum chemistry} and takes more than 30\% of the yearly cycles at \gls{NERSC}. It is a computational method to approximate the electronic structure of complex atomic and molecular systems, and achieves this at reasonable computational cost~\cite{kohn1965self,parr1989density}. Its strengths lie in its efficiency, allowing for the study of large systems, and its ability to provide valuable insights into ground-state properties. However, the accuracy of \gls{DFT} is often limited by the choice of exchange-correlation functional, which can fail to capture essential electron correlation effects required for describing van der Waals interactions~\cite{klimevs2009chemical}, electronic band gaps~\cite{mori2008localization}, and charge-transfer excitations~\cite{dreuw2004failure}. These limitations can be partially addressed by using more sophisticated exchange-correlation functionals~\cite{perdew2001jacob} which incorporate exact exchange or perturbative correlation to improve accuracy, though at a significantly higher computational cost. However, even with these enhancements, \gls{DFT} remains insufficient for solving the most challenging problems in quantum chemistry, such as strongly correlated systems or multi-reference electronic states, where more advanced wavefunction-based methods are typically required~\cite{cohen2008insights}. Quantum chemistry codes that go beyond \gls{DFT} make up another 2.6\% of the cycles. This is where quantum computers hold significant promise. By leveraging quantum mechanics, these machines could potentially tackle the exponentially complex many-body problem inherent in electronic structure calculations, offering the prospect of more accurate exchange-correlation functionals and enabling the simulation of strongly correlated materials at the precision of full configuration interaction (CI), thus overcoming key weaknesses of traditional \gls{DFT} while scaling exponentially better than classical approaches.

\textbf{Nuclear and high energy physics} make up over 18\% of the workload. Most of their computational time is dedicated to \gls{LQCD} runs. This is a computational method for studying the strong force and its associated particles, hadrons, from first principles. Its strength lies in its ability to non-perturbatively calculate properties of hadrons, such as their masses and decay constants, by discretizing spacetime onto a lattice. However, \gls{LQCD} is both computationally demanding and constrained to imaginary time, requiring vast resources to simulate realistic quark masses and large volumes, thereby limiting the precision and scope of calculations. The demands are particularly severe for computing complex observables like scattering amplitudes and real-time dynamics, since the computation is performed on Euclidean space, and thus cannot access these quantities directly. Additionally, the sign problem poses a significant barrier to the applicability of \gls{LQCD} to dense systems at finite chemical potential~\cite{Goy:2017,NAGATA:2022}. Quantum computers promise to revolutionize this field by efficiently tackling these computational challenges. Their inherent ability to simulate quantum systems could overcome the sign problem and enable simulations at physical quark masses and large volumes, ultimately leading to more accurate and comprehensive understanding of strongly-interacting matter.

In addition, there are smaller pieces of the workload, like \textbf{N-body molecular dynamics and fusion using \gls{PIC} methods}, that also stand to benefit from quantum computers. As such, more than half of the computational resources on Perlmutter are dedicated to solving quantum many-body problems, highlighting the enormous potential impact and benefit quantum computers can have on \gls{NERSC}'s mission. Collectively, we refer to all \gls{DOE} \gls{SC} mission applications that stand to benefit from large-scale \gls{QC} as the \emph{\gls{QRW}}.

We summarize this section with the following finding:
\begin{finding}
At least 50\% of \gls{NERSC} compute resources are spent on solving quantum mechanical problems relevant to \textit{materials science, quantum chemistry}, and \textit{nuclear and high energy physics}, primarily using \gls{DFT} and \gls{LQCD} codes. Quantum computers naturally have the potential to accelerate scientific discovery in these areas as they do not have to rely on the same approximations as classical algorithms to solve computational problems. We refer to these applications as the~\gls{QRW}.
\end{finding}

\chapter{A Primer on Quantum Computing}
\label{chap:qc_intro}
\myglsreset

\gls{QC}, like many scientific fields, uses its own jargon. While this enables efficient exchange of ideas among experts, it raises the bar for non-experts to understand those same ideas. In an effort to make this report broadly accessible, we provide a primer on the key \gls{QC} concepts we refer to in the remainder of this text. Our goal is to offer a broad, high‑level perspective, not a precise or comprehensive account. For detailed overviews, we refer the interested reader to~\cite{Nielsen2010,watrous2025}. Readers already familiar with \gls{QC} may consider skipping this chapter and move directly on to~\cref{chap:apps}.

\section{On qubits, gates, and circuits}
\label{sec:qubits_intro}

\gls{QC} represents a profound shift in how information can be processed, harnessing the principles of quantum mechanics to tackle problems beyond the reach of classical machines. At its foundation is the \gls{qubit}, the quantum analogue of the classical bit. While a classical bit exists strictly as 0 or 1, a \gls{qubit} can occupy a superposition of both states, $\alpha \ket{0} + \beta \ket{1}$. When \glspl{qubit} become entangled, the state of one is intrinsically correlated with the state of another, no matter the distance between them—a property that underpins much of the power of \gls{QC}. %

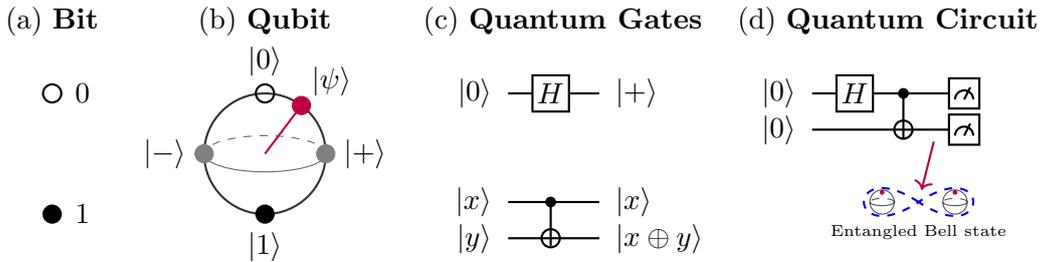
\begin{figure}[htp]
    \centering
    \begin{tikzpicture}[scale=0.8]

\def\titleY{4.5}
\def\topStateY{3.3}
\def\bottomStateY{1.3}
\def\centerY{2.3}  

\begin{scope}[shift={(0,0)}]
\node at (0, \titleY) {(a) \textbf{Bit}};
\draw[thick] (0, \topStateY) circle (0.15);
\node at (0.5, \topStateY) {$0$};
\filldraw[black] (0, \bottomStateY) circle (0.15);
\node at (0.5, \bottomStateY) {$1$};
\end{scope}

\begin{scope}[shift={(3.5,0)}]
\node at (0, \titleY) {(b) \textbf{Qubit}};

\begin{scope}[scale=1, shift={(0,\centerY)}] 
\draw[thick, opacity=0.8] (0,0) circle (1);
\draw[opacity=0.6] (-1,0) arc (180:360:1 and 0.3);
\draw[dashed, opacity=0.6] (1,0) arc (0:180:1 and 0.3);

\draw[thick] (0,1) circle (0.15);
\node[above, yshift=3pt] at (0,1) {$|0\rangle$};

\filldraw[black] (0,-1) circle (0.15);
\node[below, yshift=-3pt] at (0,-1) {$|1\rangle$};

\filldraw[gray] (1,0) circle (0.15);
\node[right, xshift=3pt] at (1,0) {$|+\rangle$};

\filldraw[gray] (-1,0) circle (0.15);
\node[left, xshift=-3pt] at (-1,0) {$|-\rangle$};

\filldraw[purple] (0.6,0.8) circle (0.15);
\node[above right] at (0.6,0.8) {$|\psi\rangle$};

\draw[thick, purple, -] (0,0) -- (0.6,0.8);
\end{scope}
\end{scope}

\begin{scope}[shift={(7.5,0)}]
\node at (0.9, \titleY) {(c) \textbf{Quantum Gates}};

\draw[thick] (0, \topStateY) -- (1.5, \topStateY);
\node[anchor=east] at (-0.1, \topStateY) {$|0\rangle$};
\node[anchor=west] at (1.6, \topStateY) {$|+\rangle$};
\draw[thick,fill=white] (0.4, 3) rectangle (1.0, 3.6);
\node at (0.7, \topStateY) {$H$};

\draw[thick] (0, 1.5) -- (1.5, 1.5);
\draw[thick] (0, 0.9) -- (1.5, 0.9);
\node[anchor=east] at (-0.1, 1.5) {$|x\rangle$};
\node[anchor=west] at (1.6, 1.5) {$|x\rangle$};
\node[anchor=east] at (-0.1, 0.9) {$|y\rangle$};
\node[anchor=west] at (1.6, 0.9) {$|x \oplus y\rangle$};
\filldraw[black] (0.7, 1.5) circle (0.08); 
\draw[thick] (0.7, 1.5) -- (0.7, 0.775);
\draw[thick] (0.7, 0.9) circle (0.15);  
\end{scope}

\begin{scope}[shift={(12.5,0)}]
\node at (1.25, \titleY) {(d) \textbf{Quantum Circuit}};

\draw[thick] (0, \topStateY) -- (2.5, \topStateY);
\draw[thick] (0, \topStateY-0.6) -- (2.5, \topStateY-0.6);

\node[anchor=east] at (-0.1, \topStateY) {$|0\rangle$};
\node[anchor=east] at (-0.1, \topStateY-0.6) {$|0\rangle$};

\draw[thick, fill=white] (0.4, \topStateY-0.3) rectangle (1.0, \topStateY+0.3);
\node at (0.7, \topStateY) {$H$};

\filldraw[black] (1.5, \topStateY) circle (0.08);
\draw[thick] (1.5, \topStateY) -- (1.5, \topStateY-0.75);
\draw[thick] (1.5, \topStateY-0.6) circle (0.15);

\begin{scope}[shift={(2.25,0)}]
\draw[thick,fill=white] (0, \topStateY-0.25) rectangle (0.5, \topStateY+0.25);
\draw[thick] (0.1, \topStateY-0.1) arc (180:0:0.15);
\draw[thick] (0.25, \topStateY-0.1) -- (0.35, \topStateY+0.1);
\draw[thick,fill=white] (0, \topStateY-0.85) rectangle (0.5, \topStateY-0.35);
\draw[thick] (0.1, \topStateY-0.7) arc (180:0:0.15);
\draw[thick] (0.25, \topStateY-0.7) -- (0.35, \topStateY-0.5);
\end{scope}

\begin{scope}[shift={(1.75,1.5)}]
\draw[thick, blue, dashed] (0,0) .. controls (-1.2,0.8) and (-1.2,-0.8) .. (0,0) 
                          .. controls (1.2,-0.8) and (1.2,0.8) .. (0,0);
                          
\begin{scope}[shift={(-0.6,0)}, scale=0.2]
\draw[opacity=0.7] (0,0) circle (1);
\draw[opacity=0.7] (-1,0) arc (180:360:1 and 0.3);
\draw[dashed, opacity=0.7] (1,0) arc (0:180:1 and 0.3);
\filldraw[purple] (0,0.8) circle (0.15);
\end{scope}

\begin{scope}[shift={(0.6,0)}, scale=0.2]
\draw[opacity=0.7] (0,0) circle (1);
\draw[opacity=0.7] (-1,0) arc (180:360:1 and 0.3);
\draw[dashed, opacity=0.7] (1,0) arc (0:180:1 and 0.3);
\filldraw[purple] (0,0.8) circle (0.15);
\end{scope}

\node[font=\tiny] at (0, -0.5) {Entangled Bell state};
\end{scope}

\draw[->, thick, purple] (2, 2.5) -- (1.8, 1.7);
\end{scope}

\end{tikzpicture}    
    \caption{(a) A bit  can be either 0 or 1; while (b) a qubit can be in a linear combination (or superposition) of $\ket{0}$ and $\ket{1}$ and its state space can be visualized by a Bloch sphere; (c) quantum gates modify the state of one or more qubits, the top panel shows a single-qubit Hadamard gate applied to a qubit in the $\ket{0}$ state and transforms it to $\ket{+}$, the bottom panel shows a two-qubit \gls{CNOT} gate that leaves the control qubit unchanged and applies a NOT to the target qubit conditioned on the state of the control; and finally (d) quantum circuits are combinations of quantum gates and measurement operations, the figure shows a two-qubit circuit
    that consists of a Hadamard gate and CNOT gate and prepares a state known as a Bell state.}
    \label{fig:overview_figure}
\end{figure}

Computation in a quantum system is achieved through \glspl{quantum gate}, which are reversible operations that transform \gls{qubit} states. These gates are assembled into \glspl{quantum circuit}, sequences of gate operations that together implement a computation. The performance and complexity of such circuits are often described in terms of \gls{quantum circuit} volume, a metric that accounts for both the number of qubits $n_Q$ and the number of gates $n_G$ of the computation.  A common characteristic of all quantum circuits is that at the end of the quantum computation (i.e. after running a \gls{quantum circuit}), some or all of the qubits are measured and classical information is extracted in the form of a bitstring that corresponds to the states in which the qubits were measured. This procedure is often referred to as taking a \gls{shot} of a quantum circuit. \Cref{fig:overview_figure} summarizes the notions of (qu)bits, quantum gates and quantum circuits (incl. quantum measurements) using a common circuit diagram notation~\cite{Nielsen2010}.

In practical hardware, \glspl{qubit} are physical devices prone to errors. Current quantum computers have up to a few hundred \glspl{physical qubit} and offer imperfect control over those qubits. Consequently, various sources of noise, including due to~\gls{SPAM}, imperfect gate operations and decoherence, limit the performance of quantum computers. This class of quantum systems has previously been dubbed \gls{NISQ} devices~\cite{Preskill2018}. When using a \gls{NISQ} device, one typically maps the \glspl{qubit} and \glspl{quantum gate} directly onto error-prone \glspl{physical qubit}. In the \gls{NISQ} paradigm, gates that entangle two \glspl{qubit}, such as the \gls{CNOT} gate (see~\cref{fig:overview_figure}), typically incur the largest error and consequently are a dominant factor determinig \gls{quantum circuit} performance. The capabilities of \gls{NISQ} hardware and applications can be extended by leveraging \gls{QEM} techniques~\cite{RevModPhys.95.045005} (more details in~\cref{sec:errors}) which typically requires taking additional samples on the quantum computer (which is also known as the sampling overhead for \gls{QEM}) to improve the quality of the result. However, there are fundamental limits on the level of improvement that is achievable and the sampling overhead typically scales exponentially~\cite{Takagi2022}. %

For large-scale, reliable computation, sufficiently good \glspl{physical qubit} can be grouped into \glspl{logical qubit} -- error-protected units formed by encoding information across many \glspl{physical qubit}. This process is central to \gls{FTQC} in which computations can proceed reliably despite the presence of noise due to an approach known as \gls{QEC}~\cite{Lidar2013, Terhal2015}. 
Fault tolerance can be implemented more easily for some gates than for others.
Clifford gates --including the Hadamard, Phase, and aforementioned \gls{CNOT} gates-- form a set that is relatively straightforward to realize. Non-Clifford gates, such as the $T$ gate or the three-qubit Toffoli gate (a quantum analog of the classical AND gate), extend the Clifford set to achieve universal quantum computation but are significantly more resource-intensive in an \gls{FTQC} setting. Implementing non-Clifford gates typically relies on the preparation of high-fidelity \emph{resource states\footnote{These are called resource states as they are consumed during the computation.}} also known as magic states. Examples include $\ket{T}$- and $\ket{CCZ}$-states which respectively can be used to implement $T$ and Toffoli gates in a fault-tolerant manner using a Clifford circuit~\cite{PhysRevA.62.052316}. High-fidelity magic states are costly to prepare, requiring a procedure known as magic state distillation and generated by magic state factories. Recent progress has significantly reduced the cost of producing magic states over a substantial range of errors (see: magic state cultivation~\cite{Gidney:2024}), nonetheless it remains a major cost for fault-tolerant quantum computation. %

Analyzing (and optimizing) the end-to-end cost of running a computation on a \gls{FTQC} system forms the topic of the field known as quantum resource estimation. In light of our previous discussion, the number of non-Clifford gates (e.g. $T$, Toffoli or continuous-angle rotation gates) is typically the most important metric considered in quantum resource estimation. In~\cref{chap:apps}, we collect and present numerous resource estimates for quantum computations within the physical sciences. %

\section{Handling errors: quantum error mitigation and correction}
\label{sec:errors}

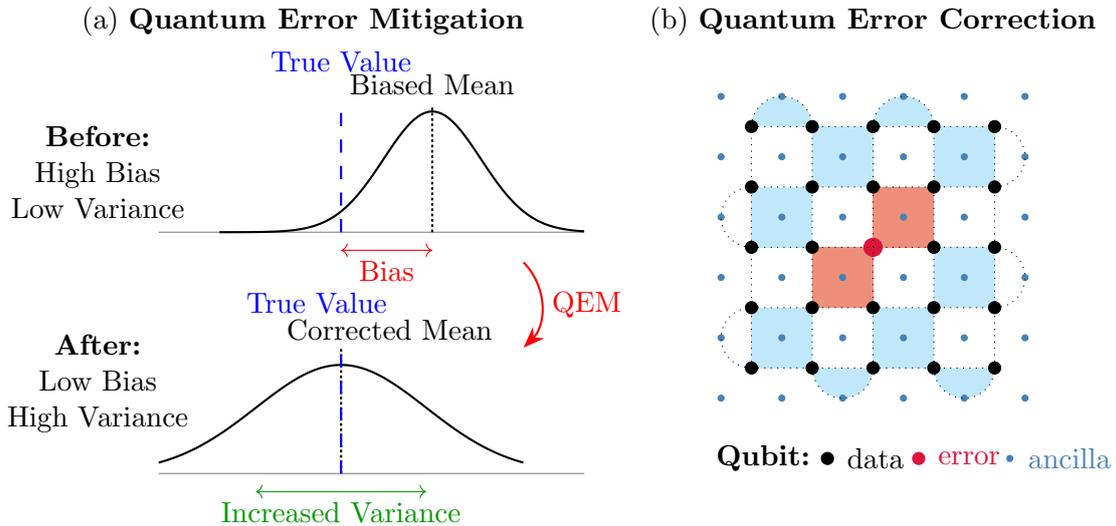
\begin{figure}[htp]
    \centering
    \begin{tikzpicture}

\def\titleY{2}
\def\topRowY{0}
\def\bottomRowY{-3}
\def\sphereScale{0.8}

\node at (-0.5, \titleY) {(a) \textbf{Quantum Error Mitigation}};

\begin{scope}[shift={(0,-4)}, scale=0.8]
    \def\truevalue{0}
    \def\biasedmean{1.5}
    \def\narrowsigma{0.8}
    \def\widesigma{1.4}
    
    \begin{scope}[yshift=4cm]
        \draw[gray] (-3,0) -- (4,0);
        
        \draw[thick, smooth] plot[domain=-2:4, samples=100] 
            (\x, {2*exp(-0.5*((\x-\biasedmean)/\narrowsigma)^2)});
        
        \draw[thick, dash pattern=on 1pt off 1pt] 
            (\biasedmean,0) -- (\biasedmean,2.1) node[above] {Biased Mean};
        
        \draw[thick, dash pattern=on 5pt off 5pt, blue] 
            (\truevalue,0) -- (\truevalue,2.1);
        \node[above, blue] at (\truevalue,2.5) {True Value};
        
        \draw[<->, red] (\truevalue, -0.3) -- (\biasedmean, -0.3) 
            node[midway, below] {Bias};
    \end{scope}
    
    \draw[-{Stealth[length=8pt]}, thick, red] 
        (3,3.5) arc[start angle=45, end angle=-45, radius=1] 
        node[midway, right] {QEM};
    
    \begin{scope}
        \draw[gray] (-3,0) -- (4,0);
        
        \draw[thick, smooth] plot[domain=-3:3, samples=100] 
            (\x, {1.8*exp(-0.5*((\x-\truevalue)/\widesigma)^2)});
        
        \draw[thick, dash pattern=on 1pt off 1pt] 
            (\truevalue,0) -- (\truevalue,2.1);
        
        \draw[thick, dash pattern=on 5pt off 5pt, blue] 
            (\truevalue,0) -- (\truevalue,2.1);
        
        \node[above, blue] at (-0.4,2.5) {True Value};
        \node[above] at (0.8,2.05) {Corrected Mean};
        
        \draw[<->, green!60!black] (-\widesigma, -0.3) -- (\widesigma, -0.3) 
            node[midway, below] {Increased Variance};
    \end{scope}
    
    \node[align=center] at (-4, 5) {\textbf{Before:}\\High Bias\\Low Variance};
    \node[align=center] at (-4, 1.5) {\textbf{After:}\\Low Bias\\High Variance};
\end{scope}
\node at (7, \titleY) {(b) \textbf{Quantum Error Correction}};

\pgfdeclarelayer{bg}
\pgfsetlayers{bg,main}

\begin{scope}[shift={(5,\bottomRowY)}, scale=0.4]

\pgfmathtruncatemacro{\dd}{5}
\pgfmathtruncatemacro{\ddd}{2*\dd}
\pgfmathtruncatemacro{\ddm}{\ddd-2}

\definecolor{mycol1}{RGB}{70, 130, 180}     
\definecolor{mycol2}{RGB}{153, 216, 247}    
\definecolor{errorcolor}{RGB}{220, 20, 60}  
\definecolor{affectedcolor}{RGB}{250, 121, 92} 

\tikzstyle{zplaq}=[fill=mycol2, opacity=0.6]
\tikzstyle{affected}=[fill=affectedcolor, opacity=0.8]

\foreach \x in {0,2,...,\ddd}{
	\foreach \y in {0,2,...,\ddd}{
		\filldraw[mycol1] (\x,\y) circle (0.1);
}}

\foreach \x in {1,3,...,\ddd}{
	\foreach \y in {1,3,...,\ddd}{
		\filldraw[black] (\x,\y) circle (0.2);
}}

\filldraw[errorcolor] (5,5) circle (0.3);
\node[errorcolor, font=\tiny\bfseries] at (5,5) {X};

\foreach \x	in {1,3,...,\ddd}{
	\draw[dotted] (\x,1) -- (\x,\ddd-1);
}
\foreach \y in {1,3,...,\ddd}{
	\draw[dotted] (1,\y) -- (\ddd-1,\y);
}

\foreach \i in {3,5,...,\dd}{
	\draw[dotted] ({2*\i-3},1) arc (-180:0:1);
	\draw[dotted] ({2*\i-5},\ddd-1) arc (180:0:1);
	\draw[dotted] (1,{2*\i-5}) arc (270:90:1);
	\draw[dotted] (\ddd-1,{2*\i-3}) arc (-90:90:1);
}

\node[anchor=west] (qbt) at (-0.5, -2) {\textbf{Qubit:}} ;
\filldraw[black] (3.5,-2) circle (0.2) node[anchor=west, xshift=3.5] {data};
\filldraw[errorcolor] (6.5,-2) circle (0.2) node[anchor=west, xshift=3.5] {error};
\filldraw[mycol1] (9.5,-2) circle (0.1) node[anchor=west, xshift=3] (anc) {ancilla};

\begin{pgfonlayer}{bg}
\foreach \x in {1,5,...,\ddm}{
	\foreach \y in {1,5,...,\ddm}{ 
		\fill[zplaq] (\x,\y) rectangle (\x+2,\y+2);
		\fill[zplaq] (\x+2,\y+2) rectangle (\x+4,\y+4);
}}

\fill[affected] (3,3) rectangle (5,5);
\fill[affected] (7,7) rectangle (5,5);

\foreach \i in {3,5,...,\dd}{
	\fill[zplaq] ({2*\i-3},1) arc (-180:0:1);
	\fill[zplaq] ({2*\i-5},\ddd-1) arc (180:0:1);
}	
\end{pgfonlayer}

\end{scope}

\end{tikzpicture}    
    \caption{(a) Quantum Error Mitigation aims to improve the result be repeating and combining multiple noisy measurement results into an improved measurement result with lower error. (b) Quantum Error Correction encodes quantum information over multiple physical qubits. When an error event occurs, as indicated by the data qubit in red, it is detected through measurement using (blue) ancilla qubits (red plaquettes) and corrective action can be taken. The entire diagram represents a single logical qubit encoded in a $d=5$ surface code.}
    \label{fig:error_figure}
\end{figure}

\gls{QEM} and \gls{QEC} share the goal of increasing the robustness of a quantum computer to errors but go about reaching this goal in different ways. \gls{QEM} accepts the inevitability of noise and errors, and compensates for these errors by applying a probabilistic correction to the \emph{distribution} of measurements. This comes at a cost of increasing the sampling overhead (i.e. multiplying the number of shots needed to reach a prescribed error target). \gls{QEC}, on the other hand, protects the quantum information against the introduction of noise by detecting and correcting for errors within \emph{individual} \glspl{shot}. This comes at a cost of increasing the encoding overhead (i.e. multiplying the number of physical qubits used to redundantly store the state of a single logical qubit).
The distinction between \gls{QEM} and \gls{QEC} is depicted graphically in~\cref{fig:error_figure}

For the remainder of this section, we will introduce the principles guiding \gls{QEM} and \gls{QEC} protocols by focusing respectively on \gls{PEC} and the surface code as prototypical examples of each. For our discussion, we assume that (1) \gls{quantum gate} errors are the dominant sources of errors in the system, thus ignoring \gls{SPAM} errors, and, (2) errors occur independently and as discrete events. If we consider $\epsilon$ as the probability an error occurs each time a \gls{quantum gate} is applied, then the overall fault rate $\lambda$ of a \gls{quantum circuit} becomes, 
\begin{equation}
\lambda = n_G \epsilon,
\label{eq:error_rate}
\end{equation}
and therefore the \gls{quantum circuit} fault rate becomes of order unity when $n_G \approx \epsilon^{-1}$~\cite{RevModPhys.95.045005}.

\gls{QEM} is most often considered in the context of producing accurate expectation values $\braket{O}$ of an observable $O$ on a \gls{NISQ} device.  \gls{PEC}~\cite{temme2017error, Endo:2018} is a widely adopted \gls{QEM} method and here we consider it as the prototypical \gls{QEM} procedure. \gls{PEC} can provide bias-free estimates of expectation values by (1) characterizing a noise channel, $\Lambda_i$, associated with a \gls{quantum gate} operation, (2) expanding it in an over-complete basis of (noisy) gates that can be executed on the quantum system, and (3) quasi-probabilistically implementing the inverse channel $\Lambda_i^{-1}$ before each \gls{quantum gate} operation to cancel out the noise. The trade off in reducing or eliminating the bias in the estimator for the observable, $\hat{O}_{\text{PEC}}$, is that the variance of the estimator is increased compared to the noisy estimator $\hat{O}_{\text{noise}}$ obtained from computing the sample mean of the noisy \gls{quantum circuit} runs. The ratio is called the sampling overhead and
it has been shown that 
for \gls{PEC} it scales approximately as~\cite{RevModPhys.95.045005,vandenBerg2023}
\begin{equation}
    C_{\text{PEC}} \approx e^{4 \lambda} = e^{4 n_{G,\text{PEC}} \epsilon},
\label{eq:pec_overhead}
\end{equation}
from which it follows that the error of the PEC estimator $\hat{O}_{\text{PEC}}$ compared to the exact observable $\braket{O}$ can be made arbitrarily small at a sampling overhead that scales exponentially in both the number of \gls{quantum gate} operations in the \gls{quantum circuit} and the error rate per \gls{quantum gate}. Rewriting~\Cref{eq:pec_overhead} as
\begin{equation}
    n_{G,\text{PEC}} \approx \frac{\ln C_{\text{PEC}}}{4 \epsilon} = \frac{\ln C_{\text{PEC}}}{4} n_{G,\text{noise}},
\label{eq:pec_gates}
\end{equation}
we see that the number of gates that can be run using PEC remains inversely proportional to the error rate but PEC can improve the pre-factor, $\ln C_{\text{PEC}}/4$, at an exponential cost in the sampling overhead.

\gls{QEC} is the hallmark of \gls{FTQC}.
The prevailing QEC scheme combines multiple physical qubits to redundantly encode information in a single logical qubit.
Within a logical qubit, some of the physical qubits are designated as `data' qubits to be used to reliably store quantum information. Others are designated as `ancilla' qubits to be used for detecting error events that would distort the quantum information.
This is loosely analogous to the use of parity bits for error correcting codes (ECC) in classical memory systems.
Measurements of the ancillary qubits can be used to detect errors without
collapsing the logical wave function.

Robust and efficient encoding of information in logical qubits is essential for QEC.
A $[\![n, k, d]\!]$-code encodes $k$ \glspl{logical qubit} in the state of $n$ \glspl{physical qubit} in such a manner that any error event that maps a valid encoded state to another valid encoded state must act on at least $d$ \glspl{physical qubit}. The value $d$ is known as the \gls{code distance} and quantifies the robustness of the code against errors; it is analogous to the number of bit errors that can be detected by a Hamming codes in classical ECC.
Typically, reaching larger code distances, and thus an increased resilience against errors, requires using more \glspl{physical qubit} per \gls{logical qubit}. 

QEC cannot be effective unless the physical error rate is below the \gls{code threshold}. For physical error rates below the \gls{code threshold}, logical errors can be (exponentially) suppressed by increasing the \gls{code distance}. For physical error rates above \gls{code threshold}, increasing the \gls{code distance} will instead worsen the logical errors due to the increased number of error-prone operations involved. 

\gls{FTQC} using a \gls{QEC} code typically progresses in \emph{code cycles} which involve measuring a subset of the \glspl{physical qubit} that make up the \gls{logical qubit}(s) in order to extract the \gls{error syndrome} which can then be used to correct any logical errors and implement \gls{quantum gate} operations on the \glspl{logical qubit}.

The \gls{surface code}~\cite{Kitaev2003,Bravyi1998,Dennis2002} is a prototypical topological error correcting code that encodes a \gls{logical qubit} in a planar, square lattice of \glspl{physical qubit} that only require nearest-neighbor connectivity. For a single \gls{logical qubit} with \gls{code distance} $d$, the encoding requires $2d^2-1$ \glspl{physical qubit} per \gls{logical qubit}, consisting of $d^2$  data qubits and $d^2-1$ ancilla qubits for \gls{error syndrome} extraction via parity check measurements. In summary, the \gls{surface code} is a $[\![n = 2d^2-1, k = 1, d]\!]$-code. A $d=5$ surface code \gls{logical qubit} is sketched in~\cref{fig:error_figure}(b). A commonly used rule-of-thumb relating the logical error rate $\epsilon_L$ to the physical device error rate $\epsilon_P$ for a \gls{code distance} $d$ \gls{surface code} is~\cite{fowler2018}
\begin{align}
\epsilon_L = 0.1 (100 \epsilon_P)^{(d+1)/2},
\label{eq:surface_code_error}
\end{align}
assuming the physical error rate is below the \gls{code threshold}. Small examples of the \gls{surface code} have already been demonstrated experimentally by several groups~\cite{Maika:2016,Marques:2022,Krinner:2022,Zhao:2022,Acharya:2023}. The \gls{surface code} offers many advantages: it is well understood, allows for a planar topology, has a high \gls{code threshold}, and has been used to propose universal quantum computer architectures~\cite{Litinski:2019b}. However, it also is notoriously resource-intensive since the number of qubits scales quadratically as a function of \gls{code distance}. Naturally, more efficient encodings have been proposed in the literature. Prominent examples include \gls{qLDPC} codes~\cite{Panteleev:2022a, Panteleev:2022b, Breuckmann_2021, Tillich_2014,Leverrier_2015,Gottesman:2014,Tremblay:2022,Leverrier:2022,Higgott:2021,Higgott:2024}; for example, \gls{qLDPC} codes based on expander graphs can in principle achieve scaling $[\![n, k = \Theta(n), d = \Theta(n)]\!]$, which is quadratically more efficient than the \gls{surface code}. 


\gls{QEC} can in principle be combined with \gls{QEM} approaches. Indeed, it has recently been shown that PEC can be beneficial in combination with \gls{QEC}~\cite{qem_Suzuki2022}, and has the potential to allow for a reduction of the \gls{code distance} by five while maintaining similar error rates at the cost of a sample overhead factor $C \leq 100$ and more intensive post-processing.

\section{Overview of quantum algorithms}
\label{sec:algorithms}

On top of these hardware and architectural foundations sit the algorithms that define \gls{QC}’s potential~\cite{QuantumAlgorithmZoo}. One of the most important classes of quantum algorithmic kernels for the physical sciences is Hamiltonian simulation~\cite{Feynman1982,Lloyd1996}, in which a quantum computer models the evolution of a quantum system for which the total energy is described by a Hamiltonian operator. This capability is invaluable for chemistry, materials science, and fundamental physics, where classical simulation quickly becomes intractable. Early approaches use product formulas also known as Trotter-Suzuki decomposition~\cite{Suzuki1991}, which break the time evolution into a sequence of simpler steps to approximate the real-time evolution as illustrated in~\cref{fig:algorithm_figure}(a). More recent methods, such as \gls{qubitization}~\cite{low2019hamiltonian}, encode the Hamiltonian operator into a structured form, and achieve optimal computational complexity: (1) the number of queries to the Hamiltonian data (oracle) scales linearly in the simulation time and as $\log(1/\epsilon)$ in the spectral error; (2) the number of qubits scales as $\log(N) + m$ with $N$ the dimension of the Hamiltonian and $m$ some comparatively small number of extra ancilla qubits. This is an exponential reduction in resources compared to a brute-force classical implementation. \gls{QSP}~\cite{low2017optimal} builds on this by enabling polynomial transformations of quantum operators through carefully orchestrated rotations, and the \gls{QSVT}~\cite{gilyen19} generalizes the technique to manipulate arbitrary operators directly—unlocking efficient algorithms for solving linear systems, performing principal component analysis, and other tasks relevant to scientific computing. Collectively, these constitute a unified framework for quantum algorithm development and discovery that is applicable to many known quantum algorithms exhibiting computational speedup~\cite{PRXQuantum.2.040203}. Additionally, \gls{QPE}~\cite{kitaev1995} and its modern variants~\cite{Ni2023, Lin22, Ding23, Dong22, Wan22, Dorner09, Higgins2007, Zhang2022}, shown in~\cref{fig:algorithm_figure}(b), use Hamiltonian simulation to compute ground state energies of quantum systems, a key primitive underpinning much of the expected \gls{QC} speedups in quantum chemistry and condensed matter physics (see~\cref{chap:apps}). %

\begin{figure}[htp]
    \centering
    \begin{tikzpicture}
\node at (3, 0.75) {(a) \textbf{Hamiltonian Simulation via Trotterization}};

\begin{scope}[shift={(0,0)},scale=0.48]
\def\numQubits{6}
\def\wireSpacing{0.8}
\def\gateWidth{0.6}
\def\gateHeight{0.5}

\foreach \q in {0,1,...,5} {
    \draw[thick] (-1, -\q*\wireSpacing) -- (1.5, -\q*\wireSpacing);
    \draw[thick] (2.25, -\q*\wireSpacing) -- (12.75, -\q*\wireSpacing);
}

\draw[thick, fill=orange!40, draw=black] (-0.75, 0.4) rectangle (1.25, -4.4);
\node at (0.25, -2) {$e^{-iHt}$};

\node at (2, -2) {\Large $=$};

\foreach \layer in {0,1,2} {
    \pgfmathsetmacro{\baseX}{3 + \layer * 3}
    
    \foreach \q in {0,1,...,5} {
        \draw[thick, fill=yellow!70, draw=black] (\baseX-\gateWidth/2, -\q*\wireSpacing-\gateHeight/2) rectangle (\baseX+\gateWidth/2, -\q*\wireSpacing+\gateHeight/2);
    }
    
    \foreach \q in {0,2,4} {
        \draw[thick, fill=blue!70, draw=black] (\baseX+1-\gateWidth/2, -\q*\wireSpacing-2*\gateHeight) rectangle (\baseX+1+\gateWidth/2, -\q*\wireSpacing-\wireSpacing+2*\gateHeight);
    }
    
    \foreach \q in {1,3} {
        \draw[thick, fill=blue!70, draw=black] (\baseX+2-\gateWidth/2, -\q*\wireSpacing-2*\gateHeight) rectangle (\baseX+2+\gateWidth/2, -\q*\wireSpacing-\wireSpacing+2*\gateHeight);
    }
}

\foreach \q in {0,1,...,5} {
    \draw[thick, fill=yellow!70, draw=black] (12-\gateWidth/2, -\q*\wireSpacing-\gateHeight/2) rectangle (12+\gateWidth/2, -\q*\wireSpacing+\gateHeight/2);
}
\end{scope}

\node at (11, 0.75) {(b) \textbf{QPE via Qubitization}};
\definecolor{mycol2}{RGB}{153, 216, 247} 
\node[anchor=west,scale=0.7] at (6.7,-1) {
\begin{quantikz}[column sep=0.15cm,row sep=0.2cm]
\lstick{$|0\rangle$} & \gate[3,style={fill=green!40}]{\chi_m} & \ctrl{3} & \qw & \qw & \qw & \qw & \qw & \cdots & \qw & \qw & \qw & \gate[3,style={fill=green!40}]{\text{QFT}^\dagger} & \meter{} \\
\lstick{$|0\rangle$} & \qw & \qw & \ctrl[open]{2} & \qw & \ctrl[open]{2} & \qw & \qw & \vdots & \qw & \qw & \qw & \qw & \meter{} \\
\lstick{$|0\rangle$} & \qw & \qw & \qw & \qw & \qw & \qw & \qw & \cdots & \ctrl[open]{1} & \qw & \ctrl[open]{1} & \qw & \meter{} \\
\lstick{} & \qw & \gate[style={fill=orange!40}]{\mathcal{W}} & \gate[style={fill=mycol2}]{\mathcal{R}} & \gate[style={fill=orange!40}]{\mathcal{W}} & \gate[style={fill=mycol2}]{\mathcal{R}} & \qw & \qw & \cdots & \gate[style={fill=mycol2}]{\mathcal{R}} & \gate[style={fill=orange!40}]{\mathcal{W}^{2^{m-1}}} & \gate[style={fill=mycol2}]{\mathcal{R}} & \qw & 
\end{quantikz}
};
\end{tikzpicture}    
    \caption{(a) When implementing Hamiltonian simulation via Trotterization, the time evolution operator $e^{-iHt}$ (orange) is approximated as a product of simpler time evolutions resulting in structured circuits with single-qubit gates (yellow) and two-qubit gates (blue) (b) Implementing Quantum Phase Estimation to compute the ground state of a Hamiltonian $H$ using  qubitization begins with a state preparation phase $\chi_m$, followed by a sequence of (controlled) \emph{walk operators} $\mathcal{W}$, which encode the spectrum of $H$ via qubitization, and reflection oracles $\mathcal{R}$, and ends with an inverse Quantum Fourier Transform. Figure adapted from~\cite{app_Babbush2018}.}
    \label{fig:algorithm_figure}
\end{figure}

While these advanced algorithms promise significant long-term benefits, today’s quantum hardware is still in the noisy intermediate-scale quantum (NISQ) era, where qubit counts are limited and error rates are high. To make progress in the near term, researchers have developed variational algorithms, which blend quantum and classical computation. In these hybrid approaches, a quantum processor prepares a parameterized quantum state, and a classical optimizer iteratively updates the parameters to minimize a cost function. Examples include the \gls{VQE}~\cite{Peruzzo2014}, used to estimate molecular ground-state energies, and the Quantum Approximate Optimization Algorithm (QAOA), designed for combinatorial optimization. These methods are more tolerant of noise and shorter circuit depths, making them practical stepping stones toward the large-scale, fault-tolerant algorithms that will eventually unlock \gls{QC}’s full potential.

\chapter{Resource Estimates for the Quantum Relevant Workload}
\label{chap:apps}
\myglsreset

In this section, we collect and compare resource estimates from the literature for more than 140 target applications across the three scientific domains identified in the previous section that make up the~\gls{QRW}. We discuss condensed matter physics and materials science in~\cref{sec:app_cm}, quantum chemistry in~\cref{sec:app_qc}, and nuclear and high energy physics in~\cref{sec:app_hep}. \Cref{sec:app_overview} provides an overview of the three domains and~\cref{sec:app_other} discusses the state of affairs for certain other quantum algorithms and application areas of interest to the \gls{DOE} \gls{SC} mission, including linear algebra, optimization, simulating differential equations, and quantum machine learning. For these other areas, we do not include resource estimates.

For each resource estimate, we report a commonly used space-time figure of merit~\cite{proctor2025benchmarking} for the \gls{quantum circuit} consisting of the number of \glspl{qubit} ($n_Q$), i.e., the space complexity, and the number of \glspl{quantum gate} ($n_G$), i.e., the time complexity. We name this quantity the $\perf$-vector (for performance) and define it as follows,
\begin{equation}
    \perf = (n_Q, n_G).
\label{eq:pvector}
\end{equation} 
We will often visualize the $\perf$-vector as an xy-coordinate on a 2D plot. The $\perf$-vector corresponds to an estimate of the volume of the \gls{quantum circuit} required to solve the application. Multiple factors can influence the $\perf$-vector required to solve an application. These factors include, but are not limited to:

\begin{enumerate}
    \item \emph{Space versus time trade-offs.} Most quantum algorithms and applications allow a trade-off between a reduction in the number of \glspl{qubit} for an increase in the number of \glspl{quantum gate} (or vice versa). A well-known example is Hamiltonian simulation, a key quantum-algorithmic kernel used in the majority of applications we discuss. Implementing a Hamiltonian simulation using a \gls{Trotter} decomposition typically requires significantly fewer \glspl{qubit} and significantly more \glspl{quantum gate} compared to a method that combines \gls{qubitization} with \gls{QSP} at a similar error level.
    \item \emph{Dominant \gls{quantum gate} complexity.} Most resource estimates only report the leading-order of gates that require the most resources to execute, such as $T$ gates, Toffoli gates, or continuous-angle rotation gates, and thus are the dominant factor determining the execution time. This assumes a cost model that is typical for the (early) \gls{FTQC} rather than \gls{NISQ} era. However, the type of \gls{quantum gate} that is most expensive might change in the future and depend on the system architecture. For example, recent work on magic state cultivation~\cite{Gidney:2024} has significantly reduced the complexity to implement a $T$ gate~\cite{Litinski:2019}; over a certain parameter range, the cost of $T$ gates is now much closer to that of a Clifford gate like a CNOT. More breakthroughs in this area can further alter this picture and in practice the dominant \gls{quantum gate} cost might differ based on the quantum computer hardware architecture.
    For our analysis, we express all resource estimates in terms of $T$ gates, which are the dominant resource constraint for many algorithms. 
    In some cases, previously published resource estimates have been reported in terms of different gate sets. In those instances, we convert the estimates into an equivalent $T$ gate cost using the following conversion rates: 
    \begin{itemize}
        \item 1 Toffoli gate = 4 $T$ gates~\cite{PhysRevA.87.022328},
        \item 1 rotation gate = 100 $T$ gates~\cite{selinger2015}.
    \end{itemize} 
    \item \emph{Topology or connectivity.} The two- and multi-qubit gates present in a \gls{quantum circuit} determine the connectivity and topology requirements on the quantum computer architecture to run the \gls{quantum circuit} without needing additional transpilation and the introduction of SWAP operations. The worst-case transpilation overhead on the number of gates to map a \gls{quantum circuit} requiring all-to-all connectivity to 2D nearest-neighbor connectivity scales as $\mathcal{O}(\sqrt{n_Q})$\footnote{Mapping to a linear topology would scale as $\mathcal{O}(n_Q$) but almost all fault-tolerant architectures require at least a 2D square grid connectivity.}. For the remainder of this report, we do not consider the effect of topology on the $\perf$-vectors reported in the literature.
    \item \emph{Circuit versus active volume.} The \gls{quantum circuit} volume (and hence the $\perf$-vector) might differ based on the quantum computer architecture. For example, \emph{active volume} architectures have been proposed for quantum computers based on photonics~\cite{active_volume}. In this setting, qubits that are idle at certain points in the \gls{quantum circuit} do not count towards the overall \gls{quantum circuit} volume. It has been estimated that an active volume architecture can reduce the number of gates to run Shor's prime factoring algorithm by an order of magnitude~\cite{active_volume} and it can also reduce the resources required for quantum chemistry applications~\cite{app_thc_bliss}. Other such architectural designs that improve the $\perf$-vector might be developed in the future.
    \item \emph{Critical depth and gate parallelization.} It is (often implicitly) assumed throughout this report that the number of gates reported represents the critical depth of the \gls{quantum circuit}. Due to the anticipated serial or small-batch generation of the resource states necessary for implementing non-Clifford gates (i.e. $T$ gates), early scalable \gls{FTQC} are not expected to support significant parallelization for non-Clifford gates. Specifically, in our discussion on execution time estimates, we do not consider the possible impact of gate parallelization.
\end{enumerate}

We conclude, based on the considerations discussed above, that the $\perf$-vector estimates should be considered as an upper bound. Furthermore,  $\perf$-vectors are also expected to reduce further by future algorithmic advances and improved implementations. We will motivate this expectation by presenting a case study from quantum chemistry in~\cref{sec:app_qc}.

\section{Condensed matter physics and materials science}
\label{sec:app_cm}

Condensed matter physics and materials science pose some of the most promising problems for quantum simulation. Strongly correlated electrons, emergent phases of matter, and exotic excitations underlie phenomena such as high-temperature superconductivity, magnetism, and topological order, areas where classical computational methods often reach their limits due to exponential scaling~\cite{anderson1987,georges1996,kotliar2006}. Accurate quantum simulations of these systems would not only provide insight into fundamental physics~\cite{auerbach1994,wen2004}, but also accelerate the discovery of practical materials for energy storage~\cite{aspuru2005,reiher2017}, catalysis~\cite{cao2019,goings2022}, and quantum technologies~\cite{bauer2020quantum,mcardle2020quantum}. As a result, condensed matter and materials science are widely viewed as domains where quantum computers could deliver some of the earliest impacts~\cite{bharti2022,aspuru2008}.

\begin{figure}[htp]
    \centering
    \includegraphics[width=0.8\textwidth]{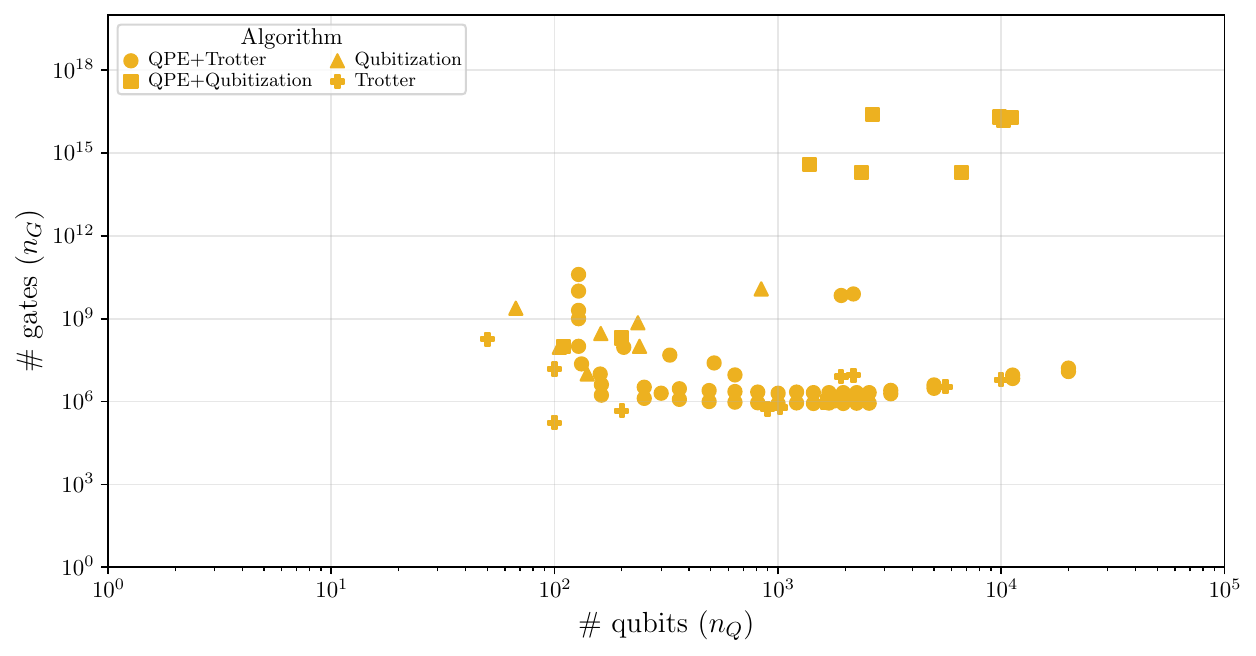}
    \caption{Overview of various resource estimates for computational problems related to materials science in terms of number of \glspl{logical qubit} ($n_Q$) and number of gates ($n_G$). Different marker types indicate the different quantum algorithms being used.}
    \label{fig:est_materials}
\end{figure}

We collected resource estimates for a total of 80 applications in condensed matter physics and materials science that span a spectrum of model complexity and are considered out of reach for classical solvers. These include simpler spin models, such as the Kitaev and Heisenberg models that can help scientists study new and interesting physical phenomena and prepare exotic phases of matter (e.g. spin glasses and quantum chaos)~\cite{app_lanl,PhysRevA.99.040301,childs18,Flannigan_2022,Yoshioka2024}; more complicated lattice models such as the Fermi-Hubbard (FH) model~\cite{app_Campbell2021,app_Babbush2018,Flannigan_2022,Yoshioka2024,baysmidt2025} that can serve as a proxy to study superconducting materials; and realistic models of materials that take into account the relevant electronic structure~\cite{Kivlichan2020}. As an example of the latter category, Refs.~\cite{app_Delgado2022,ShokrianZini2023} propose \gls{QPE}+\gls{qubitization} using first-quantized plane-wave basis to simulate materials for lithium-ion batteries such as Li$_2$FeSiO$_4$. Subsequent work~\cite{Berry2024} shows that \gls{first quantization} outperforms \gls{second quantization} for materials simulations of a Lithium-Nickel-Oxide (LNO) material.
The $\perf$-vectors of the resource estimates we collected are shown in~\Cref{fig:est_materials}. The estimates at lower \gls{quantum gate} count correspond to the simulations of spin and FH models, while the estimates for materials simulations in \gls{first quantization} show comparatively larger $\perf$-vectors.

We observe that resource estimates start at about $10^6$ gates and $10^2$ \glspl{qubit} and go up from there in both \gls{quantum gate} depth and \gls{qubit} count. We do expect condensed matter physics and materials science to be among the first scientific fields to benefit from quantum computers since many of these problems are formulated in terms of spins, which map naturally to qubits, or can be encoded in qubits with minimal resource overhead. This is the primary reason we consider them to be the earliest candidates for scientific quantum advantage among the \gls{DOE} \gls{QRW}.

\section{Quantum chemistry}
\label{sec:app_qc}

The field of quantum chemistry addresses a number of important scientific and technological challenges, from designing efficient catalysts for sustainable fuels and industrial processes to developing advanced materials for batteries, photovoltaics, and quantum information technologies~\cite{reiher2017,cao2019,bauer2020quantum}.
Quantum chemistry stands out as a leading target for \gls{QC}, as QC has the potential to revolutionize our ability to simulate molecular behavior.  Within quantum chemistry, a key distinction often lies between static and dynamic problems. Static problems focus on characterizing a molecule in equilibrium, providing a snapshot of its properties, for example determining the ground-state energy, electronic structure, and other time-independent properties. In contrast, dynamic problems explore how molecules evolve over time, such as chemical reactions, photoexcitation, charge and energy transfer, and vibrational dynamics. \gls{QC} holds the potential to provide accurate solutions to both static and dynamic quantum chemistry problems, however our survey of the literature revealed that static chemistry problems, and in particular \gls{GSEE}, have been studied in much greater detail than dynamic ones. 

\begin{figure}[ht]
    \centering
    \begin{subfigure}{0.48\textwidth}
    \includegraphics[width=\textwidth]{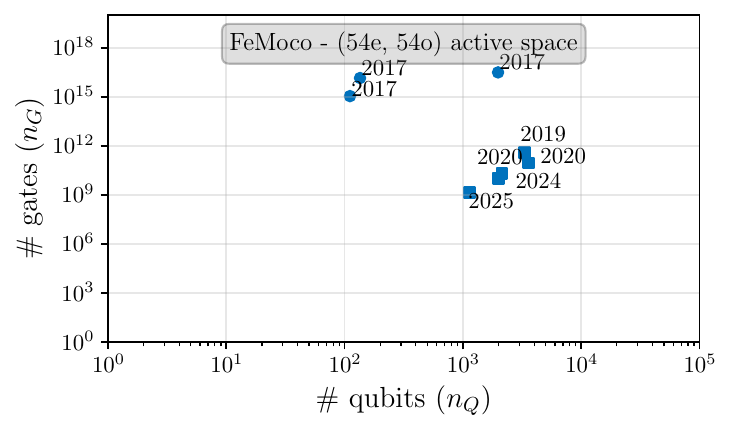}
    \caption{(54e, 54o) active space FeMoco}\label{fig:femoco_small}
    \end{subfigure}
    \hfill
    \begin{subfigure}{0.48\textwidth}
    \includegraphics[width=\textwidth]{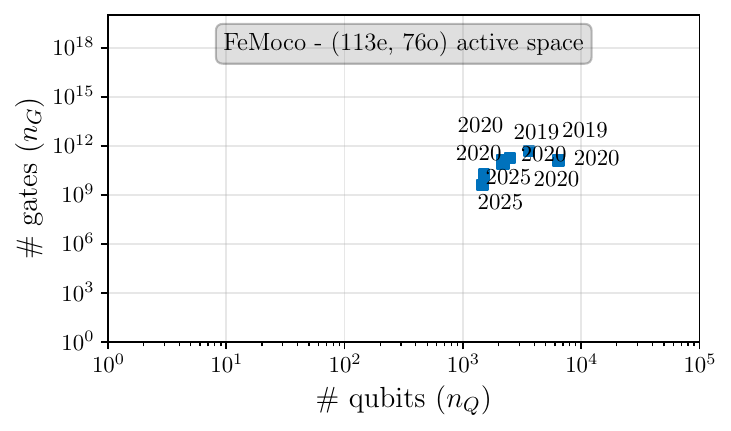}\label{fig:femoco_large}
    \caption{(113e, 76o) active space FeMoco}
    \end{subfigure}
    \caption{Overview of resource estimates for ground state energy estimation of two versions of the FeMoco system: (\emph{left}) a (54e, 54o) active space~\cite{app_Reiher2017} model, and (\emph{right}) a (113e, 76o) active space~\cite{app_Zhendong} model.}
    \label{fig:femoco}
\end{figure}

The field of quantum resource estimation for GSEE problems in quantum chemistry has matured over the past five years compared to other application areas. This is reflected by the observation that a few standardized benchmark problems have been established in the literature. This includes the FeMoco complex~\cite{app_Reiher2017} and the Cytochrome P450 system~\cite{app_goings}. These target problems have been re-analyzed as new algorithms for GSEE have been proposed and as such form an excellent yardstick to track progress in quantum algorithms for quantum chemistry. The reduction in resources required for GSEE of the FeMoco complex is visualized in~\Cref{fig:femoco} for both the original, smaller (54e, 54o) active space system, using the def2-TZVP basis~\cite{app_Reiher2017}, and the larger (113e, 76o) active space system, using the TZP-DKH basis for Fe, S, and Mo and the def2-SVP bais for the other atoms~\cite{app_Zhendong}. The dimension of the FCI space for the larger system is on the order of $10^{35}$ for the spin $S=3/2$ ground state. The quantum resources required for running \gls{QPE} on the smaller system, shown in~\Cref{fig:femoco_small}, were originally estimated using a \gls{Trotter}ization method for the time evolution~\cite{app_Reiher2017}. These estimates correspond to the data points labeled `2017' and require fewer qubits than more recent estimates at the cost of an order of magnitude more gates. 
Recent approaches all use a \gls{qubitization} of the electronic structure Hamiltonian, and distinguish themselves primarily by how they factorize and compress the two-electron integral tensor. The main approaches studied include the single factorization method~\cite{app_Berry}, the double factorization method~\cite{app_vonBurg2021}, and \gls{THC}~\cite{app_Lee, app_thc_bliss,app_Rocca2024}. In addition to these factorization strategies, spectrum amplification techniques have recently been developed to further reduce resource costs, most notably the sum-of-squares spectral amplification (SOSSA) framework~\cite{app_amplification,king2025quantum}, which builds on spectrum amplification ideas to accelerate simulations for Hamiltonians with efficiently computable sum-of-squares representations.
We observe from~\Cref{fig:femoco} that these algorithmic improvements have significantly reduced the computational resources required to solve the GSEE problem leading to roughly a 1000$\times$ reduction in the number of gates and a 5$\times$ reduction in \gls{qubit} count. Similar observations hold for other standardized benchmark problems such as the Cytochrome P450 system.

While \gls{GSEE} has received the bulk of attention, there is growing interest in developing quantum algorithms for dynamics problems, which aim to capture the time-dependent behavior of molecular systems. Classical approaches to these problems, such as Born-Oppenheimer molecular dynamics, Ehrenfest dynamics, and ab initio molecular dynamics based on time-dependent \gls{DFT}, are computationally intensive and often struggle with nonadiabatic effects or strongly correlated electrons~\cite{tully1990molecular, marx2009ab}. Quantum computers offer the potential to simulate such processes natively by evolving quantum states under a time-dependent or time-independent molecular Hamiltonian and by directly computing forces, energy gradients, and other observables needed for nuclear propagation in molecular dynamics simulations. For example, algorithms have been developed to compute molecular energy derivatives, including forces, more efficiently on quantum devices, achieving lower circuit repetition costs and even Heisenberg-limited scaling in the fault-tolerant limit~\cite{o2022efficient}.
Variational quantum algorithms have also been suggested for calculating forces within ab initio molecular dynamics frameworks~\cite{sokolov2021microcanonical}.
Algorithmic strategies for real-time evolution include \gls{Trotter}ized and qubitized time evolution~\cite{low2019hamiltonian}, variational quantum dynamics using a time-dependent \gls{VQE}~\cite{yuan2019theory}, and approaches based on linear combination of unitaries~\cite{childs2012hamiltonian} and \gls{QSP}~\cite{low2017optimal}. Ref.~\cite{ollitrault2021molecular} reviews quantum algorithms for simulating molecular quantum dynamics, highlighting both representations using \gls{first quantization} and \gls{second quantization}, and comparing variational and \gls{Trotter}-based real-time evolution methods as promising routes for capturing electron-nuclear dynamics. A number of papers~\cite{chan2023grid, rubin2024quantum} have estimated the resource requirements for quantum chemistry dynamics problems and found a need for significantly higher gate counts than \gls{GSEE}. While these studies represent important first steps in assessing the feasibility of quantum chemistry dynamics on quantum hardware, the level of effort devoted to reducing their resource requirements remains far behind that for ground-state estimation, where years of algorithmic refinements have dramatically lowered cost estimates. Consequently, we do not plot resource estimates for dynamics problems. Bridging this gap remains both a challenge and an opportunity for advancing quantum simulations of chemically relevant processes.

In addition to time evolution, significant progress has been made in developing quantum algorithms for excited state calculations, which, while not dynamical themselves, provide necessary inputs for simulating chemical dynamics. Access to accurate excitation energies and electronic couplings enables the construction of potential energy surfaces, which are needed for modeling processes like photoisomerization and energy transfer. Techniques such as state-averaged \gls{VQE}~\cite{yalouz2021state}, equation-of-motion approaches~\cite{ollitrault2020quantum}, and subspace-search methods~\cite{nakanishi2019subspace, shen2024efficient} have been proposed to efficiently target excited states on quantum hardware. These developments will be useful for quantum simulations of nonadiabatic molecular dynamics, particularly in regions near conical intersections where potential energy surfaces of different electronic states intersect and traditional approximations fail~\cite{domcke2004conical}.

Bosonic \gls{QC} platforms, particularly those using continuous-variable (CV) encodings, have also emerged as promising tools for simulating chemical dynamics. These platforms naturally represent vibrational degrees of freedom using bosonic modes, enabling compact and efficient simulations of molecular vibrations and vibronic transitions~\cite{dutta2024simulating}. Recent theoretical work has proposed using CV-based quantum processors for real-time wavepacket evolution and vibronic dynamics in anharmonic molecular potentials~\cite{malpathak2025simulating}, and hybrid bosonic-qubit~\cite{lee2024fault, liu2024hybrid} schemes are being explored to model coupled electron-vibration dynamics more efficiently. Despite these recent theoretical developments, detailed quantitative resource estimates, such as required number of bosonic modes, effective squeezing levels, gate counts, or coherence time budgets, are not yet available in the literature. Similarly, hardware roadmaps from CV-focused vendors, such as Xanadu (photonic architectures) and Alice\&Bob (cat qubits), describe platforms capable of simulating bosonic systems but as of yet provide no application specific resource benchmarks.  

We expect resource reductions and algorithmic speedups to become pervasive across all domains as quantum algorithms continue to develop and mature with future scientific advances. Indeed, this is a larger trend within algorithmic development which can be seen in the field of classical algorithms over the past decades~\cite{walker1996fourier,multigrid,Darve:2000,Garcke:2012,giles2013multilevelmontecarlomethods,Mahoney:2011}, and has historically been a major factor in guiding investments at \gls{DOE} \gls{SC}~\cite{keyes2003}.

\begin{figure}[ht]
    \centering
    \includegraphics[width=0.8\textwidth]{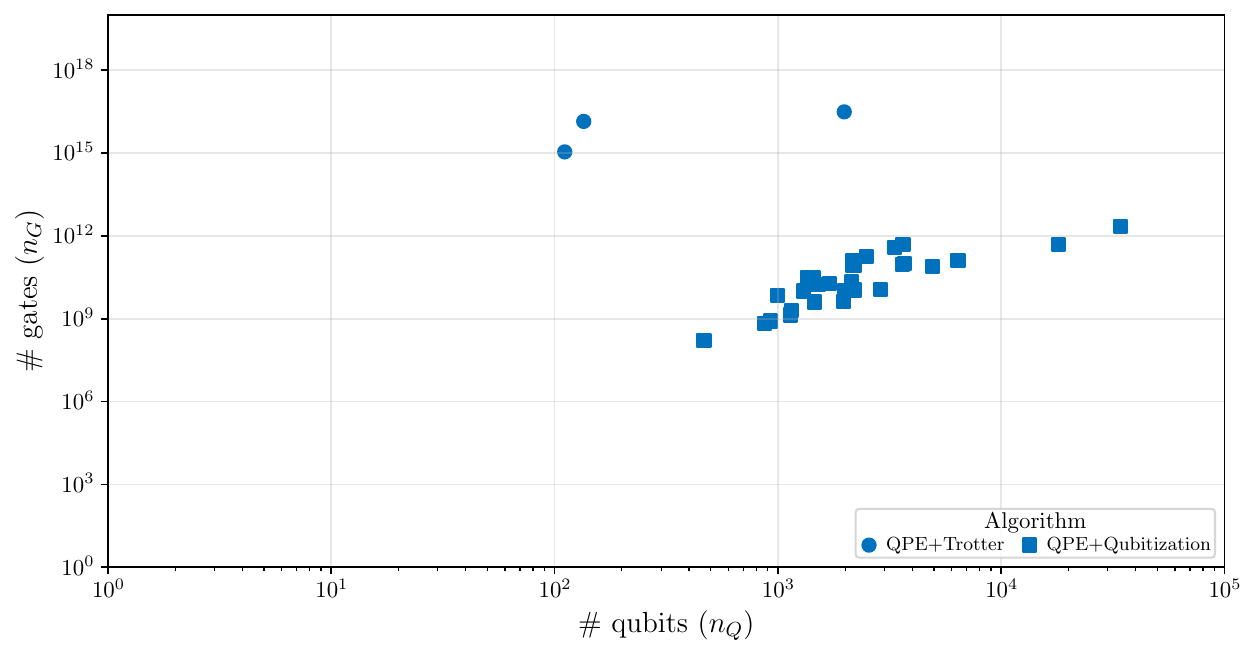}
    \caption{Overview of various resource estimates for computational problems related to quantum chemistry in terms of number of \glspl{logical qubit} ($n_Q$) and number of gates ($n_G$).  \gls{QPE}+\gls{Trotter} based applications are marked by circles, and \gls{QPE}+\gls{qubitization} by squares.}
    \label{fig:est_chemistry}
\end{figure}

\Cref{fig:est_chemistry} displays all 33 target problems we identified in the literature. These include ground state preparation and estimation problems, such as FeMoco  and Cytochrome P450, as well as computational catalysis for CO$_2$ fixation~\cite{app_vonBurg2021} and other strongly-correlated problems~\cite{Li2017,Blunt2022,elfving2020}. The figure shows that solving scientifically relevant, hard quantum chemistry problems will require the capability to run at least $10^8$ gates on about $10^3$ qubits, with estimates going up to $10^{12}$ gates and beyond, depending on the specific application and algorithm under consideration.

We summarize our findings for this section as follows:

\begin{finding}
    Resource estimation for computing ground state energies for quantum chemistry systems is among the most mature disciplines in quantum algorithm development. The workhorse algorithm has become \gls{QPE}+\gls{qubitization}. Significant progress over the last five years in terms of improved factorizations and compression of the two-electron integrals has reduced the resources required to solve key benchmark problems by orders of magnitude. We expect that future improvements to quantum algorithms will continue to reduce the resources required for applications from all domains in the \gls{DOE} \gls{QRW}.
\end{finding}

\section{Nuclear and high energy physics}
\label{sec:app_hep}

\gls{QC} holds significant potential for impact in nuclear and high energy physics. 
Despite the great success of classical lattice field theory, many phenomena remain out of reach for euclidean path-integral
Monte Carlo simulations. For instance, real time dynamics and out-of-equilibrium properties can not be currently studied with classical methods. One of the most important field theories that could take advantage of \gls{QC} is the strong force.
Interactions between quarks and gluons that make up hadrons are described by the theory of \gls{QCD}. The resulting strong force is one of the four primary computing focus areas delineated by the \gls{DOE} in the quantum information science applications roadmap~\cite{DOE_roadmap}. Hadrons are responsible for almost all the mass of matter, but the details of how hadrons form and more generally how nuclei form remain a mystery. This lack of knowledge has profound implications on our understanding of the evolution of the universe, and many experimental facilities like the Large Hadron Collider aim to help increase our knowledge for fundamental physics. These experimental searches require a strong theoretical and computational footing for progress to be made. As already mentioned, classical computing is limited to time-independent studies for only a subset of the strong force phenomenology, as first principle calculations are hindered by the exponential resource requirements~\cite{Jordan:2012, Bauer:2023}. QC will provide scientists with access to real-time dynamics and finite density, thus increasing the reach to all the phenomenology of the strong force. In field theories there is an additional complication arising from the encoding of both fermionic and bosonic degrees of freedom, which can increase the overhead and lead to non-local interactions. Other areas of interest in \gls{HEP} that could benefit from \gls{QC} are particle scattering~\cite{Chai2025fermionicwavepacket} and particle dynamic properties such as collective neutrino flavor oscillations. For a detailed summary we refer the reader to recent review articles~\cite{PRXQuantum.5.037001}.

Here we provide a collection of applications aimed at summarizing the current state of research in adopting \gls{QC}. %
The Schwinger model is one of the simplest gauge theories. A topological term in the model leads to the infamous sign problem in the classical Monte Carlo method, which means classical computational resources needed scale exponentially. As there is no sign problem for \gls{QC}, this model is a good use case for early investigations. In~\cite{app_Sakamoto2024}, the authors simulate time evolution by combining a block encoding of the Hamiltonian, after gauge fixing and mapping with the Jordan-Wigner transform, with the \gls{QSVT}. In~\cite{app_Rhodes2024}, the authors develop protocols for time evolution for U(1), SU(2), and SU(3) lattice gauge theories using a variety of methods including \gls{LCU} + \gls{qubitization} / \gls{Trotter}ization. For geometrically local Hamiltonians, one can take advantage of the HHKL algorithm~\cite{HHKL} to further reduce the cost of simulating time evolution~\cite{app_Rhodes2024}. Related results on \gls{QC} for lattice gauge theories can also be found in~\cite{kan2022,Davoudi2023}.

Other \gls{HEP} applications included in this work are collective neutrino flavor oscillations, which influence the dynamics of core-collapse supernovae and neutron star mergers, and thus terrestrial detection of these events. Neutrino processes are among the most challenging aspects of current \gls{HPC} numerical simulations of these environments, a situation complicated by the inherent quantum many-body dynamics in the problem due to neutrino-neutrino interactions, which demand a quantum treatment.  There have been recent studies on both \gls{qubit} and \gls{qutrit} processors~\cite{Spagnoli:2025} where the time evolution has been performed with first order \gls{Trotter} steps. For recent resource estimations from Los Alamos National Laboratory see~\cite{Hall:2021,app_lanl}. Here, we combine the latest estimates for the \gls{Trotter} error~\cite{Spagnoli:2025} to provide conservative bounds on the quantum resources required. The number of \gls{Trotter} steps required for a given simulation time $t$ and target overall error $\epsilon$ is $\mathcal{O}\left(\frac{t^2\mu N}{\epsilon}(\Delta\omega_\mathrm{max}+\mu)\right)$, where $\mu$ is the strength of the two-body interaction and $\Delta\omega_\mathrm{max}$ is the maximal difference between the Fourier mode frequencies. For our analysis, we estimate $\Delta\omega_\mathrm{max} \approx 100 \mu$, and $t= \mathcal{O}(\mu^{-1})$. For each \gls{Trotter} step, the number of $T$ gates is $n_G = 50 (N + N(N-1)/2)$. Using this number of gates per \gls{Trotter} step, we provide the total number of gates required for different system sizes of interest under a total error budget of $0.01$ for the full time evolution. %

\begin{figure}[ht]
   \centering
    \includegraphics[width=0.8\textwidth]{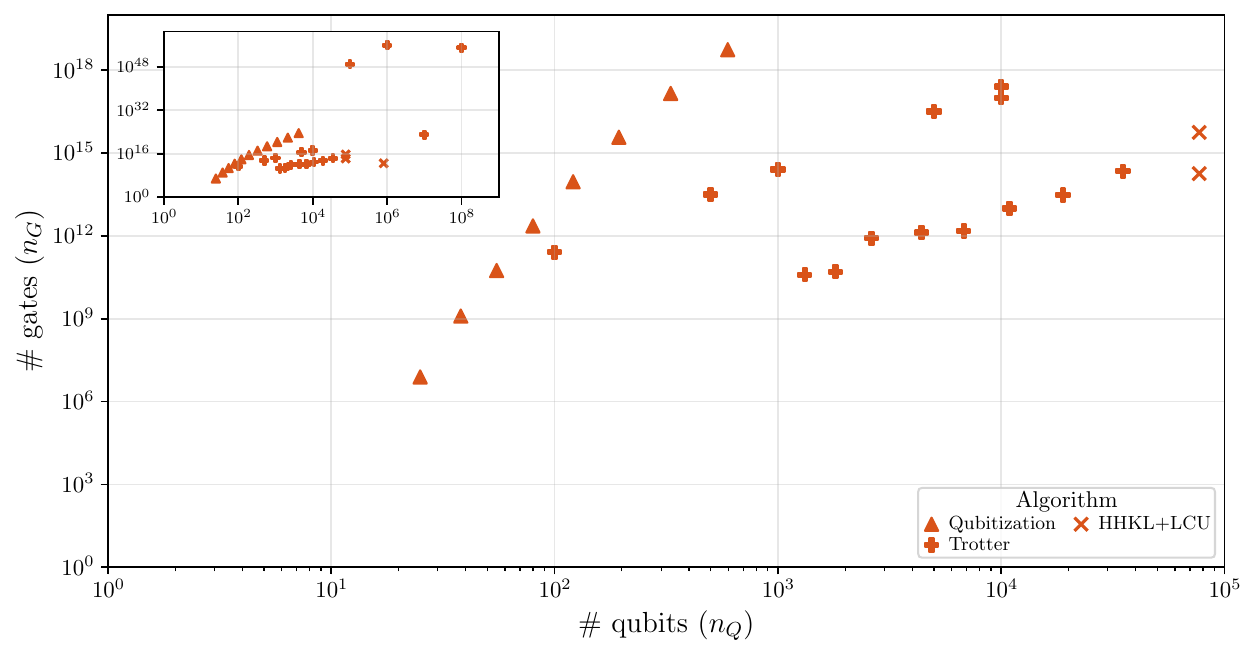}
    \caption{Overview of various resource estimates for computational problems related to nuclear and high energy physics in terms of number of \glspl{logical qubit} ($n_Q$) and number of gates ($n_G$). The inset shows all 32 $\perf$-vectors including some at very large \gls{quantum gate} and/or \gls{qubit} count.} 
    \label{fig:est_hep}
\end{figure}

We conclude by providing an overview of the resource estimates for 32 data points from both applications discussed here in~\Cref{fig:est_hep}. Compared to~\Cref{fig:est_materials,fig:est_chemistry}, we see observe that the number of gates grows significantly faster as a function of \gls{qubit} count (for an easier visual comparison see~\Cref{fig:est_overview}), which we attribute to the larger encoding overhead for applications in \gls{HEP}, and to a lesser degree to the extent to which the topic has been studied so far. Indeed, in comparison to materials science and chemistry, \gls{QC} for \gls{HEP} is less developed and recent works, as outlined here, have been focused primarily on resource estimation for simplified field theory models. We expect more progress to come in the near future as algorithmic approaches are currently being developed to reduce the quantum resources needed for both encoding the problems and for computing the dynamics~\cite{PhysRevD.101.114502,ciavarella2025}.

\section{Other application areas}
\label{sec:app_other}
Beyond materials, quantum chemistry, and high energy physics, which form the core of the \gls{DOE} \gls{QRW}, there are multiple other areas of computational science where quantum computers can impact the \gls{DOE} \gls{SC} mission. Here we summarize the state of affairs for solving problems in linear algebra, differential equations, optimization, and machine learning with quantum computers. We do not include resource estimates for these categories as they often lack standardized benchmark problems, rely on heuristics, do not describe an application with an end-to-end speedup (including input/readout), or have only examined speedups using oracle queries.

\paragraph{Linear algebra.} 
The field of quantum linear algebra originates from the work by Harrow, Hassidim, and Lloyd (HHL)~\cite{harrow09} on solving linear systems of equations on quantum computers with an exponential scaling improvement on the problem size at the cost of a polynomial increase in the dependence on problem conditioning. Subsequent work further improved the dependence on the precision~\cite{childs17} and problem conditioning~\cite{Orsucci2021,ambainis:LIPIcs.STACS.2012.636}. \gls{QSVT}~\cite{gilyen19} has become the de facto standard for solving various kinds of linear algebra problems, including linear systems and matrix functions. A key bottleneck is the block encoding input model, which can be highly resource efficient for structured data~\cite{simax_be,Sunderhauf2024} but scales exponentially for unstructured data~\cite{fable}. Other approaches reduce the complexity of the input model by trading off for an increased sample complexity~\cite{wang24}. We conclude that quantum linear algebra kernels hold great potential, mainly due to the exponential improvement in the dependence on the problem size, and can have considerable implications for a vast number of scientific codes. However, end-to-end complexities and speedups have yet to be developed and will ultimately hinge on problem characteristics (structure) and output requirements. %

\paragraph{Simulation of differential equations.}
The simulation of differential equations is ubiquitous in the physical sciences and engineering, ranging from topics like fluid dynamics and plasma physics, to structural mechanics and computational materials science. Consequently, many applications that are run at \gls{NERSC} solve a system of differential equations in some form or another. Classical algorithms for solving differential equations form a mature subfield of computational sciences and many codes and solvers have been developed. At the same time, it remains challenging to accurately simulate certain physical phenomena like turbulence, where due to the immense range of interacting lengths and time scales present, extraordinarily fine grids and small time steps are necessary and lead to computationally prohibitive costs. 

Research into quantum algorithms for simulating differential equations has seen steady progress since it was first proposed~\cite{leyton2008} shortly after the introduction of the celebrated HHL algorithm for linear systems~\cite{harrow09}. Theoretical results can be categorized in various ways, for example algorithms for \gls{ODE} systems~\cite{Krovi2023} and for \gls{PDE} systems~\cite{Childs2021}, or alternatively algorithms for linear differential equations~\cite{Shi24} and for nonlinear systems of differential equations~\cite{liu21}. 

Many approaches convert the system of differential equations to an equivalent problem for which a quantum algorithm exists. This includes mapping the system of differential equations to a system of linear equations~\cite{Berry2014,Childs2021,Childs2020,lloyd2020}, to a Hamiltonian simulation problem by introducing auxiliary variables~\cite{Shi24,wright24}, or to Fourier space using the quantum Fourier transform (QFT) that leads to simple \gls{quantum circuit} designs~\cite{lubasch2025}. Nonlinear problems require careful consideration as solving them on a quantum computer is intrinsically difficult due to the linear nature of quantum mechanics (see Ref.~\cite{Tennie2025} for a review). One approach is to use linearization~\cite{liu21,tennie2024} to transform a nonlinear system to a linear one. Another method is to use a hybrid classical-quantum workflow~\cite{lubasch19, kryiienko2021, pool24} that translates the differential equation to a minimization problem, which can be evaluated on a quantum computer and variationally minimized using a classical computer. Similar to the case of quantum linear algebra, simulating differential equations on quantum computers holds great potential, but showing evidence of end-to-end speedups remains an open problem.

\paragraph{Optimization and search.} 
Quantum algorithms for combinatorial optimization have been widely studied after the initial introduction of Grover's search algorithm~\cite{grover}. Grover's algorithm provides an asymptotic quadratic speedup for unstructured search problems. Recent work studying Max-$k$-SAT ~\cite{Cade2023quantifyinggrover} suggests that practical end-to-end quantum speedups may be challenging to achieve once constant factors are taken into consideration. Furthermore, the overhead required for achieving fault-tolerance using a \gls{surface code} encoding may outweigh the advantages a quadratic speedup offers~\cite{PRXQuantum.2.010103}. This does not rule out that heuristic and approximate quantum approaches~\cite{farhi2014,PhysRevLett.134.160601} to optimization can offer substantial benefits. Moreover, recent work suggests that certain optimization problems may be amenable to beyond-quadratic speedups, for example~\cite{PhysRevX.15.021077} suggests a quartic speedup for the planted noisy $k$ XOR problem.

Scheduling algorithms, specifically load-balancing algorithms, are a category of optimization problems of great relevance to the \gls{HPC} community. Modern \gls{HPC} distributed memory problems have increased substantially in complexity, both technically and scientifically. Multi-threading, \gls{GPU} performance, code coupling, multi-physics, multi-scale, I/O methodologies, digital twin techniques, and workflow applications have all increased the complexity of optimizing the use of \gls{HPC} compute resources for large-scale, dynamically variable simulations. As Moore’s Law reaches its end and “re-compile and go faster” performance gains disappear, it’s critical for distributed memory \gls{HPC} applications to use available resources as efficiently as possible to maintain scientific scalability with next-generation architectures. 

Classical load-balancers use approximation algorithms due to time constraints — dynamic in-situ load-balancing cannot substantially increase the time-to-solution. So, simple load balancers typically use a broad overview of application performance and aim to do “good enough, fast enough”. High-dimensional optimization solutions have the potential to provide more accurate descriptions of applications, and therefore create more accurate solutions that greatly improve resource utilization. Quantum optimization algorithms that can quickly solve a complex scheduling problem would be of board interest to a variety of distributed memory \gls{HPC} simulations.

This category of quantum algorithm research has started to be explored using quantum annealing~\cite{SC24LB} as has the similar field of task scheduling~\cite{tiger2020}.

\paragraph{Machine learning and artificial intelligence.}
The integration of \gls{QC} with \gls{ML} and \gls{AI} is poised to overcome several fundamental computational barriers faced by classical methods. \glspl{QPU} offer the potential to address key deficiencies in classical \gls{ML}, particularly in the efficient handling of high-dimensional feature spaces, combinatorial optimization, and probabilistic sampling. For example, quantum kernel methods can map data into exponentially large Hilbert spaces, enabling the creation of classifiers and regressors that are classically infeasible~\cite{schuld2022quantum}. In the context of optimization -- a central task in \gls{AI} and \gls{ML} -- \glspl{VQA} leverage quantum parallelism to explore complex solution landscapes, providing a pathway to escape local minima that often trap classical heuristics. This has implications for the training of generative models, solving large-scale combinatorial problems, and optimizing policies in reinforcement learning~\cite{cerezo2021variational}. Additionally, quantum algorithms can accelerate sampling from complex probability distributions, directly impacting the scalability of Bayesian inference and other probabilistic modeling tasks~\cite{liu2021rigorous}. As quantum hardware matures, these algorithmic advances could enable \gls{ML} and \gls{AI} models to tackle problems previously out of reach for classical computation.

A key challenge, however, is the quantum data loading problem: encoding large classical datasets into quantum states can require resources that scale polynomially with dataset size, potentially eliminating any quantum speedup. As a result, most practical approaches currently focus on hybrid quantum-classical models, where classical computers handle data-intensive tasks and QPUs are used for select, computationally hard subroutines~\cite{schuld2022quantum, cerezo2021variational}. Another challenge lies in the large regions in the parameter space of quantum circuits where the cost function gradient becomes exponentially small, making optimization extremely difficult. This phenomenon is commonly referred to barren plateaus~\cite{McClean:2018,Garcia:2023}.

The relationship between ML and quantum computing is also reciprocal, as ML techniques are crucial for advancing quantum hardware. Machine learning methods, especially neural networks, are increasingly used for \gls{QEC}, learning to decode and correct errors in quantum systems more efficiently than traditional methods~\cite{harper2021fast}. Furthermore, reinforcement learning agents have demonstrated success in automating the calibration and control of qubits, leading to high-fidelity \gls{quantum gate} operations and improved system stability~\cite{niu2019universal, sivak2022modelfree}. This interplay is accelerating progress toward scalable \gls{FTQC}.

\section{Summary}
\label{sec:app_overview}

\begin{figure}[htp]
    \centering
    \includegraphics[width=0.8\linewidth]{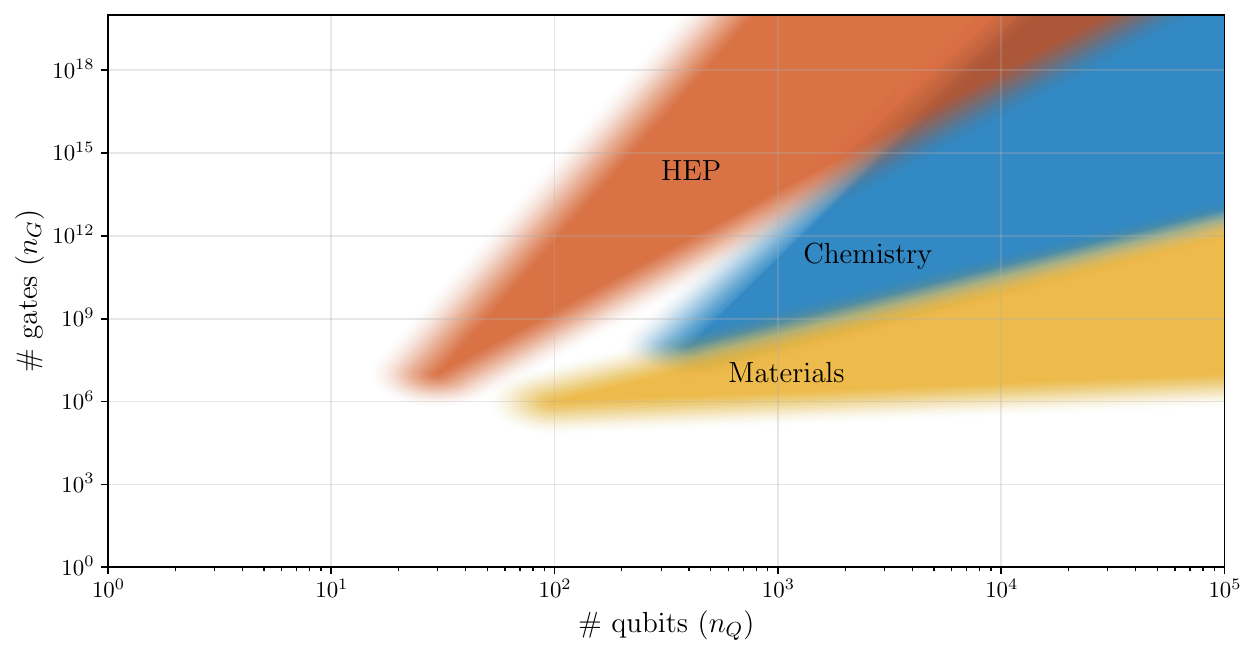}
    \caption{Qualitative overview of the different regions and scaling of $\perf$-vectors for applications in condensed matter and materials science (yellow), quantum chemistry (blue), and nuclear and high energy physics (red).}
    \label{fig:est_overview}
\end{figure}

We summarize the qualitative differences in $\perf$-vectors for condensed matter/materials science, quantum chemistry, and \gls{HEP} in~\Cref{fig:est_overview}. We observe that in general the condensed matter problems require comparatively the fewest number of gates for a given \gls{qubit} count (which we use as a proxy for problem size) and that they exhibit the most advantageous scaling behavior. We attribute this to the fact that (1) these problems either map 1-to-1 to qubits or with minimal encoding overhead, and (2) the fields of condensed matter physics and materials science have spent significant efforts to study, develop, and optimize quantum algorithms and applications. The $\perf$-vectors of quantum chemistry applications fall in the intermediate region in between materials and \gls{HEP}. This is due to the greater \gls{qubit} encoding overhead compared to condensed matter physics and the greater maturity of the field compared to \gls{HEP}. Finally, resource estimates for \gls{HEP} require the largest $\perf$-vectors and the poorest scaling with problem size as they have both a high encoding overhead and the applications studied are not as mature. Based on these observations, we expect that condensed matter physics and materials science problems will be among the first domains of interest to \gls{DOE} \gls{SC} mission to benefit from quantum computers with chemistry and \gls{HEP} likely following later. We summarize this with the following finding.

\begin{finding}
    Condensed matter and materials science applications are expected to be impacted first as they map naturally from spin to qubit systems and require the fewest resources. Quantum chemistry applications require an intermediate amount of resources and have seen steady algorithmic improvements. \gls{HEP} applications require the most resources among the problems that were considered and likely offer opportunities for significant algorithmic improvements.
\end{finding}

With regard to algorithms, we expect Hamiltonian simulation, either through \gls{Trotter}ization or \gls{QSVT}, and determination of ground state properties to dominate the span across the \gls{QRW}.

\begin{finding}
    All known quantum applications within the physical sciences for which evidence of exponential speedups exist fall in two categories: dynamical properties (correlation functions, response functions, adiabatic state preparation, ...) and static properties (\gls{GSEE}, excited states, ...). Consequently, the main algorithmic primitives are Hamiltonian simulation, either using \gls{Trotter}ization (minimal \gls{qubit} requirements) or \gls{qubitization}/\gls{QSVT} (optimal \gls{quantum gate} requirements), and \gls{QPE} which relies on Hamiltonian simulation as a subroutine. For quantum chemistry in particular, the majority of resource estimates focus on \gls{GSEE} with \gls{QPE}; dynamical problems are comparatively understudied.
\end{finding}

Our analysis has centered around the $\perf$-vector which represents the volume of a \gls{quantum circuit}. As we will discuss in \cref{chap:vendorroadmaps}, the $\perf$-vector determines which quantum computer can run which application. However, depending on the nature of the end-to-end application, a different number of circuit repetitions or \glspl{shot} may be required. We write the number of shots as $N_s$ and also refer to this quantity as the sampling complexity. The combination of the $\perf$-vector and the number of shots $N_s$ will determine our execution time estimate presented in~\cref{chap:exec_time_throughput}.

\begin{table}[ht]
\caption{Three prototypical sampling complexities for different use cases considered in this report.}\label{tab:sampling_complexity}
\centering
\begin{tabular}{>{\columncolor{gray!20}}lc}
\toprule\\
\textbf{Application} & \textbf{$N_s$}\\
\midrule
Ground state energy via \gls{QPE} & $10^{\phantom{1}}$\\
Measuring an observable $\braket{\psi |O| \psi}$ (\gls{VQE}, \gls{Trotter}, ...) & $10^3$\\
Dynamical simulations requiring multiple configurations & $10^5$\\
\bottomrule
\end{tabular}
\end{table}

\Cref{tab:sampling_complexity} lists three different prototypical sampling complexities that we use for the remainder of this report, and in particular in~\cref{chap:exec_time_throughput}. For applications that use \gls{QPE} to determine a ground state energy, multiple trials may be required when the reference state is not the exact ground state. However, we expect \gls{QPE} to be used only when a good reference state is available and use $N_s = 10$. For applications that require measuring an observable (e.g. magnetization) $\braket{\psi |O| \psi}$, with $\ket{\psi}$ prepared using \gls{VQE}, \gls{Trotter}, or some other state preparation procedure, we assume a larger $N_s = 10^3$ as many samples are required to reduce the variance of the estimator according to the standard quantum limit. Finally, we contemplate a workflow where the quantum computer is used to support dynamical simulations of some quantum many-body system which requires simulating many different configurations of the many-body system. An example includes calculating the binding affinity of a drug to a target protein which can demand millions of energy calculations using different molecular configurations~\cite{PhysRevResearch.4.043210}. Optimistically, we assume $N_s = 10^5$ for a similar but smaller workflow.

\chapter{Vendor technology roadmaps}
\label{chap:vendorroadmaps}
\myglsreset

In this section, we collect and unify the public technology roadmaps of ten vendors active in the quantum computing space. We focus on the three technological platforms that are currently leading development efforts: superconducting qubits, trapped ions, and neutral atoms. The different vendors report their roadmap milestones according to different metrics, including:
\begin{itemize}
    \item The number of qubits in the system, $n_Q$ (\glspl{physical qubit} and/or \glspl{logical qubit});
    \item The error rate of the system, $\epsilon$;
    \item The number of quantum gates the system can reliably execute, $n_G$;
    \item The \gls{quantum circuit} depth the system can reliably execute, $d_C$.
\end{itemize}

Based on the milestone data provided, we uniformize the roadmaps against the volumetric $\perf$-vector introduced in~\Cref{eq:pvector}. We use the following rules-of-thumb to convert, respectively, error rates and \gls{quantum circuit} depths to an equivalent number of gates:
\begin{align}
n_G = \epsilon^{-1},\quad \text{and} \quad n_G = n_Q \cdot d_C / 2.    
\end{align}
 For vendors that report a maximum \gls{quantum circuit} depth $d_C$, we convert to \gls{quantum gate} depth by assuming a maximally dense \gls{quantum circuit} of $3 n_Q / 4$ gates per circuit layer. We stress that our rules-of-thumb are a first order approach to obtain estimates that can be reasonably compared to each other. A more precise analysis would require more detailed information that is often publicly unavailable or perhaps still unknown.

The different vendor roadmaps report the number of qubits, the first metric, at their technology milestones either by specifying the number of \glspl{physical qubit}, \glspl{logical qubit}, or both, or remaining ambiguous. 
The distinction between \glspl{physical qubit} and \glspl{logical qubit} becomes relevant as the field matures from \gls{NISQ} systems~\cite{Preskill2018} to (early) \gls{FTQC}~\cite{shor96}. In a \gls{NISQ} system the (noisy) \glspl{physical qubit} are used directly as a platform for logical (but noisy) quantum operations, while in a \gls{FTQC} many imperfect \glspl{physical qubit} work together in a carefully orchestrated manner to represent fewer \glspl{logical qubit} with improved performance. We observe that different milestones either specify that the proposed system is a \gls{NISQ} system or a (early) \gls{FTQC} system, or do not specify which category the system falls into. In the latter case, we assume that systems that can run $10^4$ gates or fewer are \gls{NISQ} systems (labeled \textit{N}), systems that can reliably run between $10^4$ to $10^6$ gates are early \gls{FTQC} systems (labeled \textit{EF}), and systems that can reliably run more than $10^6$ gates are \gls{FTQC} systems (labeled \textit{F}). The number of qubits $n_Q$ that we report in the $\perf$-vector is the number of \glspl{physical qubit} for \gls{NISQ} systems and the number of \glspl{logical qubit} for (early) \gls{FTQC} systems. The $\perf$-vector then estimates an upper bound on the performance of the technology that each vendor is developing. Any quantum application that requires at most $n_Q$ qubits and at most $n_G$ gates will be able to run on the forthcoming system. We denote this as $\perf^{\text{app}} \leq \perf^{\text{qc}}$.

It follows from our discussion on \gls{QEM} and \gls{QEC} in~\cref{sec:errors} that these protocols can improve the figures of merit and performance bounds if we allow to take more samples and increase the classical post processing cost. For the purposes of our analysis, we assume that \gls{QEM} can at most double the number of gates that can be executed on a \gls{NISQ} device, i.e.~$n_{G,\text{QEM}} = 2 n_G$. In the case of \gls{PEC}, we get from~\Cref{eq:pec_gates} that this requires a sampling overhead of $C_{\text{PEC}} = e^8 \approx 3,000$, which is at the limit of what we deem practical. For roadmap milestones pertaining to \gls{NISQ} devices, we present two values for the number of gates: the value based on the information provided by the vendor, and twice this value when assuming \gls{QEM} is used. The second, higher number $\perf_{\text{QEM}}^{\text{qc}} = (n_Q, 2n_G)$ will be considered the \emph{region of extended capabilities} of the \gls{NISQ} device and will be plotted as a light shaded area. We remark that for \gls{NISQ} milestones there is no leeway in the $n_Q$ parameter of the figure of merit as every individual \gls{physical qubit} is directly used for computation. For \gls{FTQC} milestones, we consider that \gls{QEC} and \gls{QEM} may be used in conjunction and allow for a reduction of the code distance $d$ by five while maintaining the same error rate at the cost of a sample overhead factor $C \leq 100$ and more intensive post-processing~\cite{qem_Suzuki2022}. This leads to a figure of merit $\perf^{\text{qc}}_{\text{QEM+QEC, 1}} = (\hat{n}_Q, n_G)$ with $\hat{n}_Q > n_Q$ and the exact increase depends on the error correction code and \gls{code distance}. Similarly, at fixed \gls{code distance}, one could reach error rates similar to \gls{code distance} $d+5$ leading to a figure of merit $\perf^{\text{qc}}_{\text{QEM+QEC, 2}} = (n_Q, \hat{n}_G)$ with $\hat{n}_G > n_G$. We use these assumptions to estimate a region of extended capabilities for \gls{FTQC} milestones that will be plotted as a light shaded area. In this case, both $n_Q$ (by reducing the \gls{code distance}) \emph{and} $n_G$ (by keeping the \gls{code distance} constant but decreasing the effective error rate) can be improved using \gls{QEC}+\gls{QEM}. %

As vendors support different underlying \glspl{physical qubit}, their error correction strategies will likely differ as well. By default, we will assume \glspl{surface code} for error correction of operations and PEC for error mitigation. Whenever more detailed information about the future architecture is available, we aim to incorporate it in our estimates. Depending on the hardware characteristics, and the sometimes partial information provided by vendors, we fill in the gaps based on our assumptions stated above. %

\section{Overview of individual vendor roadmaps}
\label{sec:roadmaps}

In this section we provide an overview of all public roadmap data we identified at the time of writing. We categorize the vendors according to the core qubit technology of their hardware platforms:
\begin{itemize}
    \item \cref{ssec:superconducting}: superconducting circuits, which include transmon, coaxmon, and cat qubit types;
    \item \cref{ssec:ions}: trapped ions;
    \item \cref{ssec:atoms}: neutral atoms;
    \item \Cref{ssec:other}: other technologies and vendors that have not released official roadmaps. 
\end{itemize}   
Within each section, vendors are listed in alphabetical order. We refer to~\cite{ianreppel} for a different overview of quantum vendor roadmaps. %

As our goal is to glean the expected capabilities of near-term quantum systems that the industry is pursuing, we strive to cover as many public roadmaps as possible across different qubit modalities, but we do not claim that our overview is complete. Furthermore, as the field of quantum computing is rapidly progressing, existing roadmaps are updated regularly and new roadmaps are announced frequently. The data we collected reflects a moment in time and is subject to rapid change.

For each vendor, we provide a brief overview of the key concepts and ideas behind their technology and approach to building a scalable quantum computer. We summarize each roadmap in a two panel figure, focusing on recent and future milestones. On the left hand side, a table shows data provided by the vendor in black font and inferred from our assumptions in blue font; the table columns indicate the year for each milestone, the name of the milestone where available, the type of milestone (N for \gls{NISQ}, EF for early \gls{FTQC}, or F for \gls{FTQC}), the number of qubits $n_Q$ (both \glspl{physical qubit} and \glspl{logical qubit}), and a metric that can be related to the number of gates (i.e. the error, the number of gates, or the \gls{quantum circuit} depth). On the right hand side, a scatter plot shows the $\perf$-vector for each milestone, a dark shaded bounding box highlights the milestone with the largest $\perf$-vector, and a light shaded area indices the region of extended performance, based on \gls{QEM} for \gls{NISQ} milestones and \gls{QEC}+\gls{QEM} for \gls{FTQC} milestones.

\subsection{Qubits based on superconducting circuits}
\label{ssec:superconducting}

\paragraph{Alice \& Bob.}

Alice \& Bob is a start-up based out of Paris, France that is developing quantum computers based on \emph{cat qubits}~\cite{MIRRAHIMI2016}. Cat qubits represent an innovative approach to building more stable quantum computers by encoding information in superpositions of classical-like states of microwave light, analogous to Schr\"{o}dinger's cat being in two states at once. The physical realization of cat states is typically achieved within superconducting microwave resonators, which act as cavities to confine microwave photons that make up the coherent states. Josephson junctions introduce a non-linearity in the superconducting circuit that is required to control the state and perform quantum operations.

An advantage of cat qubits is an inherent, hardware-level protection against bit-flip errors, one of the two main types of quantum errors; this protection becomes exponentially stronger as the size (average photon number $\bar n$) of the cat state increases. While this comes at the cost of a smaller linear increase in phase-flip errors, the resulting biased noise profile -- where one error type is far less likely than the other -- significantly simplifies the complex task of \gls{QEC}. The \gls{QEC} reduces from a 2D problem, possibly requiring a \gls{surface code} to correct both bit- and phase-flips, to a (nearly) 1D problem which can be stabilized by a $[\![2d-1, 1, d]\!]$ repetition code leading to a linear reduction in qubits compared to a \gls{surface code} of similar \gls{code distance}. In this case, we can estimate the logical phase-flip error rate as~\cite{Ruiz:2025},
\begin{align}
\epsilon_L = 0.056 \times \left( \frac{\bar{n}^{0.86} k_1/k_2}{0.013} \right)^{(d+1)/2} +(d-1)e^{-2\bar{n}}, 
\label{eq:cat_rep_code}
\end{align}
where $k_1$ is the single photon loss rate, and $k_2$ is the two-photon dissipation rate. Based on~\cite{Ruiz:2025}, we assume that $k_1/k_2\approx 10^{-4}$ and $\bar{n}=11$, that the encoding ratio of \glspl{logical qubit} to \glspl{physical qubit} is $2d-1$, and that there is hardware-level protection against bit-flips leading to a logical bit-flip error rate smaller than $10^{-9}$ for average photon number $\bar{n}=11$. 

\begin{figure}[htp]
    \centering
    \begin{minipage}{0.5\textwidth}
        \centering
        \raisebox{1.5\height}{ 
        \small
        \begin{tabular}{>{\columncolor{gray!20}}lccccc}
            \toprule
            \rowcolor{gray!40} \textbf{Year}  & \textbf{Name} & \textbf{Type} & \multicolumn{2}{c}{\textbf{$n_Q$}} & \textbf{error} \\
            \rowcolor{gray!40} & & & Phys. & Log. & \\
            \midrule
            2024 & Boson 4 & N & 1 & 0 & -- \\
            2025  & Helium & EF & 16 & 1 & $10^{-2}$\\
           \estimate{2027} & Lithium & EF & 48 & 4 & $10^{-3}$ \\
           \estimate{2029} & Beryllium & EF & 250 & 5 & $10^{-4}$\\
            2030  & Graphene & F & 2,000 & 100 & $10^{-6}$\\
            \bottomrule
        \end{tabular}}
        \label{tab:rm_ab}
    \end{minipage}%
    \hfill
    \begin{minipage}{0.5\textwidth}
        \centering
        \includegraphics[width=\textwidth]{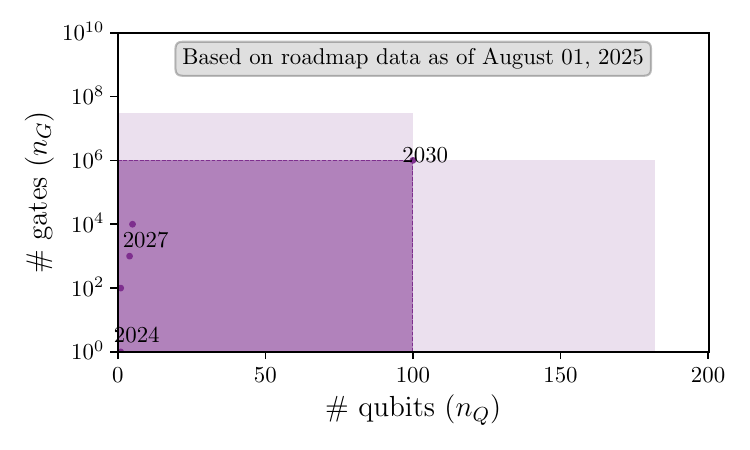}
        \label{fig:rm_ab}
    \end{minipage}
    \caption{Roadmap data for: Alice \& Bob~\cite{roadmap_alicebob} -- \textit{Cat qubits}. \textit{Left}: Table with technology milestones. The Lithium and Beryllium systems have no target year on the roadmap and we interpolate the dates based on the target years for the Helium and Graphene systems. \textit{Right}: Plot with equivalent technology milestones on number of qubits versus number of gates axes. The light shaded area sketches the performance boundary using a quantum error mitigation approach on the largest-scale system. \label{fig:alicebob}}
\end{figure}

\Cref{fig:alicebob} summarizes the roadmap Alice \& Bob released in December 2024~\cite{roadmap_alicebob}. It lays out 5 technology milestones starting with the Boson 4 system released in 2024, which consists of a single cat qubit and supports a limited (non-universal) set of single-qubit gates, and culminates in the Graphene system in 2030 which is projected to have 100 \glspl{logical qubit} at an error rate of $10^{-6}$. Based on the available data, we assume Alice \& Bob to pursue an architecture inspired by a repetition code to suppress phase flip errors as their main strategy until the logical phase-flip error rate becomes comparable to the bit-flip error. We thus use~\Cref{eq:cat_rep_code} to estimate the region of extended performance. Around the 2030 milestone, Alice \& Bob's roadmap suggests they will start using sparse LDPC codes to further reduce the logical error rates.

\paragraph{Google Quantum AI.}

\begin{figure}[htp]
    \centering
    \begin{minipage}{0.52\textwidth}
        \centering
        \raisebox{1.25\height}{ 
        \footnotesize
        \begin{tabular}{>{\columncolor{gray!20}}lccccc}
            \toprule
            \rowcolor{gray!40} \textbf{Year} & \textbf{Name} & \textbf{Type} & \multicolumn{2}{c}{\textbf{$n_Q$}} & \textbf{error} \\
            \rowcolor{gray!40} & & & Phys. & Log. & \\
            \midrule
            2019   & Sycamore  & N & 53 & -- & -- \\
            2023   & Sycamore-2  & N & 72 & -- & -- \\
            2024   & Willow  & N & 105 & -- & $3\times10^{-3}$ \\
            \estimate{2027}  & Milestone 3 & EF & $10^{3}$ & \estimate{5} &  $10^{-6}$\\
            \estimate{2030}  & Milestone 4  & EF & $10^{4}$ & \estimate{50} &  $10^{-6}$\\
            \estimate{2033}  & Milestone 5  & F & $10^{5}$ & \estimate{502} &  $10^{-6}$\\
            \estimate{2036}  & Milestone 6  & F & $10^{6}$ & \estimate{1545} &  $10^{-13}$\\
            \bottomrule
        \end{tabular}}
        \label{tab:rm_google}
    \end{minipage}%
    \hfill
    \begin{minipage}{0.48\textwidth}
        \centering
        \includegraphics[width=\textwidth]{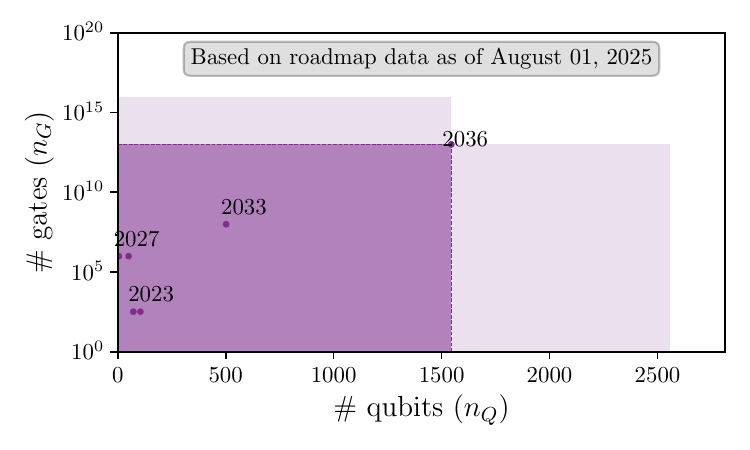}
        \label{fig:rm_google}
    \end{minipage}
    \caption{Roadmap data for: Google Quantum AI~\cite{roadmap_google} -- \textit{Superconducting qubits}. \textit{Left}: Table with technology milestones. \textit{Right}: Plot with equivalent technology milestones on number of qubits versus number of gates axes. The light shaded area sketches the performance boundary using a quantum error mitigation approach on the largest-scale system. \label{fig:google}}
\end{figure}

Google Quantum AI is building quantum processors based on superconducting transmon qubits arranged in a square grid topology where each qubit has four nearest-neighbors. This naturally matches the connectivity requirements for a \gls{surface code}~\cite{Acharya:2023}. Google's current 105 qubit system named Willow was announced in late 2024~\cite{Acharya2024}, while the previous system named Sycamore was originally announced in 2019 with 53 qubits~\cite{Arute2019} and upgraded to 70 qubits in 2023~\cite{Morvan2024}. Google Quantum AI presents their technology roadmap~\cite{roadmap_google} in terms of 6 milestones and we summarize them in~\Cref{fig:google}. The first milestone, labeled \emph{beyond classical}, was achieved in 2019 on the Sycamore device~\cite{Arute2019}. Milestone 2, labeled \emph{Quantum Error Correction}, was achieved in 2023 on the second generation Sycamore-2 device~\cite{Acharya:2023}. Future milestones 3 through 6 continuously improve both the scale and the quality of the system and Google reports the expected number of \glspl{physical qubit} and the logical error rates in their roadmap. Based on this information, we estimate the \gls{code distance} using \Cref{eq:surface_code_error}, which in turns allows us to estimate the number of \glspl{logical qubit}. As their milestones have no estimates on when they will become available, we allow for a 3 year period between consecutive milestones, extrapolating from the announcement of Willow. We assume a physical error rate of $10^{-3}$ up to milestone 4, and $5\times 10^{-4}$ afterwards.

\paragraph{IBM Quantum.}

\begin{figure}[htp]
    \centering
    \begin{minipage}{0.5\textwidth}
        \centering
        \raisebox{1.4\height}{
        \scriptsize
        \begin{tabular}{>{\columncolor{gray!20}}lccccc}
            \toprule
            \rowcolor{gray!40} \textbf{Year} & \textbf{Name} & \textbf{Type} & \multicolumn{2}{c}{\textbf{$n_Q$}} & \textbf{$n_G$} \\
            \rowcolor{gray!40} & &  & Phys. & Log. & \\
            \midrule
            2020 & Falcon & N & 27 & -- & -- \\
            2022 & Eagle & N & 127 & -- & -- \\
            2024 & Heron & N & 133 & -- & 5,000 \\
            2025 & Nighthawk & N & 120 & -- & 5,000 \\
            2026 & Nighthawk & N & 120 & -- & 7,500 \\
            2027 & Nighthawk & N & 120 & -- & 10,000\\
            2028 & Nighthawk & N & 120 & -- & 15,000\\
            2029 & Starling & F & \estimate{8,000} & 200 & $10^8$\\
            2033+ & Blue Jay & F & \estimate{58,000} & 2,000 & $10^9$\\
            \bottomrule
        \end{tabular}}
        \label{tab:rm_ibm}
    \end{minipage}%
    \hfill
    \begin{minipage}{0.5\textwidth}
        \centering
        \includegraphics[width=\textwidth]{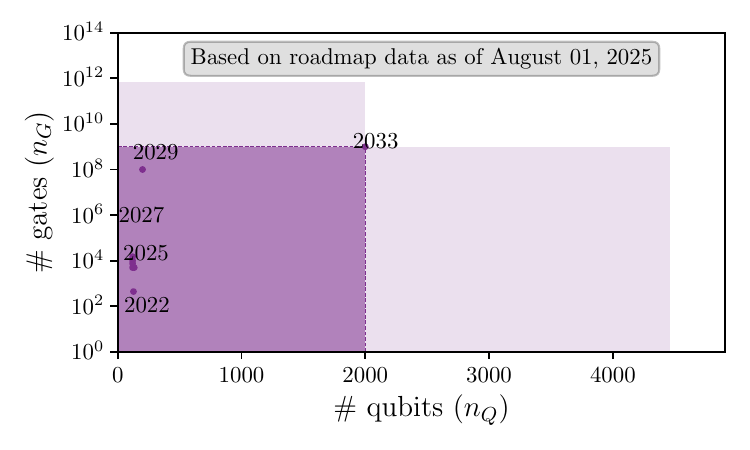}
        \label{fig:rm_ibm}
    \end{minipage}
    \caption{Roadmap data for: IBM Quantum~\cite{roadmap_ibm} -- \textit{Superconducting qubits}. \textit{Left}: Table with technology milestones. \textit{Right}: Plot with equivalent technology milestones on number of qubits versus number of gates axes. The light shaded area sketches the performance boundary using a quantum error mitigation approach on the largest-scale system. \label{fig:ibm}}
\end{figure}

IBM Quantum is building superconducting transmon quantum processors with the qubits arranged in a heavy-hex lattice in recent systems. The company has most recently updated their roadmap~\cite{roadmap_ibm,roadmap_ibm2} in June 2025 and \Cref{fig:ibm} summarizes the data in our standardized format. Up until 2028, IBM Quantum reports the number of \glspl{physical qubit} and gates. Starting from the Starling system in 2029, IBM Quantum is expecting a \gls{FTQC} device and the company reports the number of \glspl{logical qubit} and number of gates for the \gls{FTQC} systems on their roadmap.

Recent research on \gls{FTQC} architectures from IBM Quantum suggests that the company is pursuing a \gls{QEC} strategy based on \gls{qLDPC} codes~\cite{Cross:2024,Bravyi:2024}. More precisely, a modular architecture based on bivariate bicycle codes, dubbed the \emph{gross} code~\cite{Yoder:2025} appears to be a leading candidate for future systems on the IBM Quantum roadmap. This type of architecture will require modifications to the current heavy-hex topology by the introduction of (long-range) couplers~\cite{Bravyi2022}. The company has presented first experimental results that demonstrate entanglement generation across such a coupler~\cite{heya2025}. Based on~\cite{Yoder:2025}, we use a 10$\times$ reduction in the ratio of \glspl{physical qubit} to \glspl{logical qubit} over a \gls{surface code} \gls{logical qubit} with the same error rate as a rule of thumb. For Starling and Blue Jay time frames, we assume a physical error rate of approximately $10^{-3}$.

\paragraph{IQM.}

\begin{figure}[htp]
    \centering
    \begin{minipage}{0.5\textwidth}
        \centering
        \raisebox{1.5\height}{
        \begin{tabular}{>{\columncolor{gray!20}}lcccc}
            \toprule
            \rowcolor{gray!40} \textbf{Year}  & \textbf{Type} & \multicolumn{2}{c}{\textbf{$n_Q$}} & \textbf{error} \\
            \rowcolor{gray!40} & & Phys. & Log. & \\
            \midrule
            2024 & N    & 54 & -- & $1\times10^{-3}$ \\
            2025 & N   & 150 & -- &  $8 \times 10^{-4}$ \\
            2026 & N    & 300 & -- &  $6 \times 10^{-4}$ \\
            2027 & EF    & $10^{3}$ & 36 &  $10^{-5}$\\
            2028 & EF    & $5 \times 10^{3}$ & 180 &  $10^{-6}$\\
            2030 & F    & $4 \times 10^{4}$ & 720 &  $10^{-7}$\\
            2031 & F    & $10^{5}$ & 1,800 &  $10^{-8}$\\
            2033 & F    & $10^{6}$ & 7,200 &  $10^{-9}$\\
            \bottomrule
        \end{tabular}}
        \label{tab:rm_iqm}
    \end{minipage}%
    \hfill
    \begin{minipage}{0.5\textwidth}
        \centering
        \includegraphics[width=\textwidth]{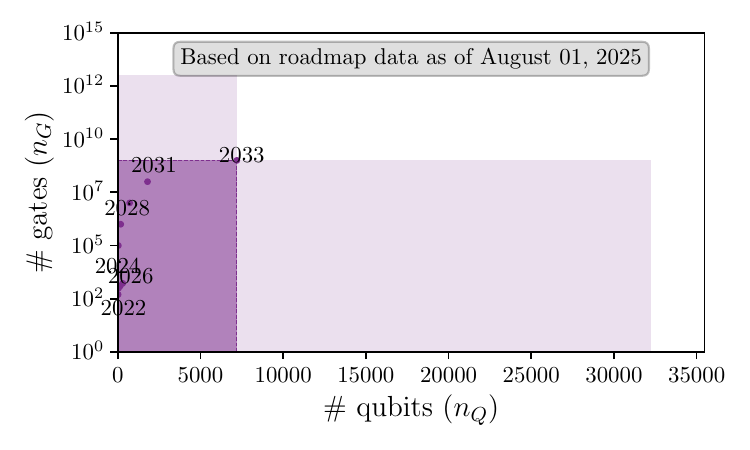}
        \label{fig:rm_iqm}
    \end{minipage}
    \caption{Roadmap data for: IQM~\cite{roadmap_iqm} -- \textit{Superconducting qubits}. \textit{Left}: Table with technology milestones. \textit{Right}: Plot with equivalent technology milestones on number of qubits versus number of gates axes. The light shaded area sketches the performance boundary using a quantum error mitigation approach on the largest-scale system. \label{fig:iqm}}
\end{figure}

IQM, founded in 2018, is a company headquartered in Finland that is developing quantum computers based on superconducting transmon qubits. They have released a public roadmap which includes milestones through 2033~\cite{roadmap_iqm}. At present, IQM develops systems in two different topologies: a \emph{crystal} topology implementing a 2D square lattice compatible with a \gls{surface code} encoding and a \emph{star} topology with one central qubit connected to all other qubits. According to their roadmap, these two technologies will be combined into a single approach in 2027. We conservatively assume a \gls{surface code} encoding for their later milestones given the high number of \glspl{physical qubit} to \glspl{logical qubit}. IQM's roadmap is summarized in~\Cref{fig:iqm}.

\begin{figure}[htp]
    \centering
    \begin{minipage}{0.57\textwidth}
        \centering
        \raisebox{1.25\height}{ 
        \scriptsize
        \begin{tabular}{>{\columncolor{gray!20}}lcccccc}
            \toprule
            \rowcolor{gray!40} \textbf{Year}  & \textbf{Name} & \textbf{Type} & \multicolumn{2}{c}{\textbf{$n_Q$}} & \textbf{error} & \textbf{$f$ [MHz]}\\
            \rowcolor{gray!40} & & & Phys. & Log. & & \\
            \midrule
            2021   & Sophia & N & 4 & -- & -- & -- \\
            2022   & Lucy  & N & 8  & -- &  -- & --\\
            2024   & Toshiko  & N & 32 & -- &  $10^{-2}$ & --\\
            2025   & Genesis  & EF & 16 & 16 &  $10^{-3}$ & --\\
            2028   & Titan & EF & 2,000 & 200 & $10^{-6}$  & 1\\
            2031   & Athena & F & 75,000 & 5,000 & $10^{-9}$ & 3\\
            2034   & Atlas  & F & 1,000,000 & 50,000 &  $10^{-12}$ & 10\\
            \bottomrule
        \end{tabular}}
        \label{tab:rm_oqc}
    \end{minipage}%
    \hfill
    \begin{minipage}{0.43\textwidth}
        \centering
        \includegraphics[width=\textwidth]{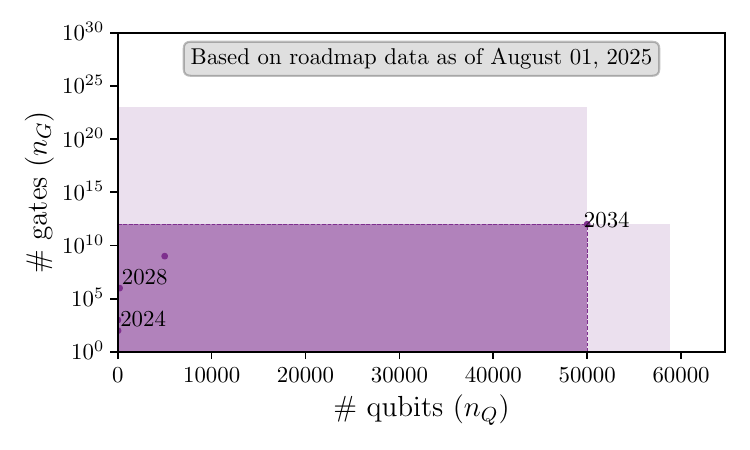}
        \label{fig:rm_oqc}
    \end{minipage}
    \caption{Roadmap data for: Oxford Quantum Computing~\cite{roadmap_oqc} -- \textit{Superconducting qubits}. \textit{Left}: Table with technology milestones. \textit{Right}: Plot with equivalent technology milestones on number of qubits versus number of gates axes. The light shaded area sketches the performance boundary using a quantum error mitigation approach on the largest-scale system. \label{fig:oqc}}
\end{figure}

\paragraph{Oxford Quantum Circuits.}
Oxford Quantum Circuits is a quantum computing startup founded in 2017 and headquartered in the United Kingdom. They are building quantum computers based on superconducting qubits with a coaxmon chip design, a 3D architecture in which the qubit and resonator are fabricated on opposing sides of the chip, allowing the separation of the control and readout wiring to different sides of the chip~\cite{coaxmon}. The company released a technology roadmap in June 2025~\cite{roadmap_oqc}, which we summarize in~\Cref{fig:oqc}. Oxford Quantum Computing is actively developing multi-mode qubits~\cite{wills2022a} and in particular dimon qubits~\cite{dimon}, where a single dimon qubit is made up of two Josephson junctions and allows for a dual rail encoding of the quantum information~\cite{dualrail} which is potentially more error resistant, in particular against erasure errors. We remark that Oxford Quantum Circuits is one of the few companies reporting clock speeds for their future Titan, Athena, and Atlas systems.

\paragraph{Rigetti Computing.}

\begin{figure}[htp]
    \centering
    \begin{minipage}{0.5\textwidth}
        \centering
        \raisebox{1.5\height}{ 
        \scriptsize
        \begin{tabular}{>{\columncolor{gray!20}}lccccc}
            \toprule
            \rowcolor{gray!40} \textbf{Year}  & \textbf{Name} & \textbf{Type} & \multicolumn{2}{c}{\textbf{$n_Q$}} & \textbf{error} \\
            \rowcolor{gray!40} & & & Phys. & Log. & \\
            \midrule
            2022   & Aspen-M-X & N & 80 & -- & $5\times10^{-2}$ \\
            2023   & Ankaa-1 & N & 84 & -- & $5\times10^{-2}$ \\
            2023   & Ankaa-2  & N & 84 & -- &  $2 \times 10^{-2}$ \\
            2024   & Ankaa-3  & N & 84 & -- &  $1 \times 10^{-2}$ \\
            2025   & --       & N & 108 & -- &  $5 \times 10^{-3}$\\
            \bottomrule
        \end{tabular}}
        \label{tab:rm_rigetti}
    \end{minipage}%
    \hfill
    \begin{minipage}{0.5\textwidth}
        \centering
        \includegraphics[width=\textwidth]{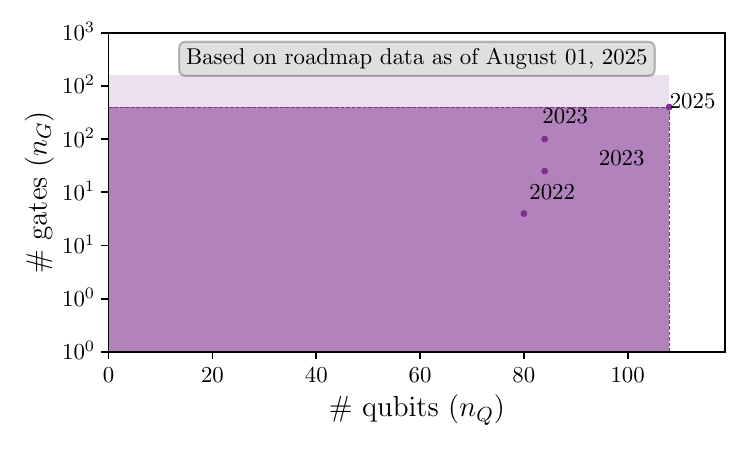}
        \label{fig:rm_rigetti}
    \end{minipage}
    \caption{Roadmap data for: Rigetti Computing~\cite{roadmap_rigetti} -- \textit{Superconducting qubits}. \textit{Left}: Table with technology milestones. \textit{Right}: Plot with equivalent technology milestones on number of qubits versus number of gates axes. The light shaded area sketches the performance boundary using a quantum error mitigation approach on the largest-scale system. \label{fig:rigetti}}
\end{figure}

Rigetti Computing is a full-stack quantum computing company founded in 2013, headquartered in Berkeley, California, and developing superconducting quantum processors based on transmons. While Rigetti Computing has not released a multi-year technology roadmap, their investor presentation from March 2025~\cite{roadmap_rigetti} contained details about recent systems as well as a system expected later in 2025. The milestones are summarized in~\Cref{fig:rigetti}; we remark that all listed devices are \gls{NISQ} devices and we do not infer an error correction strategy.

\subsection{Qubits based on trapped ions}
\label{ssec:ions}

\paragraph{IonQ.}

IonQ, founded in 2015 and headquartered in College Park, Maryland, is building quantum computers based on linear ion traps. An advantage of this approach is that two-qubit gates between two ions are mediated through the collective motion of the chain which allows for a native all-to-all connectivity within a trap. 

\begin{figure}[htp]
    \centering
    \begin{minipage}{0.5\textwidth}
        \centering
        \raisebox{1.5\height}{ 
        \small
        \begin{tabular}{>{\columncolor{gray!20}}lcccc}
            \toprule
            \rowcolor{gray!40} \textbf{Year} & \textbf{Type} & \multicolumn{2}{c}{\textbf{$n_Q$}} & \textbf{error} \\
            \rowcolor{gray!40} & & Phys. & Log. & \\
            \midrule
            2025 & N & 64 & -- & $10^{-4}$\\
            2026 & EF & 256 & 12 & $10^{-7}$\\
            2027 & EF & 10,000 & 800 & $10^{-7}$\\
            2028 & EF & 20,000 & 1,600 & $10^{-7}$\\
            2029 & F & 200,000 & 8,000 & $10^{-12}$\\
            2030 & F & 2,000,000 & 80,000 & $10^{-12}$\\
            \bottomrule
        \end{tabular}}
        \label{tab:rm_ionq}
    \end{minipage}%
    \hfill
    \begin{minipage}{0.5\textwidth}
        \centering
        \includegraphics[width=\textwidth]{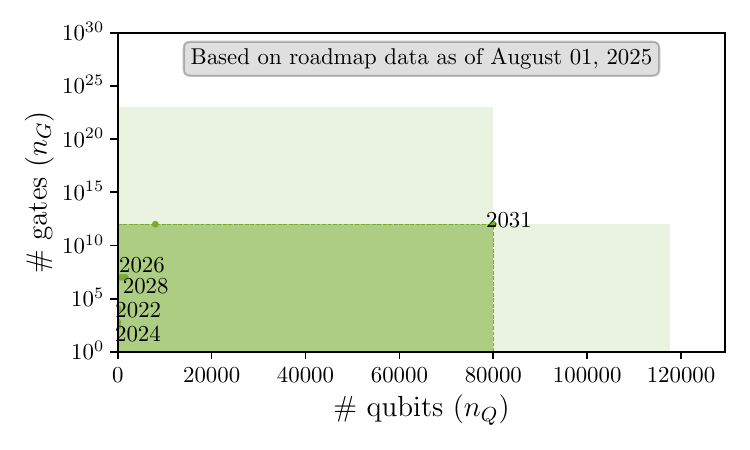}
        \label{fig:rm_ionq}
    \end{minipage}
    \caption{Roadmap data for: IonQ~\cite{roadmap_ionq} -- \textit{Trapped ions}. \textit{Left}: Table with technology milestones. \textit{Right}: Plot with equivalent technology milestones on number of qubits versus number of gates axes. The light shaded area sketches the performance boundary using a quantum error mitigation approach on the largest-scale system. \label{fig:ionq}}
\end{figure}

In a recent press release from June 13, 2025~\cite{roadmap_ionq}, IonQ stated updated milestones out to 2030 which include plans for early and large-scale \gls{FTQC} systems. We summarize these most recent projections in~\Cref{fig:ionq}. IonQ is pursuing a modular architecture which links together multiple ion traps via photonic interconnects. For its 2027 milestone, IonQ wants to achieve 10,000 \glspl{physical qubit} on a single chip before moving to a modular system in 2028 where two chips will be interconnected leading to a system with 20,000 \glspl{physical qubit}. IonQ envisions their architecture to be compatible with multiple error correction schemes. However, no further details are provided about their error correction strategy but given (1) the high degree of connectivity within a trap and (2) a modest ratio of 12.5 up to 25 \glspl{physical qubit} per \glspl{logical qubit} for their technology milestones, one could expect an approach based on sparse LDPC codes.

\paragraph{Quantinuum.}

\begin{figure}[htp]
    \centering
    \begin{minipage}{0.5\textwidth}
        \centering
        \raisebox{1.5\height}{
        \small
        \begin{tabular}{>{\columncolor{gray!20}}lccccc}
            \toprule
            \rowcolor{gray!40} \textbf{Year}  & \textbf{Name} & \textbf{Type} & \multicolumn{2}{c}{\textbf{$n_Q$}} & \textbf{error} \\
            \rowcolor{gray!40} & & & Phys. & Log. & \\
            \midrule
            2025     & Helios & EF & 96 & 50 & $1\times10^{-4}$ \\
            2027   & Sol & EF & 192 & 100 &  $1 \times 10^{-5}$ \\
            2029    & Apollo & F & 5,000 & \estimate{500} &  $1 \times 10^{-10}$ \\
            \bottomrule
        \end{tabular}}
        \label{tab:rm_quantinuum}
    \end{minipage}%
    \hfill
    \begin{minipage}{0.5\textwidth}
        \centering
        \includegraphics[width=\textwidth]{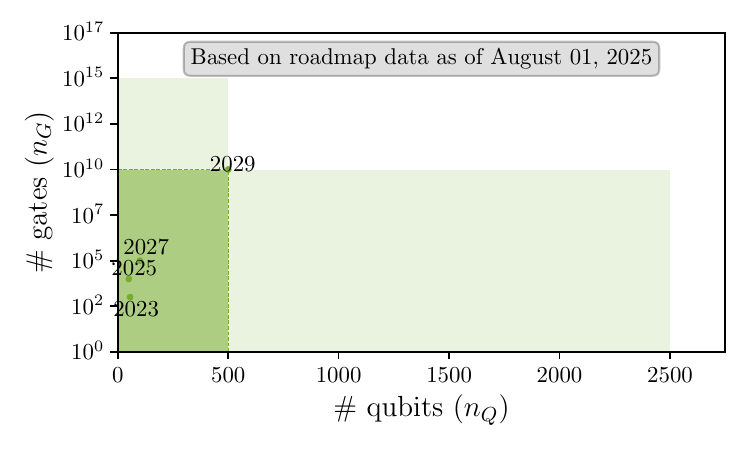}
        \label{fig:rm_quantinuum}
    \end{minipage}
    \caption{Roadmap data for: Quantinuum~\cite{roadmap_quantinuum} -- \textit{Trapped ions}. \textit{Left}: Table with technology milestones. \textit{Right}: Plot with equivalent technology milestones on number of qubits versus number of gates axes. The light shaded area sketches the performance boundary using a quantum error mitigation approach on the largest-scale system. \label{fig:quantinuum}}
\end{figure}

Quantinuum is a quantum computing company headquartered in Cambridge, United Kingdom and Broomfield, Colorado that was formed after the merger of Cambridge Quantum and Honeywell Quantum Solutions. Quantinuum is building trapped ion quantum computers based on \gls{QCCD} architecture. Their H-series systems are currently industry-leading~\cite{quantinuumvolume} in terms of the quantum volume benchmark for \gls{NISQ} systems. The company has announced their roadmap~\cite{roadmap_quantinuum} of ion trap devices up to the Apollo system in 2029 and specifies milestones in terms of qubits and error rates. Note that Quantinuum reports the Apollo system to have ``hundreds'' of \glspl{logical qubit}; we choose 500 as a proxy. The Quantinuum roadmap is summarized in~\Cref{fig:quantinuum}. One could expect \gls{qLDPC} or many-hypercube codes~\cite{Goto:2024} as their strategy for error correction due to high qubit connectivity.

\subsection{Qubits based on neutral atoms}
\label{ssec:atoms}

\begin{figure}[htp]
    \centering
    \begin{minipage}{0.5\textwidth}
        \centering
        \raisebox{1.5\height}{ 
        \begin{tabular}{>{\columncolor{gray!20}}lccccc}
            \toprule
            \rowcolor{gray!40} \textbf{Year} & \textbf{Type} & \multicolumn{2}{c}{\textbf{$n_Q$}} & \textbf{error} & \textbf{$d_C$} \\
            \rowcolor{gray!40} & & Phys. & Log. & & \\
            \midrule
            2024  & N & 1600 & 2 & $5 \times 10^{-3}$ & --\\
            2026  & EF & 8000 & 10 & -- & $10^3$\\
            2028  & EF & 40000 & 100 & -- & $10^6$ \\
            \bottomrule
        \end{tabular}}
        \label{tab:rm_infleqtion}
    \end{minipage}%
    \hfill
    \begin{minipage}{0.5\textwidth}
        \centering
        \includegraphics[width=\textwidth]{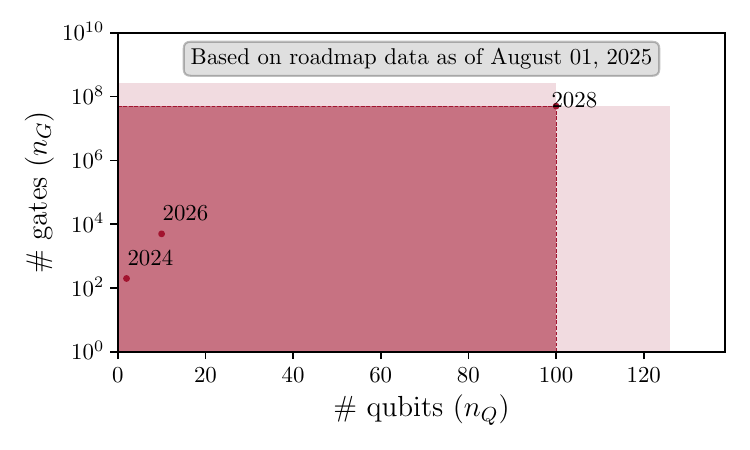}
        \label{fig:rm_infleqtion}
    \end{minipage}
    \caption{Roadmap data for: Infleqtion~\cite{roadmap_infleqtion} -- \textit{Neutral atoms}. \textit{Left}: Table with technology milestones. \textit{Right}: Plot with equivalent technology milestones on number of qubits versus number of gates axes. The light shaded area sketches the performance boundary using a quantum error mitigation approach on the largest-scale system. \label{fig:infleqtion}}
\end{figure}

\paragraph{Infleqtion.}
Infleqtion\cite{Bedalov:2024,roadmap_infleqtion}, based out of Boulder, Colorado, is building quantum computers out of dual-species arrays of neutral atoms and reports the number of \glspl{logical qubit} and a combination of error rates and \gls{quantum circuit} depths on their roadmap that is summarized in~\Cref{fig:infleqtion}. Their leading \gls{QEC} strategy revolves around \gls{qLDPC} codes, and they have recently introduced an open source software library for constructing and analyzing \gls{qLDPC} codes~\cite{Perlin:2023}. On the experimental and architecture side, Infleqtion is exploring techniques for non-destructive readout and tightly focused laser beams that can address single atoms to reduce \gls{quantum gate} execution times by removing the need to shuttle atoms~\cite{radnaev2024}.

\paragraph{Pasqal.}

\begin{figure}[htp]
    \centering
    \begin{minipage}{0.55\textwidth}
        \centering
        \raisebox{1.5\height}{ 
        \scriptsize
        \begin{tabular}{>{\columncolor{gray!20}}lcccccc}
            \toprule
            \rowcolor{gray!40} \textbf{Year} & \textbf{Name} & \textbf{Type} & \multicolumn{2}{c}{\textbf{$n_Q$}} & \textbf{error} &  \textbf{$f$ [Hz]}\\
            \rowcolor{gray!40} & & & Phys. & Log. & &  \\
            \midrule
            2026 & Orion-$\gamma$ & N-EF  & 200 & 2 & $2 \times 10^{-2}$ & $10$\\
            2027 & Vela & EF & 1000 & 2 & $10^{-3}$ & $10$\\
            2028 & Centaurus & EF & 1000 & 20 & $10^{-3}$ & $10$\\
            2029 & Lyra & F & 10000 & 200 & $10^{-5}$ & $>100$\\
            \bottomrule
        \end{tabular}}
        \label{tab:rm_pasqal}
    \end{minipage}%
    \hfill
    \begin{minipage}{0.45\textwidth}
        \centering
        \includegraphics[width=\textwidth]{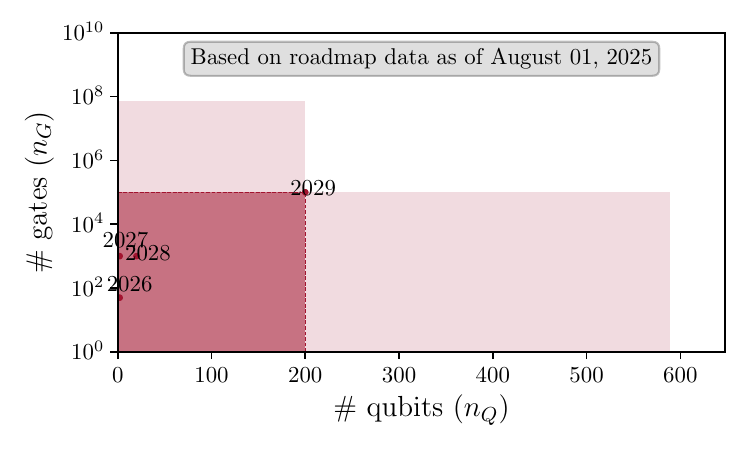}
        \label{fig:rm_pasqal}
    \end{minipage}
    \caption{Roadmap data for: Pasqal~\cite{roadmap_pasqal} -- \textit{Neutral atoms}. \textit{Left}: Table with technology milestones. \textit{Right}: Plot with equivalent technology milestones on number of qubits versus number of gates axes. The light shaded area sketches the performance boundary using a quantum error mitigation approach on the largest-scale system. \label{fig:pasqal}}
\end{figure}

Pasqal is a quantum computing startup based in France that is developing quantum computers based on arrays of Rubidium atoms. The company has recently demonstrated the capability to load large arrays of over a 1,000 atoms~\cite{PhysRevApplied.22.024073} and announced their technical roadmap in June of 2025~\cite{roadmap_pasqal} which we summarize in~\cref{fig:pasqal}. They aim to combine analog with \gls{FTQC} approaches in future hybrid systems such as Centaurus and Lyra. Although we were not able to identify specific details on this hybrid approach, recent research~\cite{PhysRevA.107.042602} suggest that analog mode may be used for preparing an initial many-body state that is subsequently used by a quantum algorithm running in digital mode.  We remark that Pasqal is one of the few companies reporting clock speeds for their future systems.

\subsection{Vendors without public technology roadmaps}
\label{ssec:other}

In this section we provide an overview of vendors active in the quantum computing space that have not released a technology roadmap that allows us to infer a $\perf$-vector based on our prior assumptions. We do not claim that our overview is exhaustive, since the quantum computing space is vast and evolves constantly. We do remark that while we arguably covered the most mature qubit modalities in the first part of this section (superconducting, ions, and neutral atoms), many different approaches are currently under investigation and development; some alternative technologies include photonic quantum computers, topological quantum computing, silicon spin qubits, and color centers in diamonds.

\paragraph{Alpine Quantum Technologies.}
Alpine Quantum Technologies is a company focused on building quantum computers based on trapped-ion technology. Alpine Quantum Technologies is based out of Innsbruck, Austria and currently offers a 20 qubit rack-mounted system. At the time of writing, Alpine Quantum Technologies has not released a technology roadmap that allows us to infer a $\perf$-vector.

\paragraph{Atom Computing.}
Atom Computing is a startup based in Berkeley, California and was founded in 2018. The company is developing quantum computers based on atomic arrays of optically-trapped neutral atoms. They encode the qubit state in nuclear spin states of the atom, which leads to qubits that interact weakly with their environment and have a long coherence time~\cite{Barnes2022}. Recent news stories by the company include the announcement of the second generation system with over a 1,000 \glspl{physical qubit}~\cite{atomcomputingsystem}, a demonstration of universal gates on a neutral atom system~\cite{Muniz2025}, and an announcement of a collaboration with Microsoft~\cite{microsoft_atom}. Atom Computing has not released a technology roadmap that allows us to infer a $\perf$-vector.

\paragraph{Equal1.}
Equal1 is a quantum startup founded in 2017 in Dublin, Ireland. Their technology is rack-mounted
silicon based spin qubits~\cite{Bashir:2020,Staszewski:2022,Bashir:2025}. One of their main goals is to make it easy to integrate with \gls{HPC}-class environments. One of the key advantages of this technology lies in the compatibility with standard
semiconductor manufacturing, enabling, among other capabilities, co-integration with control and readout electronics,
thus improving scalability and reducing interconnect complexity~\cite{Kriekouri:2025}. The company has not released a public technology roadmap.

\paragraph{Microsoft.}
Microsoft is pursuing quantum computers based on hardware-protected topological qubits. The company recently announced the Majorana 1 chip~\cite{majorana1,microsoft2025} and has proposed plans to scale up quantum computational systems based on topological qubits to large-scale fault-tolerant systems~\cite{arxiv.2502.12252}. At the time of writing, Microsoft has not released technology milestones that allow us to infer a $\perf$-vector.

\paragraph{PsiQuantum.}
The mission statement of PsiQuantum is to build and deploy the first useful quantum computers. To this end, the company is pursuing an approach based on photonics and leveraging the semiconductor manufacturing industry~\cite{psiquantum2025}. PsiQuantum aims to implement their \glspl{physical qubit} via a dual-rail encoding based on single photons, relying on single photon sources, single photon detectors, and optical switches. Recent research suggests that they are pursuing an approach known as \emph{fusion-based quantum computing} which relies on entangling measurements (or fusions) on the qubits of constant-sized resource states~\cite{Bartolucci2023}, and can be used in an \emph{active volume} architecture~\cite{active_volume} which holds the potential to significantly reduce the $\perf$-vector for many scientifically relevant applications. At the time of writing, PsiQuantum has not released a public technology roadmap.

\paragraph{Quantum Circuits Inc.}
Quantum Circuits Inc. was founded in 2015 out of research by Yale University. The company is building quantum computers based on superconducting circuitry. Their qubits use a dual-rail encoding which allows for native, high-fidelity detection of erasure errors due to single photon loss, a known, dominant error channel for superconducting qubits~\cite{quantumcircuitsTechnology}. At the time of writing, Quantum Circuits has not released a public technology roadmap.

\paragraph{QuEra Computing.}
QuEra Computing is a startup based in Boston, Massachusets that is building neutral atom quantum computers. Their current roadmap~\cite{roadmap_quera} only specifies milestones in terms of number of qubits and does not provide any information from which we can infer the number of gates and we thus do not provide estimates for a $\perf$-vector. Their milestones include: (i) ten \glspl{logical qubit} on at least 256 \glspl{physical qubit} in 2024, (ii) 30 \glspl{logical qubit} on at least 3,000 \glspl{physical qubit} in 2025, and (iii) 100 \glspl{logical qubit} on at least 10,000 \glspl{physical qubit} in 2026. In light of recent publications~\cite{Zhou:2024}, we can expect \gls{qLDPC} codes to be their target quantum error correction strategy.

\paragraph{Xanadu.}
Xanadu is a quantum computing company founded in 2016 and headquartered in Toronto, Canada. Xanadu is building quantum computers based on photonics. Their qubits are encoded in optical states called Gottesman-Kitaev-Preskill states which offer a universal \gls{quantum gate} set based on Gaussian operations~\cite{Bourassa2021}. Recent experimental results demonstrate the generation of such GKP states~\cite{Larsen2025} and a prototype of a modular, networked photonic quantum computer~\cite{AghaeeRad2025}. The company has not released a public technology roadmap.

\section{Overview of all vendor roadmaps}

\begin{figure}[htp!]
    \centering
    \includegraphics[width=0.8\linewidth]{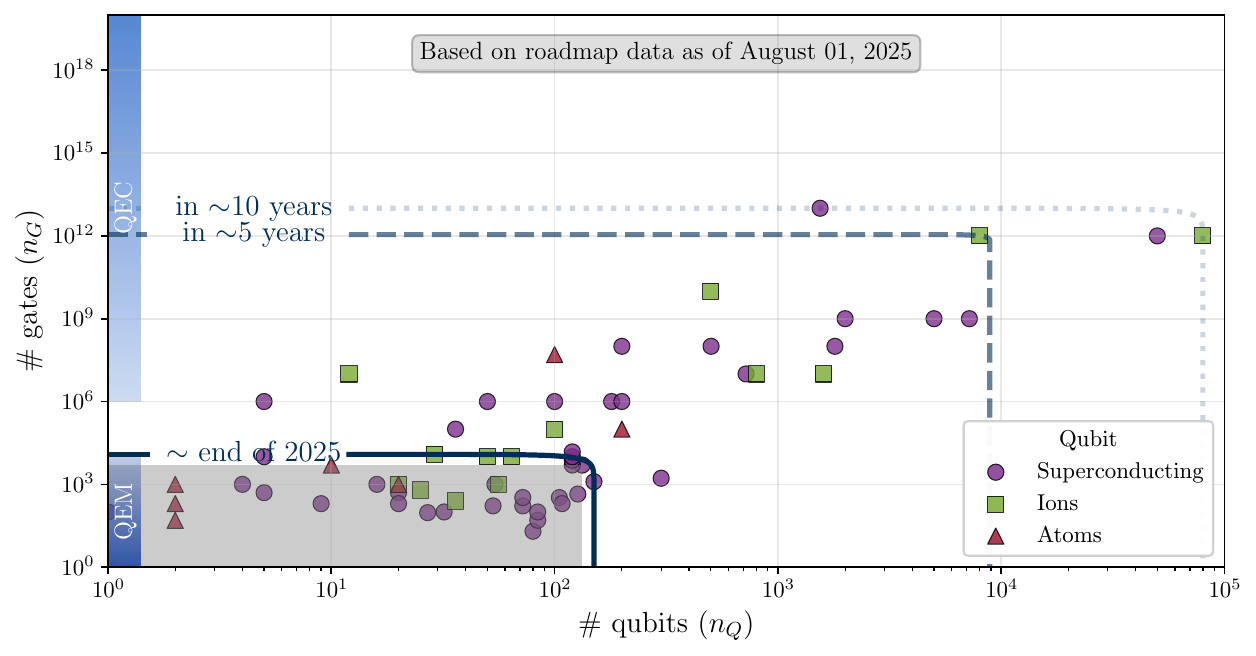}
    \caption{Overview of all technology milestones presented in~\cref{sec:roadmaps}. Purple circles indicate milestones for superconducting quantum computers, green squares indicate trapped ion systems, and red triangles correspond to neutral atom systems. The current limits on quantum computer performance correspond to the area in gray. The dark blue line shows the expected performance by the end of 2025, the medium blue line shows the expected performance in five years from the time of writing, and the light blue line shows the expected performance ten years from today.}
    \label{fig:roadmaps}
\end{figure}

Next, we combine and consolidate the milestones from all ten vendor roadmaps in~\Cref{fig:roadmaps}. The markers show individual milestones sorted by qubit type, the grayed-out area marks the performance bounds of current quantum systems, and the solid lines show the \emph{expected} performance bounds by the end of the current calendar year, and in five and ten years from today. On the y-axis, we list \gls{quantum gate} depths up to $10^4$ that are achievable by \gls{QEM} methods and \gls{quantum gate} depths larger than $10^6$ that are expected to require \gls{QEC} approaches. The intermediate region $[10^4, 10^6]$ will likely be unreachable by \gls{QEM} only, but perhaps \gls{QED} with post-selection can bridge this region. Nonetheless, to achieve the most ambitious five year technology milestones, which predict an increase in \gls{quantum gate} depth by \emph{9 orders of magnitude} and in \gls{qubit} count by \emph{3 orders of magnitude}, it is indisputable that \gls{FTQC} will be necessary.

We summarize our conclusions from evaluating the various roadmaps in the following finding:

\begin{finding}
    The different vendor roadmaps outlining their projected technological progress over the next five to ten years show reasonable alignment in terms of the $(n_Q, n_G)$ figure-of-merit. All vendors project substantial improvements in the figure-of-merit, often by many orders of magnitude. We do foresee that the transition to \gls{FTQC} will be necessary within the next five years to realize this upscaling.
\end{finding}

\section{Vendor roadmaps versus application requirements}
\label{sec:roadmaps_v_apps}

\begin{figure}[htp]
    \centering
    \includegraphics[width=0.8\linewidth]{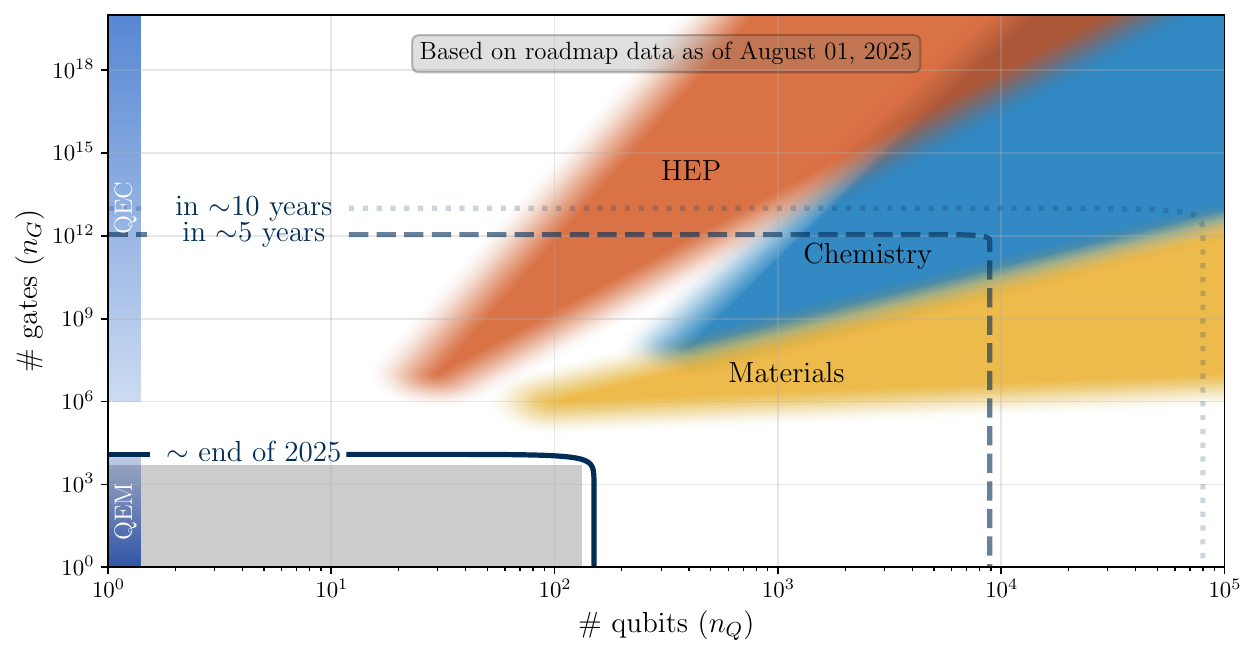}
    \caption{Overview of vendor milestones expected by the end of 2025, in five years from today, and in ten years from today (without the individual milestones shown in~\Cref{fig:roadmaps}) together with the regions of application resource requirements for high energy physics in red, quantum chemistry in blue, and materials science in yellow shown previously in~\Cref{fig:est_overview}. We observe a significant overlap between the capabilities of future quantum computers listed on the vendor roadmaps and requirements for scientific quantum advantage relevant to \gls{DOE} \gls{SC} mission.}
    \label{fig:roadmaps_all_applications}
\end{figure}

Finally, in \Cref{fig:roadmaps_all_applications} we combine the results displayed in \Cref{fig:est_overview,fig:roadmaps} and plot the observed scaling of scientific applications in materials science, chemistry, and \gls{HEP} together with the hardware performance bounds we derived from vendor projections. \Cref{fig:roadmaps_all_applications} is a more detailed version of~\Cref{fig:roadmap_regions}.

As the plots show, with time progression, we can expect the hardware technology, coupled with error mitigation, to provide support for many scientific applications in the \gls{QRW} within the next five years. This finding is summarized below.

\begin{finding}
    Under our assumptions, vendor technology roadmaps and \gls{DOE} \gls{SC} application resource requirements to support the \gls{QRW} are expected to converge, meet, and overlap within the next five to ten years.
\end{finding}

\chapter{Execution times and system throughput}
\label{chap:exec_time_throughput}
\myglsreset

Our analysis thus far has centered around the volumetric figure of merit $\perf$ introduced at the beginning of the report. The $\perf$-vector only indicates whether a given quantum computer can run a particular application. It does not provide an indication on the execution time of said application. Execution times are crucial both in terms of the end user experience and to understand the value of a quantum computer in terms of the rate at which it produces scientific results. An improved understanding of execution times is essential to make informed procurement, planning and operational decisions when hosting \gls{QC} systems.

In~\Cref{sec:exec_time} we discuss the expected clock speeds for (early) \gls{FTQC} systems based on results in the literature and describe key assumptions made. Based on this discussion we define eight regions of feasibility for quantum application execution time. Next, we proceed to introduce a simple model, called \gls{SQSP}, for estimating system throughput in~\Cref{sec:throughput}. Our model uses a benchmark set of six applications which represent the~\gls{QRW} and nine different quantum computer configurations, informed by~\Cref{chap:apps,chap:vendorroadmaps} respectively. This conceptual model allows us to reason quantitatively about a range of different scenarios.

\section{Execution times}
\label{sec:exec_time}
Execution times on \gls{FTQC} systems will be highly relevant, both for economical and practical considerations. For example, if an application has a runtime of 1 month and the quantum system has a production life cycle of 3 years, or 36 months, then the value of solving this application should be at least 1/36$^{\text{th}}$, or roughly 3\%, of the total cost of ownership of the quantum system. On the other hand, increasingly long execution times rapidly become impractical for end users, certainly if the quantum computation is part of a larger computational workflow which requires multiple iterations or \glspl{shot}. 

Accurately estimating execution times requires deep insights into the quantum computer architecture and structure of the application. The result of \gls{quantum circuit} compilation, which is an NP-complete problem~\cite{Botea_Kishimoto_Marinescu_2021}, can strongly influence the \gls{quantum gate} depth of a circuit when mapped to hardware. Other factors that play an important role include:
\begin{itemize}
    \item The duration for implementing each \gls{quantum gate} from a universal \gls{quantum gate} set (including both the leading order type of \glspl{quantum gate} as well as the lower order ones). This step is typically rate-limited by the measurement time required to extract the \gls{error syndrome}; 
    \item The speed of the \gls{QEC} decoder that analyzes the \gls{error syndrome} data and plans corrective actions can slow down the quantum computer~\cite{camps24};
    \item The overhead incurred by topology, routing, and qubit shuttling when mapping an application to a quantum computer.
\end{itemize}
Collectively, these determine the \emph{logical clock speed} or \emph{frequency} $f$ of the quantum computer. As discussed in~\cref{sec:roadmaps}, only two vendors, Oxford Quantum Computing (1--10 MHz) and Pasqal (10--100+ Hz), provide data on the logical clock speeds as part of their technology roadmaps. 

As detailed performance projections are scarce, we provide a high-level overview of execution times. Our discussion is founded on the two data points available from vendors and supplemented with results from the literature, a level of detail we consider adequate for the purpose of reasoning about our findings. %

We use the following ranges of clock speed for each technology type:
\begin{itemize}
    \item \emph{Superconducting}: $f \in [100~\text{kHz}, 10~\text{MHz}]$. This range is based on clock speed estimates from Oxford Quantum Computing and~\cite{Bravyi2022,caune2024,PhysRevLett.127.130501,Moskalenko2022}.
    \item \emph{Trapped ions}: $f \in [1~\text{kHz}, 1~\text{MHz}]$. This range is based on clock speed estimates from~\cite{Schfer2018,ballance16}.
    \item \emph{Neutral atoms}: $f \in [10~\text{Hz}, 100~\text{kHz}]$. This range is based on clock speed estimates from Pasqal and~\cite{radnaev2024,Graham:23,sinclair25}.
    \item \emph{Photons}: $f \in [100~\text{MHz}, 10~\text{GHz}]$. This range is based on clock speed estimates from~\cite{Rudolph_2023}.
\end{itemize}
We observe that these clock speeds span nine orders of magnitude and it is natural to assume that execution times and throughput metrics will span a comparibly large range.

We consider the logical clock speed as the rate at which quantum gates can be executed. So an application which requires $n_G$ \glspl{quantum gate} of the dominant type per run and is repeated for $N_s$ times incurs a total \gls{quantum gate} cost $\mathcal{C}_{\text{tot}} = N_s n_g$ and has an estimated execution time of, 
\begin{equation}
t_{\text{exec}} = \mathcal{C}_{\text{tot}} / f.
\label{eq:texec}
\end{equation}
We remark that according to this model, the execution time estimate is independent of the number of \glspl{logical qubit} required by an application. This assumption may approximately be satisfied if resource states to implement a dominant gate can be generated and applied to any \glspl{qubit} in the circuit in a time that does not depend on $n_Q$. More detailed models must take into account more fine-grained aspects of the \gls{FTQC} architecture, which may vary significantly between different vendors and is left for future analyses.

\begin{figure}[htp!]

\centering
\includegraphics[width=0.8\linewidth]{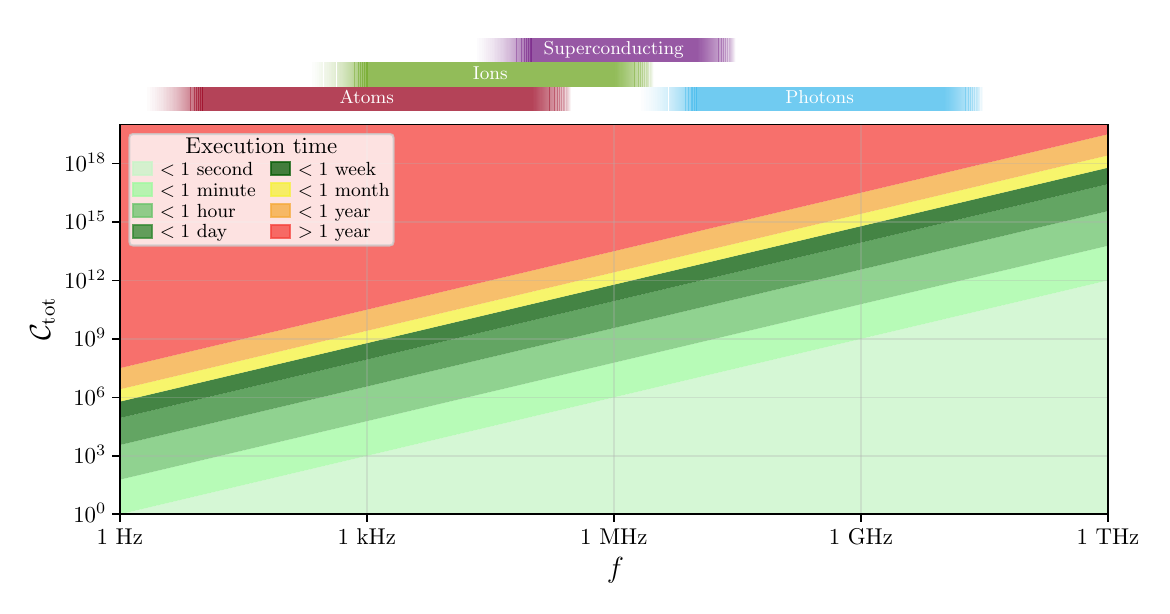}
\caption{Execution time as a function of logical clock speed~$f$ and total cost $\mathcal{C}_{\text{tot}} = N_s n_g$. The execution time is estimated as $\mathcal{C}_{\text{tot}}/f$. Regions where the execution times are less than 1 second, 1 minute, 1 hour, 1 day, and 1 week are shown in gradually darker shades of green. Execution times less than 1 month are shown in yellow, less than 1 year in orange, and more than 1 year in red. The horizontal strips at the top of the figure show the range of clock frequencies for each technology type as described in the main text. The strips are stacked vertically in order to visually distinguish them.}
 \label{fig:execution_times}
\end{figure}

Under these assumptions, \Cref{fig:execution_times} shows eight different regions of feasibility for the range of clock speeds $f$ starting at 1~Hz and going up to 1~THz (informed by the previous discussion) and for a total cost $\mathcal{C}_{\text{tot}}$ up to $10^{20}$ (informed by the application requirements discussion in~\cref{chap:apps}). These color-coded regions include execution times of: 
\begin{itemize}
    \item \emph{Less than 1 second}: in this case the application runs in near real-time from an end user perspective, it is easy to run applications in interactive mode, and it is practical to integrate the quantum application into a larger \gls{HPC} workflow with limited scheduling concerns.
    \item \emph{Less than 1 minute}: in this case the time to execute the application will become noticeable for the end user, running the application in an interactive environment starts to become cumbersome, and scheduling concerns are relevant for integrated workflows.
    \item \emph{Less than 1 hour}: in this case it is impractical to run the application in interactive mode and an approach using batch jobs is desired, co-scheduling \gls{HPC} resources in concert with the quantum system will be necessary for workflow-style jobs.
    \item \emph{Less than 1 day}: this is approaching the limit of what we deem practical for a single application at \gls{NERSC}; at this timescale scheduling might be less of a concern as multiple other jobs can be run while the quantum system is active.
    \item \emph{Less than 1 week}: this is at the limit of what we deem practical, this timescale should be reserved for valuable production runs only.
    \item \emph{Less than 1 month}: this timescale should be considered only in extraordinary circumstances for important applications, for example when it resolves a scientific question of significant value (~3\% of cost of ownership of the quantum computer); it is unlikely that this will become a relevant use case at \gls{NERSC}.
    \item \emph{Less than 1 year}: this timescale should be considered only in the most extraordinary circumstances for high-impact applications where the value of the solution is on par with the total cost of ownership of the quantum computer; it is unlikely that this will become a relevant use case at \gls{NERSC}.
    \item \emph{More than 1 year}: we do not consider this to be a useful timescale. 
\end{itemize}

The current time limit on jobs submitted to the Perlmutter system at \gls{NERSC} is 48h. Due to the different nature and speed of \gls{FTQC}, it is plausible that longer running jobs will be supported on future quantum systems (i.e. up to 1 week) for valuable production runs. We estimate that within a 1 week period an \gls{FTQC} system can run an application with a total cost of about $10^7$ for the slowest clock speed estimates up to more than $10^{16}$ for the fastest clock speed estimates.

\section{Workload and system throughput}
\label{sec:throughput}

In this section, we use data collected thus far to construct a simple quantum workload model that we subsequently optimize for 9 different quantum system specifications. Our model is conceptual in nature and lacks sufficient detail both on the applications and quantum system specifications to draw precise conclusions. Nonetheless, we consider this an important stepping stone towards modeling the value and performance of \gls{FTQC} systems. System level performance models have historically been used by \gls{NERSC} as part of procurement processes for \gls{HPC} systems. Examples include the \gls{SSP}~\cite{Kramer2005ssp} and \gls{SSI}~\cite{Austin18ssi} metrics which are respectively absolute and relative system-level metrics for a benchmark workload relevant to \gls{NERSC}. 

\subsection{Prototypical workload}

To define a quantum version of a system level performance model, we start by proposing a collection of prototypical benchmark applications that represent the~\gls{QRW}. As there is no historical quantum workload data at \gls{NERSC}, we use a set of six applications based on the discussion in~\Cref{chap:apps}. The six applications~(A--F) are summarized in~\Cref{tab:quantum_apps}. The values for the number of shots $N_s$ are based on~\cref{tab:sampling_complexity}.

\begin{table}[htbp]
\centering
\small
\caption{Six quantum computing applications (A--F), determined by number of qubits $n_Q$, number of gates $n_G$, number of samples/repetitions ($N_s$), total cost ($\mathcal{C}_{\text{tot}}$), and prototypical use case.}
\label{tab:quantum_apps}
\begin{tabular}{lccccl}
\toprule
\rowcolor{gray!40}  \textbf{App.} & \textbf{$n_Q$} & \textbf{$n_G$} & \textbf{$N_s$} & \textbf{$\mathcal{C}_{\text{tot}}$} & \textbf{Prototypical use case} \\
\midrule
\textcolor{colorA}{\rule{0.8em}{0.8em}} A & 200 & $10^{6}$ & $10^{1}$ & $10^{7}$ & Condensed matter physics (static) \\
\textcolor{colorB}{\rule{0.8em}{0.8em}} B & 200 & $10^{6}$ & $10^{3}$ & $10^{9}$ & Condensed matter physics (dynamics) \\
\textcolor{colorC}{\rule{0.8em}{0.8em}} C & 1,000 & $10^{9}$ & $10^{1}$ & $10^{10}$ & Quantum chemistry (static) \\
\textcolor{colorD}{\rule{0.8em}{0.8em}} D & 1,000 & $10^{9}$ & $10^{5}$ & $10^{14}$ & Quantum chemistry (dynamics)\\
\textcolor{colorE}{\rule{0.8em}{0.8em}} E & 2,000 & $10^{11}$ & $10^{1}$ & $10^{12}$ & Quantum chemistry (static, large system)\\
\textcolor{colorF}{\rule{0.8em}{0.8em}} F & 250 & $10^{12}$ & $10^{1}$ & $10^{13}$ & High energy physics (static) \\
\bottomrule
\end{tabular}
\end{table}

Applications A and B listed in~\Cref{tab:quantum_apps} are proxies for condensed matter physics and materials science problems discussed in~\cref{sec:app_cm} and both have a gate depth of $10^6$ which was the minimal gate depth identified for this category. Application A assumes ten repetitions, which we deem a reasonable number for a static problem such as \gls{GSEE} provided that the reference wavefunction has sufficient overlap with the ground state. Application B on the other hand assumes $10^3$ repetitions which represents the sampling overhead required to measure the observable of interest after a quantum dynamics simulation.  

Applications C, D and E are all inspired by the quantum chemistry discussion in~\cref{sec:app_qc}; C and E both represent static problems like \gls{GSEE} for respectively a smaller and larger system, they both assume $N_s=10$ but differ in gate depth between $10^9$ and $10^{11}$. Application D on the other hand represents some dynamics problem. As was mentioned in~\cref{sec:app_qc}, resource data for dynamics problems is scarce in the literature. Hence, we use the same $n_G$ as in Application C, but with drastically increased sampling overhead of $N_s = 10^5$ to reflect the larger number of measurements typically required for a dynamics problem. 

Finally, Application F represents the \gls{HEP} applications discussed in~\cref{sec:app_hep} as demonstrated by the highest gate depth of $10^{12}$ among all six problems. 

Within each scientific domain, we erred on selecting applications on the cheaper side since (1) faster execution times do lead to higher throughput and thus we expect our results to reflect an optimistic scenario, and (2) we expect a general trend  of resource reduction through algorithmic advances. Overall, the suite of benchmark applications in~\cref{tab:quantum_apps} spans seven orders of magnitude from $10^{7}$ up to $10^{14}$ in total cost and six orders of magnitude in gate depth. We remark that, while we list data for $n_Q$ for these benchmark applications, the execution time estimates are independent of $n_Q$ based on our prior assumptions and modeling approach.

\subsection{Prototypical quantum computers}

\Cref{tab:quantum_systems} describes nine different possible quantum system specifications varying in maximal gate depth $n_g \in [10^6, 10^9, 10^{12}]$ (in line with the vendor roadmaps), also dubbed megaquop~\cite{Preskill_2025} (for million quantum operations), gigaquop and teraquop \gls{FTQC} systems, and clock speed $f \in [1$ kHz, 1 MHz, 1 GHz$]$ (in line with~\cref{fig:execution_times}). We remark that we do not list $n_Q$ for these reference systems as the execution time estimate is independent of $n_Q$ and we make the blanket assumption that the number of qubits is never the constraining factor that determines whether or not an application can run.

\begin{table}[htbp]
\centering
\caption{Nine different quantum computer configurations as determined by the maximum number of gates they can reliably run ($n_G$) and clock frequency $f$.}
\label{tab:quantum_systems}
\begin{tabular}{@{}l*{3}{c}@{}}
\toprule
& \multicolumn{3}{c}{\textbf{$f$}} \\
\cmidrule(l){2-4}
\textbf{$n_G$} & 1\,kHz & 1\,MHz & 1\,GHz \\
\midrule
$10^6$ (Megaquop) & \tikz[baseline=-0.5ex]\node[circle,draw,inner sep=1pt,minimum size=1.5em]{1}; & \tikz[baseline=-0.5ex]\node[circle,draw,inner sep=1pt,minimum size=1.5em]{2}; & \tikz[baseline=-0.5ex]\node[circle,draw,inner sep=1pt,minimum size=1.5em]{3}; \\
$10^9$ (Gigaquop) & \tikz[baseline=-0.5ex]\node[circle,draw,inner sep=1pt,minimum size=1.5em]{4}; & \tikz[baseline=-0.5ex]\node[circle,draw,inner sep=1pt,minimum size=1.5em]{5}; & \tikz[baseline=-0.5ex]\node[circle,draw,inner sep=1pt,minimum size=1.5em]{6}; \\
$10^{12}$ (Teraquop) & \tikz[baseline=-0.5ex]\node[circle,draw,inner sep=1pt,minimum size=1.5em]{7}; & \tikz[baseline=-0.5ex]\node[circle,draw,inner sep=1pt,minimum size=1.5em]{8}; & \tikz[baseline=-0.5ex]\node[circle,draw,inner sep=1pt,minimum size=1.5em]{9}; \\
\bottomrule
\end{tabular}
\end{table}

\subsection{Execution times of benchmark workload}

\begin{figure}[htp!]
\centering
\includegraphics[width=0.8\linewidth]{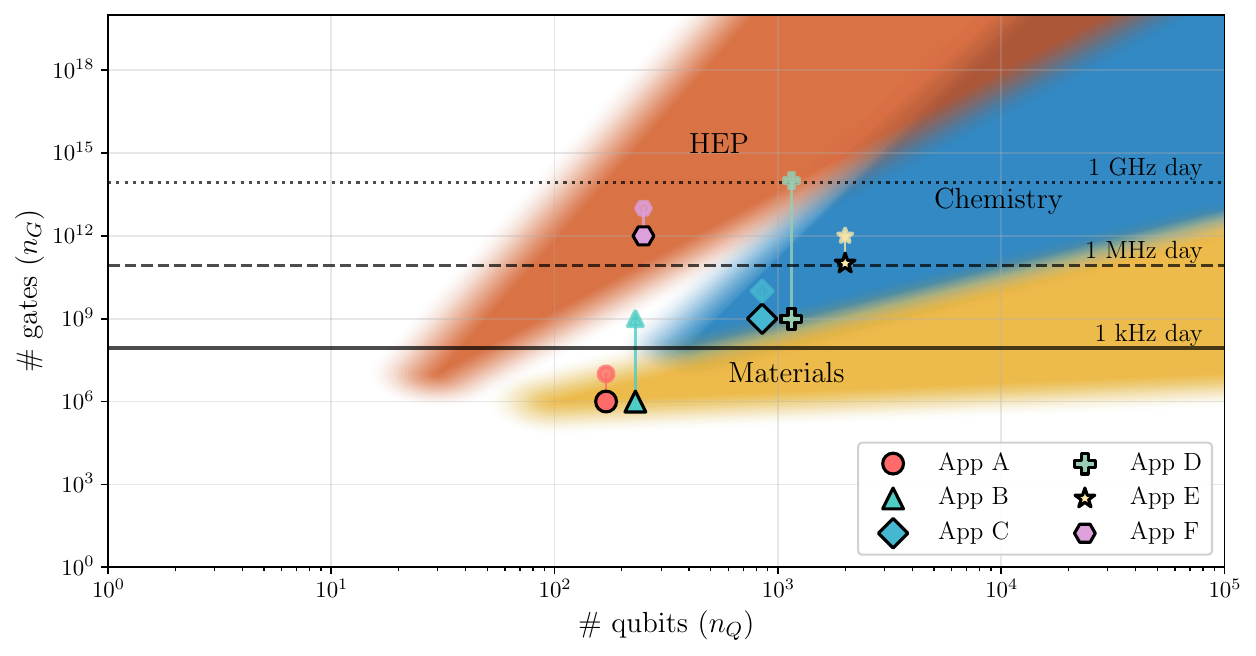}
\caption{Domain specific resource requirements introduced in~\Cref{fig:est_overview} together with $\perf$-vectors for the six benchmark applications (A--F) introduced in~\cref{tab:quantum_apps} as markers with black edges. The corresponding $(n_Q, \mathcal{C}_\text{tot})$-vector for each application is plotted as an edge-less marker of the same type connected by a vertical line to the $\perf$-vector. The solid, dashed, and dotted horizontal lines show the largest application that can be run within a day period on respectively 1 kHz, 1 MHz, and 1 GHz devices considered in~\cref{tab:quantum_systems}}
 \label{fig:apps_speed}
\end{figure}

\Cref{fig:apps_speed} shows both the $\perf = (n_G, n_Q)$ and  $(n_G, \mathcal{C}_\text{tot})$ vectors for the six benchmark applications introduced in~\cref{tab:quantum_apps} together with the domain-specific regions identified in~\cref{fig:est_overview}. This shows that within each domain, we selected $\perf$-vectors on the lower end of their respective range. When we consider $\mathcal{C}_\text{tot}$ instead of $n_G$, the cost increases multiplicatively by a factor of $N_s$ but this does not affect whether a given system is able to run an application reliably, it merely increases the total execution time on the quantum system.

\Cref{tab:execution_times} summarizes the execution time estimates based on~\cref{eq:texec} for each combination of application (A--F) and quantum system (1--9) introduced in the previous section. Execution time estimates are listed in terms of the eight ranges introduced in~\cref{fig:execution_times}. Applications C--F and E--F cannot run on respectively the megaquop and gigaquop systems as their number of gates $n_G$ lies beyond the limits of the devices.

The main conclusion we draw from~\cref{tab:execution_times} is that there is a clear separation into three categories of applications that impose increasingly demanding requirements on the quantum systems: 
\begin{enumerate}
    \item Application A and B: These have the lowest total cost $\mathcal{C}_\text{tot}$ at respectively $10^7$ and $10^9$ and can run with acceptable performance on each of the nine quantum systems considered. Naturally, faster clock speeds lead to faster quantum calculations, but even at a frequency of 1~kHz applications of this size become feasible.
    \item Application C and E: These have an intermediate total cost of respectively $10^{10}$ and $10^{12}$ and require a processing speed of 1~MHz to achieve feasible execution times. Application E has a gate depth beyond $10^9$ and thus is only deemed feasible on systems 8 and 9.
    \item Application D and F: These have the highest total cost at respectively $10^{14}$ and $10^{13}$. At this level, only the GHz-rate systems result in a feasible execution time estimate. Application F has a gate depth beyond $10^9$ and thus is only deemed feasible on system 9.
\end{enumerate}


\begin{table}[htbp]
\centering
\small
\caption{Execution times for quantum applications A--F on nine quantum computing systems. Colors indicate time ranges: \textcolor{timecolor1}{\rule{0.8em}{0.8em}} $<1$s, \textcolor{timecolor2}{\rule{0.8em}{0.8em}} $<1$min, \textcolor{timecolor3}{\rule{0.8em}{0.8em}} $<1$h, \textcolor{timecolor4}{\rule{0.8em}{0.8em}} $<1$day, \textcolor{timecolor5}{\rule{0.8em}{0.8em}} $<1$wk, \textcolor{timecolor6}{\rule{0.8em}{0.8em}} $<1$mo, \textcolor{timecolor7}{\rule{0.8em}{0.8em}} $<1$yr, \textcolor{timecolor8}{\rule{0.8em}{0.8em}} $>1$yr. Invalid combinations (i.e. when the number of gates in the application is larger than the number of gates the system can reliably run) are marked with $\times$.}
\label{tab:execution_times}
\setlength{\tabcolsep}{8pt} 
\renewcommand{\arraystretch}{1.6} 
\begin{tabular}{@{}lcc*{6}{c}@{}}
\toprule
& & & \multicolumn{6}{c}{\textbf{Quantum Applications}} \\
\cmidrule(l){4-9}
\textbf{System Type} & \textbf{System} & $f$ & \textcolor{colorA}{\rule{0.8em}{0.8em}}~\textbf{A} & \textcolor{colorB}{\rule{0.8em}{0.8em}}~\textbf{B} & \textcolor{colorC}{\rule{0.8em}{0.8em}}~\textbf{C} & \textcolor{colorD}{\rule{0.8em}{0.8em}}~\textbf{D} & \textcolor{colorE}{\rule{0.8em}{0.8em}}~\textbf{E} & \textcolor{colorF}{\rule{0.8em}{0.8em}}~\textbf{F} \\
\midrule
\multirow{3}{*}{\parbox{2.5cm}{\centering\textbf{Megaquop}\\$n_G = 10^6$}} & \tikz[baseline=-0.5ex]\node[circle,draw,inner sep=1pt,minimum size=1.5em]{1}; & 1 kHz & \roundcell{timecolor4}{\textcolor{white}{$<1$day}} & \roundcell{timecolor6}{\textcolor{black}{$<1$mo}} & $\times$ & $\times$ & $\times$ & $\times$ \\
& \tikz[baseline=-0.5ex]\node[circle,draw,inner sep=1pt,minimum size=1.5em]{2}; & 1 MHz & \roundcell{timecolor2}{\textcolor{black}{$<1$min}} & \roundcell{timecolor3}{\textcolor{white}{$<1$h}} & $\times$ & $\times$ & $\times$ & $\times$ \\
& \tikz[baseline=-0.5ex]\node[circle,draw,inner sep=1pt,minimum size=1.5em]{3}; & 1 GHz & \roundcell{timecolor1}{\textcolor{black}{$<1$s}} & \roundcell{timecolor2}{\textcolor{black}{$<1$min}} & $\times$ & $\times$ & $\times$ & $\times$ \\
\midrule
\multirow{3}{*}{\parbox{2.5cm}{\centering\textbf{Gigaquop}\\$n_G = 10^9$}} & \tikz[baseline=-0.5ex]\node[circle,draw,inner sep=1pt,minimum size=1.5em]{4}; & 1 kHz & \roundcell{timecolor4}{\textcolor{white}{$<1$day}} & \roundcell{timecolor6}{\textcolor{black}{$<1$mo}} & \roundcell{timecolor7}{\textcolor{white}{$<1$yr}} & \roundcell{timecolor8}{\textcolor{white}{$>1$yr}} & $\times$ & $\times$ \\
& \tikz[baseline=-0.5ex]\node[circle,draw,inner sep=1pt,minimum size=1.5em]{5}; & 1 MHz & \roundcell{timecolor2}{\textcolor{black}{$<1$min}} & \roundcell{timecolor3}{\textcolor{white}{$<1$h}} & \roundcell{timecolor4}{\textcolor{white}{$<1$day}} & \roundcell{timecolor8}{\textcolor{white}{$>1$yr}} & $\times$ & $\times$ \\
& \tikz[baseline=-0.5ex]\node[circle,draw,inner sep=1pt,minimum size=1.5em]{6}; & 1 GHz & \roundcell{timecolor1}{\textcolor{black}{$<1$s}} & \roundcell{timecolor2}{\textcolor{black}{$<1$min}} & \roundcell{timecolor2}{\textcolor{black}{$<1$min}} & \roundcell{timecolor5}{\textcolor{white}{$<1$wk}} & $\times$ & $\times$ \\
\midrule
\multirow{3}{*}{\parbox{2.5cm}{\centering\textbf{Teraquop}\\$n_G = 10^{12}$}} & \tikz[baseline=-0.5ex]\node[circle,draw,inner sep=1pt,minimum size=1.5em]{7}; & 1 kHz & \roundcell{timecolor4}{\textcolor{white}{$<1$day}} & \roundcell{timecolor6}{\textcolor{black}{$<1$mo}} & \roundcell{timecolor7}{\textcolor{white}{$<1$yr}} & \roundcell{timecolor8}{\textcolor{white}{$>1$yr}} & \roundcell{timecolor8}{\textcolor{white}{$>1$yr}} & \roundcell{timecolor8}{\textcolor{white}{$>1$yr}} \\
& \tikz[baseline=-0.5ex]\node[circle,draw,inner sep=1pt,minimum size=1.5em]{8}; & 1 MHz & \roundcell{timecolor2}{\textcolor{black}{$<1$min}} & \roundcell{timecolor3}{\textcolor{white}{$<1$h}} & \roundcell{timecolor4}{\textcolor{white}{$<1$day}} & \roundcell{timecolor8}{\textcolor{white}{$>1$yr}} & \roundcell{timecolor6}{\textcolor{black}{$<1$mo}} & \roundcell{timecolor7}{\textcolor{white}{$<1$yr}} \\
& \tikz[baseline=-0.5ex]\node[circle,draw,inner sep=1pt,minimum size=1.5em]{9}; & 1 GHz & \roundcell{timecolor1}{\textcolor{black}{$<1$s}} & \roundcell{timecolor2}{\textcolor{black}{$<1$min}} & \roundcell{timecolor2}{\textcolor{black}{$<1$min}} & \roundcell{timecolor5}{\textcolor{white}{$<1$wk}} & \roundcell{timecolor3}{\textcolor{white}{$<1$h}} & \roundcell{timecolor4}{\textcolor{white}{$<1$day}} \\
\bottomrule
\end{tabular}
\end{table}

\subsection{Sustained quantum system performance}

In this section, we combine the data generated in~\cref{tab:execution_times} into a single system level metric that  we call the~\gls{SQSP} metric and is an extension of the \gls{SSP} metric previously introduced by \gls{NERSC}~\cite{Kramer2005ssp} to quantum systems. The \gls{SSP} metric for a supercomputer is defined as the geometric mean of the flop rate of each application in a properly defined benchmark suite on said supercomputer. In absence of the capability to run a representative benchmark suite on a large-scale quantum computer, we use the execution time estimates in~\cref{tab:execution_times} to compute the expected throughput $T_i$ of application $i$ over a one year time window, i.e.,
\begin{equation}
     T_i = 1 \text{yr}\, / \,t_{\text{exec}, i}.
\end{equation}
The one year time period is chosen as it corresponds to the typical cycle at which projects are allocated computer resources at \gls{NERSC}.

We then define \gls{SQSP} as the geometric mean of the throughput values of every application in a suite of $n$ benchmarks:
\begin{equation}
\text{SQSP} = \left(\Pi_{i = 1}^{n} T_i\right)^{1/n}.\end{equation}
Applications that cannot run on a particular quantum system ($\times$ in~\cref{tab:execution_times}) have infinite execution time and zero throughput, and the SQSP for such a system will be zero.
The value of SQSP metric thus depends critically on the suite of benchmarks, and SQSP values for different suites cannot be compared to each other.

\Cref{tab:system_throughput} shows the results of the \gls{SQSP} metric based on the data collected in~\cref{tab:execution_times} for the nine quantum systems we consider. 
The SQSP for each system type is evaluated using a different benchmark suite consisting of only the subset of applications from Table~\ref{tab:execution_times} that can run on that architecture.
We do observe that the \gls{SQSP} metric spans about eight orders of magnitude. 
We conclude from our analysis that (1) faster \gls{FTQC} devices naturally lead to high \gls{SQSP} results, (2) at megaquop-scale all systems lead to \gls{SQSP} of at least $\mathcal{O}(10^3)$ megaquop-jobs/year, (3) at gigaquop-scale, the throughput of the kHz system drops below $10$ gigaquop-jobs/year, indicating that higher frequencies will be a necessity, and (4) at teraquop-scale, clock speeds will need to reach at least MHz-range in order to achieve meaningful throughput.

\begin{table}[htbp]
\centering
\caption{\gls{SQSP} for a one year window for every system defined in~\cref{tab:quantum_systems} with respect to a benchmark suite which is a subset of the workload defined in~\cref{tab:quantum_apps}.}
\label{tab:system_throughput}

\begin{tabular}{@{}llccl@{}}
\toprule
\textbf{System Type} & \textbf{Suite} & \textbf{System} & $f$ & \textbf{SQSP}\hspace{6cm} \\
\midrule
\multirow{3}{*}{\parbox{2.75cm}{\centering\textbf{Megaquop}\\$n_G = 10^6$}} & 
\multirow{3}{*}{A--B}
& \tikz[baseline=-0.5ex]\node[circle,draw,inner sep=1pt,minimum size=1.5em]{1}; & 1 kHz & \multirow{9}{5.5cm}{\throughputplot} \\
& & \tikz[baseline=-0.5ex]\node[circle,draw,inner sep=1pt,minimum size=1.5em]{2}; & 1 MHz & \\
& & \tikz[baseline=-0.5ex]\node[circle,draw,inner sep=1pt,minimum size=1.5em]{3}; & 1 GHz & \\
\midrule
\multirow{3}{*}{\parbox{2.75cm}{\centering\textbf{Gigaquop}\\$n_G = 10^9$}} & 
\multirow{3}{*}{A--D}
& \tikz[baseline=-0.5ex]\node[circle,draw,inner sep=1pt,minimum size=1.5em]{4}; & 1 kHz & \\
& & \tikz[baseline=-0.5ex]\node[circle,draw,inner sep=1pt,minimum size=1.5em]{5}; & 1 MHz & \\
& & \tikz[baseline=-0.5ex]\node[circle,draw,inner sep=1pt,minimum size=1.5em]{6}; & 1 GHz & \\
\midrule
\multirow{3}{*}{\parbox{2.75cm}{\centering\textbf{Teraquop}\\$n_G = 10^{12}$}} & 
\multirow{3}{*}{A--F}
& \tikz[baseline=-0.5ex]\node[circle,draw,inner sep=1pt,minimum size=1.5em]{7}; & 1 kHz & \\
& & \tikz[baseline=-0.5ex]\node[circle,draw,inner sep=1pt,minimum size=1.5em]{8}; & 1 MHz & \\
& & \tikz[baseline=-0.5ex]\node[circle,draw,inner sep=1pt,minimum size=1.5em]{9}; & 1 GHz & \\
\\
\bottomrule
\end{tabular}

\end{table}

The rudimentary performance model expressed by \cref{eq:texec} does not account for details of the \gls{FTQC} architecture (such as qubit connectivity, parallel gate operations, or multi-qubit gates) that have the potential to dramatically alter system performance. Practical use of \gls{SQSP} would avoid this limitation by relying on measured runtimes or more detailed device-specific models. Nevertheless, the current analysis is capable of summarizing anticipated application-level  capabilities across the \gls{QC} industry.

\chapter{Conclusion}
\label{chap:conclusion}
\myglsreset

\gls{NERSC} is exploring and evaluating alternative compute technologies
that can deliver increased performance in a more power-efficient manner in the 2030's time-frame for the next-next HPC generation. One key direction in this regard is represented by quantum computers that can lead to exponential speedups for important scientific problems that classical HPC cannot solve. Based on the \gls{QIS} Applications Roadmap of the \gls{DOE}~\cite{DOE_roadmap}, we have focused in this report on materials science, quantum chemistry, and high energy physics, and collected over 140 end-to-end resource estimates for benchmark problems from the scientific literature. These domains of science cover over 50\% of the \gls{DOE} \gls{SC} production workload, which aggregates the computational needs of more than 12,000 \gls{NERSC} users across the \gls{DOE} landscape. 

The space of quantum hardware development has seen major improvement in recent years, especially through noise reduction and early tests of error correction~\cite{Maika:2016,Marques:2022,Krinner:2022,Zhao:2022,Acharya:2023,Acharya2024} across different technology propositions such as neutral atoms, ions, and superconducting qubits. Given these developments, we see very ambitious roadmaps across the whole sector, as this is a high-risk high-reward environment when considering the disruptive power of quantum computing. On the government side, we notice a similar fast approach as can be seen from the DARPA  Quantum Benchmarking Initiative (QBI), which targets \gls{FTQC} in under 10 years~\cite{darpa:qbi}. This bullish approach will require risk mitigation as the technology is not yet mature and recent successes, while very promising, have been in the realm of experimental laboratories and the production chain is currently under development. The market is exploring many hardware technologies with superconducting, neutral atoms, and ions currently taking the lead. Alternative technologies like photonics, spin qubits, topological qubits, etc are also currently under exploration by various vendors. At the present, it remains too early to assess which approach will stand the test of time, and the coming years will be crucial in technology development and the transition from laboratory to production scale. 

We have superimposed the resources needed for the scientific applications collected from the literature with the vendors roadmaps and see convergence in the near future, particularity so considering algorithmic developments.  Indeed, on the algorithmic front, we notice resource requirements being greatly reduced with every development (see FeMoco in quantum chemistry). The combination of these two trends promises early fault tolerant quantum relevant workloads in the next five to ten years. Additionally, there is a high likelihood that new and heuristic impactful applications will be developed as better quantum systems become available to researchers.

Whether this utility will cover a reasonable fraction of \gls{HPC} workloads (e.g. around 50\%), or become a dedicated resource for certain workloads, remains to be seen, and further algorithmic development and hardware validation are needed in this direction. To get an initial estimate of timing and throughput, we selected a few applications from condensed matter, chemistry, and HEP and computed their runtime for quantum hardware clock speeds between 1 KHz and 1 GHz. We proposed the \gls{SQSP} as a metric to compare system-level throughput for a quantum system. Perhaps not surprisingly, the number of tasks completed varies wildly between different hardware specifications. This is an indication of the need for careful selection of applications for early \gls{FTQC}, where resources are expected to be limited.

In addition, the classical resources needed for error correction and hybrid algorithms with both quantum and classical resources in the loop, might require tight coupling with low latencies between \gls{HPC} and \gls{QC}~\cite{camps24}. This inevitably adds another layer of complication to the proposed vendor roadmaps. A tight collaboration between the public and private sector is key in overcoming these challenges and unlocking the quantum computing potential.


\chapter*{Acknowledgments}
\addcontentsline{toc}{chapter}{Acknowledgments}
We thank Akel Hashim for feedback on the manuscript. This research was supported by the U.S. Department of Energy (DOE) under Contract No. DEAC02-05CH11231, through the National Energy Research Scientific Computing Center (NERSC), an Office of Science User Facility located at Lawrence Berkeley National Laboratory.

\clearpage
\chapter*{List of Acronyms}
\label{sec:acronyms}
\addcontentsline{toc}{chapter}{List of Acronyms}
\begingroup
\renewcommand*{\glossarysection}[2][]{%
  \section*{#2}%
}
\printglossary[type=\acronymtype,title={}]
\endgroup

\clearpage
\chapter*{Glossary of Terms}
\label{sec:glossary}
\addcontentsline{toc}{chapter}{Glossary of Terms}
\begingroup
\renewcommand*{\glossarysection}[2][]{%
  \section*{#2}%
}
\printglossary[title={}]
\endgroup

\printbibliography[heading=bibintoc]

\end{document}